\documentclass[times,usenatbib]{mn2e}
\RequirePackage{graphicx}
\RequirePackage{subfigure}
\RequirePackage{rotating}

\newcommand{\hiir}{H~{\scshape ii}~region}
\newcommand{\hiirs}{H~{\scshape ii}~regions}
\newcommand{\um}{$\mu$m}
\newcommand{\degree}{^{\circ}}

\newcommand{\bubblecount}{5,106}

\title[Milky Way Project DR1]{The Milky Way Project First Data Release:  A Bubblier Galactic Disk\footnote{This publication has been made possible by the participation of more than 35,000 volunteers.}}

\author[Simpson et al.]
{\parbox{\textwidth}{R. J. Simpson$^{1}$\thanks{Email: robert.simpson@astro.ox.ac.uk},
M. S. Povich$^{2, 3}$,
S. Kendrew$^{4}$,
C. J. Lintott$^{1, 5}$,
E. Bressert$^{6,7,8}$,
K. Arvidsson$^{5}$,
C. Cyganowski$^{8,13}$,
S. Maddison$^{12}$
K. Schawinski$^{10, 11, 14}$,
R. Sherman$^{9}$,
A. M. Smith$^{1, 5}$,
G. Wolf-Chase$^{5, 9}$}\vspace{0.8cm}\\
\parbox{\textwidth}{
$^{1}$Oxford Astrophysics, Denys Wilkinson Building, Keble Road, Oxford, OX1 3RH, UK \\
$^{2}$Department of Astronomy and Astrophysics, The Pennsylvania State
University, \\ 525 Davey Laboratory, University Park, Pennsylvania, 16802, USA \\
$^{3}$NSF Astronomy and Astrophysics Postdoctoral Fellow \\
$^{4}$Max Planck Institut f\"{u}r Astronomie, K\"{o}nigstuhl 17, 69117 Heidelberg, Germany \\
$^{5}$Astronomy Department, Adler Planetarium, 1300 S. Lake Shore Drive, Chicago, IL 60605, USA \\
$^{6}$ESO, Karl-Schwarzschild-Strasse 2, D87548, Garching, Germany \\
$^{7}$School of Physics, University of Exeter, Stocker Road, Exeter EX4 4QL, UK \\
$^{8}$Harvard-Smithsonian Center for Astrophysics, 60 Garden St., Cambridge MA 02138, USA \\
$^{9}$Dept. of Astronomy \& Astrophysics, University of Chicago, 5640 S. Ellis Ave., Chicago, IL 60637, USA \\
$^{10}$Department of Physics, Yale University, P.O. Box 208121, New Haven, CT 06520, USA \\
$^{11}$Yale Center for Astronomy and Astrophysics, Yale University, P.O. Box 208121, New Haven, CT 06520, USA \\
$^{12}$Centre for Astrophysics \& Supercomputing, Swinburne University, H39, PO Box 218, Hawthorn, VIC 3122, Australia \\
$^{13}$NSF Astronomy and Astrophysics Postdoctoral Fellow \\
$^{14}$Einstein Fellow }}

\date{Draft form \today .}

\pagerange{000--000}

\begin{document}

\label{firstpage}

\maketitle

\begin{abstract}

We present a new catalogue of \bubblecount\ infrared bubbles created through visual classification via the online citizen science website `The Milky Way Project'. Bubbles in the new catalogue have been independently measured by at least 5 individuals, producing consensus parameters for their position, radius, thickness, eccentricity and position angle. Citizen scientists -- volunteers recruited online and taking part in this research -- have independently rediscovered the locations of at least 86\% of three widely-used catalogues of bubbles and \hiirs\, whilst finding an order of magnitude more objects. 29\% of the Milky Way Project catalogue bubbles lie on the rim of a larger bubble, or have smaller bubbles located within them, opening up the possibility of better statistical studies of triggered star formation. Also outlined is the creation of a `heat map' of star-formation activity in the Galactic plane. This online resource provides a crowd-sourced map of bubbles and arcs in the Milky Way, and will enable better statistical analysis of Galactic star-formation sites.

\end{abstract}

\section{Introduction}

\hiirs\ ionised by young, O and B-type stars provide the most readily observable tracers of star formation in the Milky Way and other galaxies. Ionised gas produces strong emission in optical and infrared (IR) recombination lines, forbidden lines and thermal (free-free) radio continuum. Dust mixed with ionised gas and heated by the hard radiation field makes \hiirs\ bright sources of thermal IR emission.

In the Milky Way, the spatial morphology of individual \hiirs\ can generally be resolved, revealing complex structures which are often shaped by the newly formed stars \citep{Anderson+11}. One particularly common and informative morphology is the presence of a rounded `bubble' of emission from gas excited presumably by a central source. 

\citet[hereafter CP06, CWP07]{CP06,CWP07} catalogued nearly 600 IR `bubbles' in the inner $130\degree$ of the Galactic plane by visually searching for ring-shaped structures (complete or partially broken) in 3.6--8.0~\um\ images from the Galactic Legacy Infrared Survey Extraordinaire \citep[GLIMPSE;][]{GLIMPSE,GLIMPSErev}. In the GLIMPSE-I survey area, CP06 catalogued 322 bubbles, of which 25\% corresponded to previously catalogued radio \hiirs\ and 75\% were attributed to late B-type stars with insufficient ionising luminosity to produce radio-bright \hiirs . \citet{Bania+10} later selected 24~\um\ diffuse emission sources from {\it Spitzer}/MIPSGAL survey images \citep{MIPSGAL} with spatially coincident 21 cm continuum emission for a radio recombination line survey with the Green Bank Telescope, and discovered several hundred new Galactic \hiirs\ in the region $-16\degree \le l \le 67\degree$ and $|b| \le 1\degree$, including those associated with 65 of the CP06 bubbles.

GLIMPSE images are particularly helpful in identifying bubbles as emission from polycyclic aromatic hydrocarbons (PAHs) at 8~\um\ demarcates the bubble rims in {\it Spitzer Space Telescope} images, while a second peak, arc, or torus of 24~\um\ emission from warm dust is frequently observed inside the bubble with a morphology that closely traces the radio continuum emission \citep[CP06;][]{WP08, WHM+10}. This qualitative pattern of bright 8~\um\ emission shells surrounding regions of bright 24~\um/radio emission is also characteristic of giant Galactic \hiirs\ and extragalactic star-forming regions \citep[e.g.][]{Povich+07,Bendo+08,Relano+Kennicutt09,Flagey+11}.

Nevertheless, existing surveys are not sufficient to establish the true relationship between bubbles and HII regions. Both CP06 and CWP07 stressed that their bubble catalogues were far from complete, particularly with regard to bubbles with large (${>}10'$) and small (${<}2'$) angular diameters, and the majority of previously catalogued, bright radio \hiirs\ \citep{Paladini+03} in the GLIMPSE survey area were not associated with catalogued bubbles. None of the large bubbles identified by \citet{Rahman+Murray10} were included in the CP06 and CWP07 catalogues, nor was the large bubble associated with the well-studied giant \hiir\ M17 identified by \citet{Povich+09}. These large bubbles were missing from the catalogues because they are generally faint, broken, and confused with smaller, brighter \hiirs.

In this paper, we present a new catalogue of \bubblecount\ IR bubbles identified via visual inspection of the GLIMPSE and MIPSGAL survey images as part of the citizen science Milky Way Project\footnote{http://www.milkywayproject.org} (MWP). Our catalogue expands the CP06 and CPW07 catalogues by nearly an order of magnitude and hence represents a far more complete sample of Galactic \hiirs. Three key advances in the bubble selection process enabled the identification of bubbles that were missed by CP06 or CWP07: (1) We enlisted ${>}35,000$ volunteers to examine the images rather than relying on only a handful of experts; (2) we incorporated  MIPSGAL 24~\um\ data, which greatly facilitates the identification of bubbles compared to relying primarily on 8.0 \um\ data; and (3) we relaxed the criteria of what defines a `bubble' to include more incomplete shells and arc-shaped structures. The remainder of this paper is organised as follows: In sections 2 and 3 we describe the MWP and the construction of our catalogue. Some initial results from the catalogue are presented in section 4 and discussed in section 5. We summarise our conclusions and directions for future work in section 6.

\section{The Milky Way Project}
\label{mwp-desc}

Galaxy Zoo \citep{Lintott+08, Lintott+11} and the larger suite of Zooniverse\footnote{http://www.zooniverse.org} projects \citep{2011MNRAS.412.1309S}, have successfully built a large community of volunteers\footnote{Over 480,000 registered volunteers at time of writing.} eager to participate in scientific activities. 

The Zooniverse has shown that enlisting `citizen scientists' via the internet is a powerful way to analyze large amounts of data. Human brains excel at pattern recognition tasks, and most people will reach a level of accuracy as high as any expert after a brief introduction. By enlisting citizen scientists, researchers can extend visual classification to large samples of images, having each image examined by a large number of independent classifiers. This allows researchers to tap into the `wisdom of the crowd' effect where the consensus of a group of non-experts is often more accurate than the testimony of a single expert.

\begin{figure}
\includegraphics[width=0.48\textwidth]{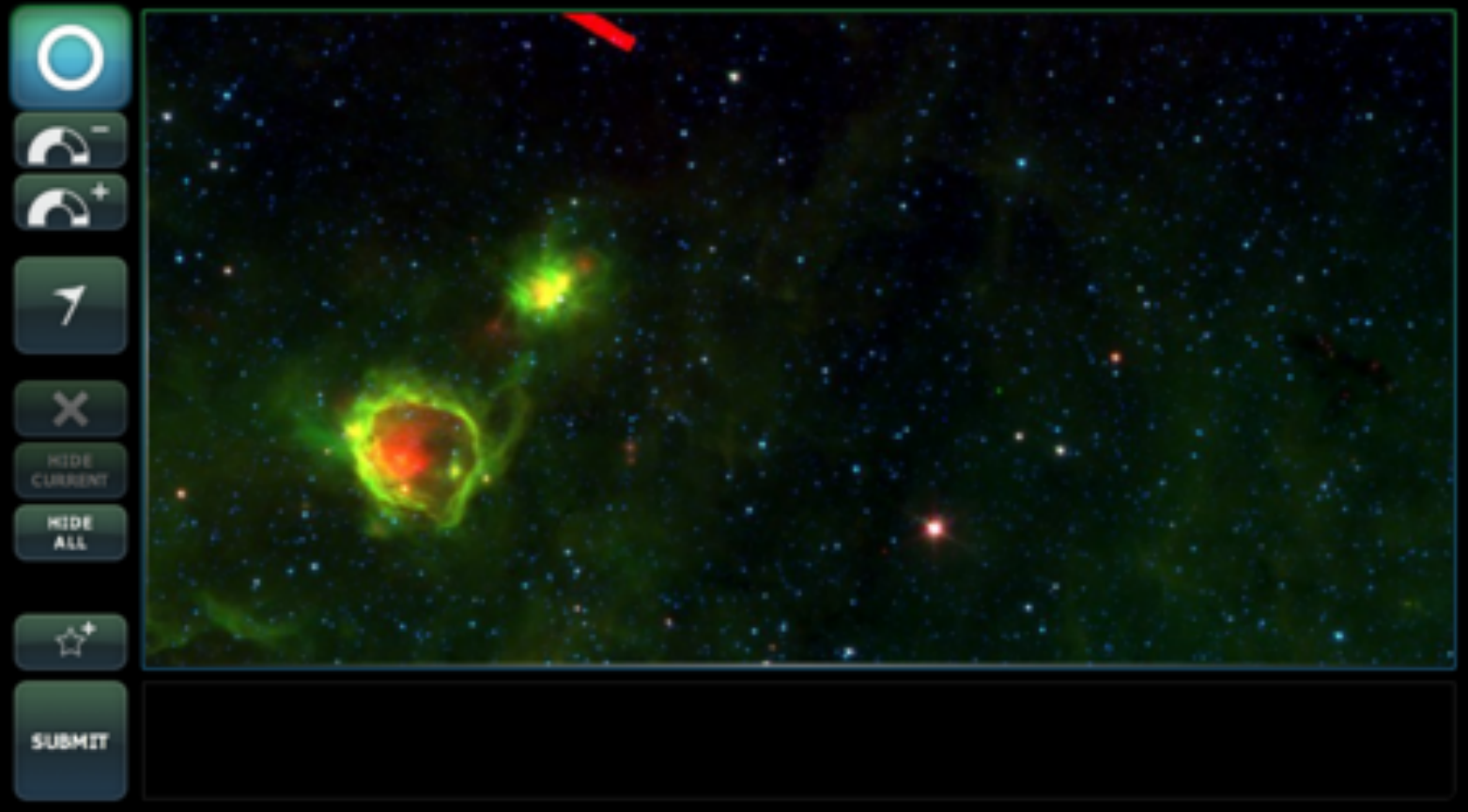}
\caption{Screenshot of the Milky Way Project user interface. Colour figure available online.}
\label{user-interface}
\end{figure}

Another advantage of enlisting human classifiers is their ability to recognize unusual objects which computer search algorithms  may be unable to spot. This has been shown by the serendipitous discovery of Hanny's Voorwerp \citep{2009MNRAS.399..129L} and the case of the Galaxy Zoo `Green Peas' \citep{2009MNRAS.399.1191C}. Spitzer GLIMPSE data is ideally suited to classification by citizen scientists as the amount of data is large and the images contain complex, overlapping structures that are impossible to disentangle using automated algorithms. The task of recognising bubbles may eventually be handled by advanced machine-learning algorithms \citep[e.g.][]{Beaumont+11} but in the meantime the community of Zooniverse users are keen to contribute to astronomy and science \citep{2010AEdRv...9a0103R}.

The MWP is the ninth online citizen science project created using the Zooniverse Application Programming Interface (API) toolset. The Zooniverse API is the core software supporting the activities of all Zooniverse citizen science projects. Built originally for Galaxy Zoo 2, the software is now being used by 11 different projects. The Zooniverse API is designed primarily as a tool for serving up a large collection of `assets' (for example, images or video) to an interface, and collecting back user-generated interactions with these assets.

The assets in the MWP are multiband, false-colour JPEG images, created by gridding the {\it Spitzer} GLIMPSE and MIPSGAL mosaics into smaller images at three different zoom levels. The highest zoom level provides users with tiles of 0.3$^{\circ}\times$0.15$^{\circ}$, and at a resolution of 800$\times$400 pixels these tiles nearly reproduce the $1.2''$ pixel scale of the GLIMPSE survey images. Larger tile sizes of 0.75$^{\circ}\times$0.375$^{\circ}$ and 1.5$^{\circ}\times$0.75$^{\circ}$ were also generated. The tiles were plotted in an overlapping grid to allow all parts of the inner Galactic plane ($|l|\le 65\degree$, $|b|\le 1\degree$) to be viewed by the MWP users, at all zoom levels. To provide an optimal representation of the dynamic range within each tile, each of 3 single-band images was independently scaled to a square-root stretch function (with the faintest 5\% of image pixels clipped to black and the brightest 0.2\% clipped to white), assigned to a colour channel (red=24~\um, green=8.0~\um, blue=4.5~\um), and finally composited into a 3--colour image. The MIPSGAL 24~\um\ mosaics frequently saturate in regions of bright nebulosity, and saturated 24~\um\ pixels were set to maximum red to preserve the visual appeal of the images and to avoid presenting MWP users with saturation artefacts. The resulting composite images allow visual identification of both bright and faint features within a given image tile.

The MWP user interface (see Figure~\ref{user-interface}) was built using Flash, based upon the pre-existing Moon Zoo interface \citep{Joy+11}. Volunteers are primarily encouraged to draw ellipses onto the image to mark the locations of bubbles. A short, online tutorial shows how to use the tool, and examples of prominent bubbles are given\footnote{http://www.milkywayproject.org/tutorial}. As a secondary task, users can also mark rectangular areas of interest, which can be labelled as small bubbles, green knots, dark nebulae, star clusters, galaxies, fuzzy red objects or `other'. Examples of these are also given in a tutorial on the website, and these are discussed further in Section~\ref{other-objects}. Users can add as many annotations as they wish before submitting the image, at which point they are given another image for annotation. Each image's annotations are stored in a database as a classification, and users can see the images they have classified in a part of the site called `My Galaxy'. Users can only classify a given image once.

When marking bubbles, users place a circular annulus that can be scaled in size and stretched into an elliptical annulus. As they first draw out an object, they are able to control the position and size of the bubble. Once the bubble has been drawn they can edit these initial parameters as well as the bubble's ellipticity, annular thickness and rotation. In this way users can attempt to match the bubbles they see in each image to give an accurate representation. Users are also able to mark regions of the annulus where there is no obvious emission -- as in the case of broken or partial bubbles. These `cut-outs' are created by erasing (and then re-filling if necessary) segments of the annulus. Of all the bubbles drawn, 75\% have thicknesses other than the default or minimum values, 50\% are non-circular (of which 56\% have been rotated) and only 12\% were drawn with cut-outs.

When marking another area of interest on an image (e.g. star clusters, green knots, etc) users simply draw rectangles. Since these objects are secondary to the main bubble-finding task, the site was designed so that they should be simple and quick to mark. Simple rectangles allow us to record the positions and approximate sizes of any interesting objects.

Citizen scientists can discuss and share objects and images via the `Talk' interface\footnote{http://talk.milkywayproject.org} where more heavily discussed objects trend upwards as they do, for example, in a news aggregator. The `Talk' web application is open source\footnote{https://github.com/zooniverse/Talk} and was developed by the Zooniverse team for general use on its projects, including the MWP. Through the use of `Talk', interesting objects float to the top of discussion and are identified as interesting to the MWP scientists. Thus, by harnessing the social nature of the MWP, we can extract additional information from the classification process.

Principally the objects highlighted by volunteers using `Talk' are visually interesting, unusual or defy classification in the primary interface. One example of such a feature are the `yellowballs' - named by users because of their compact, rounded, yellow appearance in MWP images. It is believed that these represent a type of ultra-compact \hiir\ and they are the subject of a future paper currently being prepared. They were an unintended consequence of our colour scheme and thus no flag was provided in the main interface to enable volunteers to mark their presence.

\subsection{Users' favourite bubbles}

\begin{figure}
\includegraphics[width=0.48\textwidth]{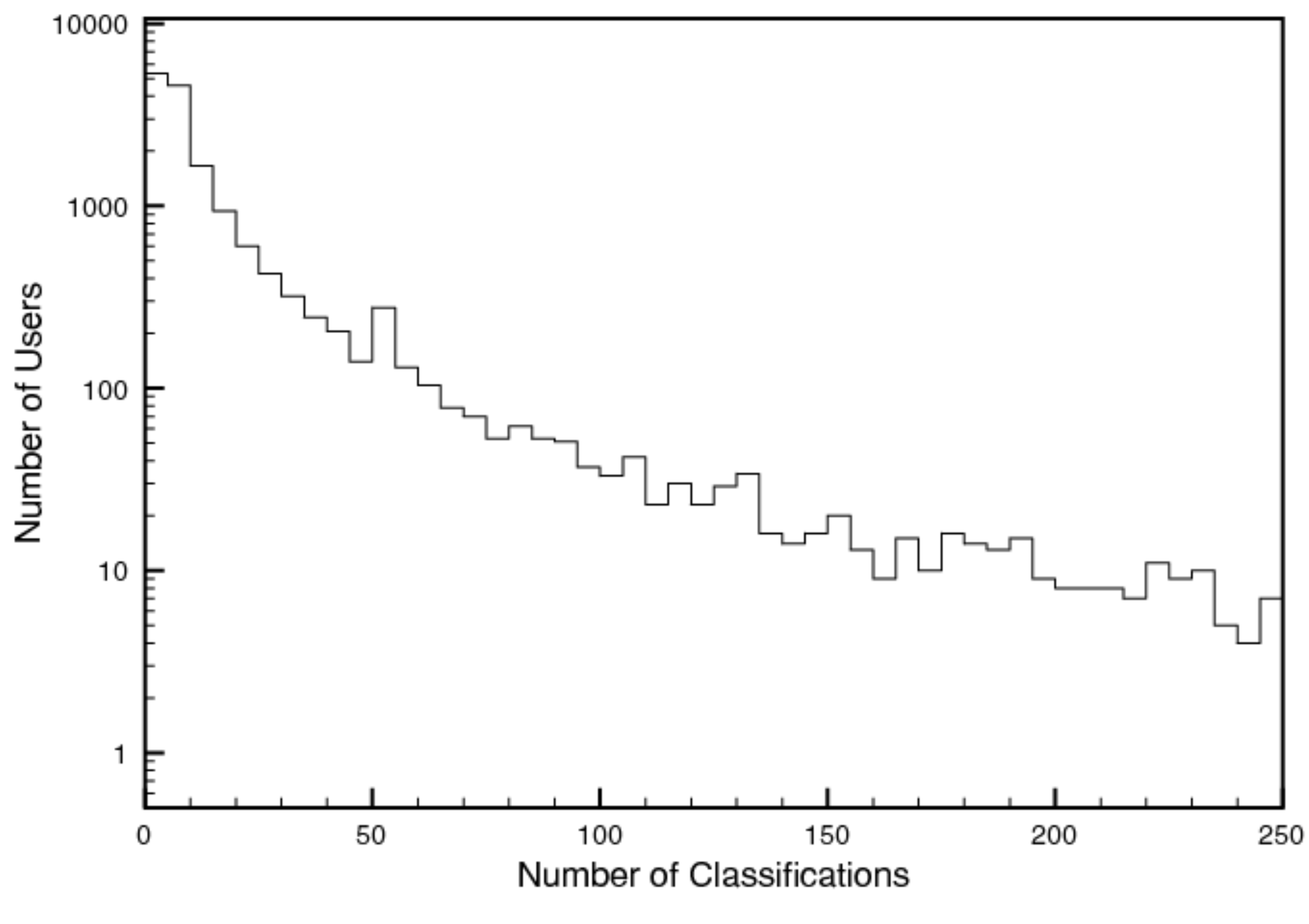}
\caption{Distribution of users with number of classifications.}
\label{user-classifications}
\end{figure}

\begin{figure*}
\begin{center}
\subfigure[$l$=18.8$^{\circ}$, $b$=-0.13$^{\circ}$, $zoom$=1]{
\includegraphics[width=0.45\textwidth]{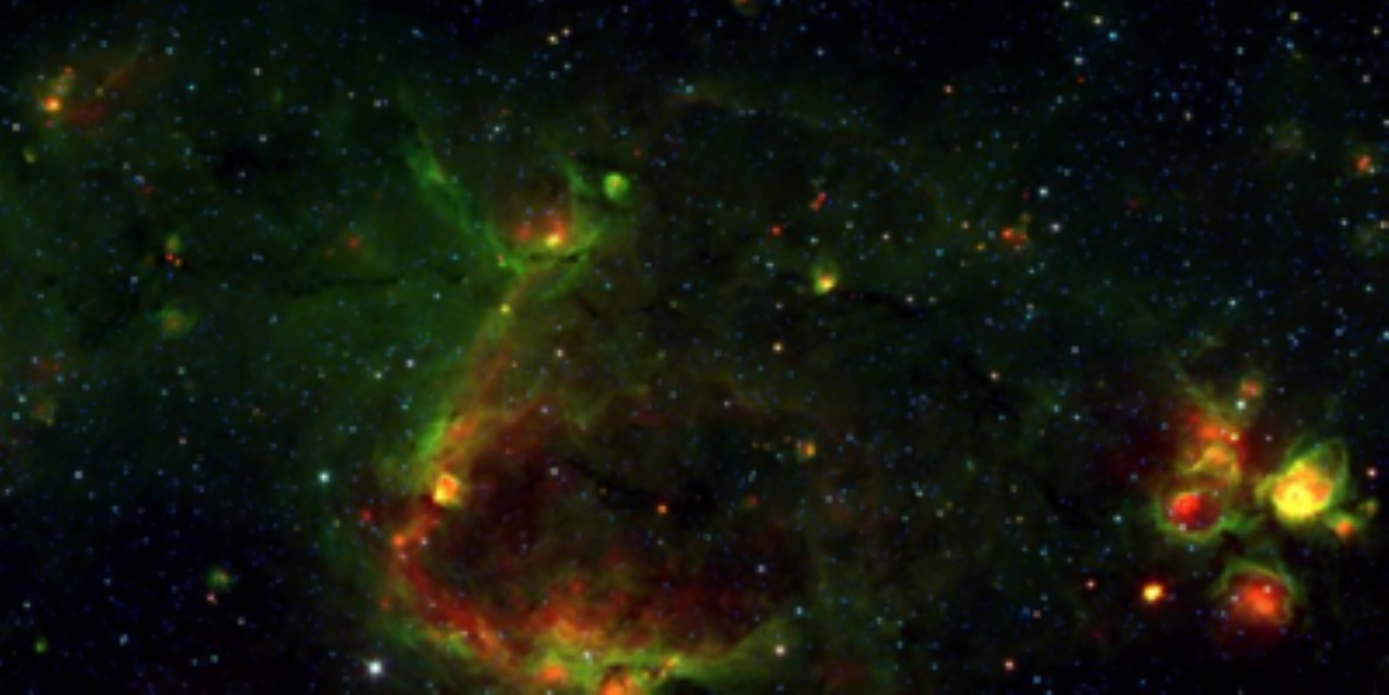}}
\subfigure[$l$=16.83$^{\circ}$, $b$=0.69$^{\circ}$, $zoom$=2, the Eagle Nebula]{
\includegraphics[width=0.45\textwidth]{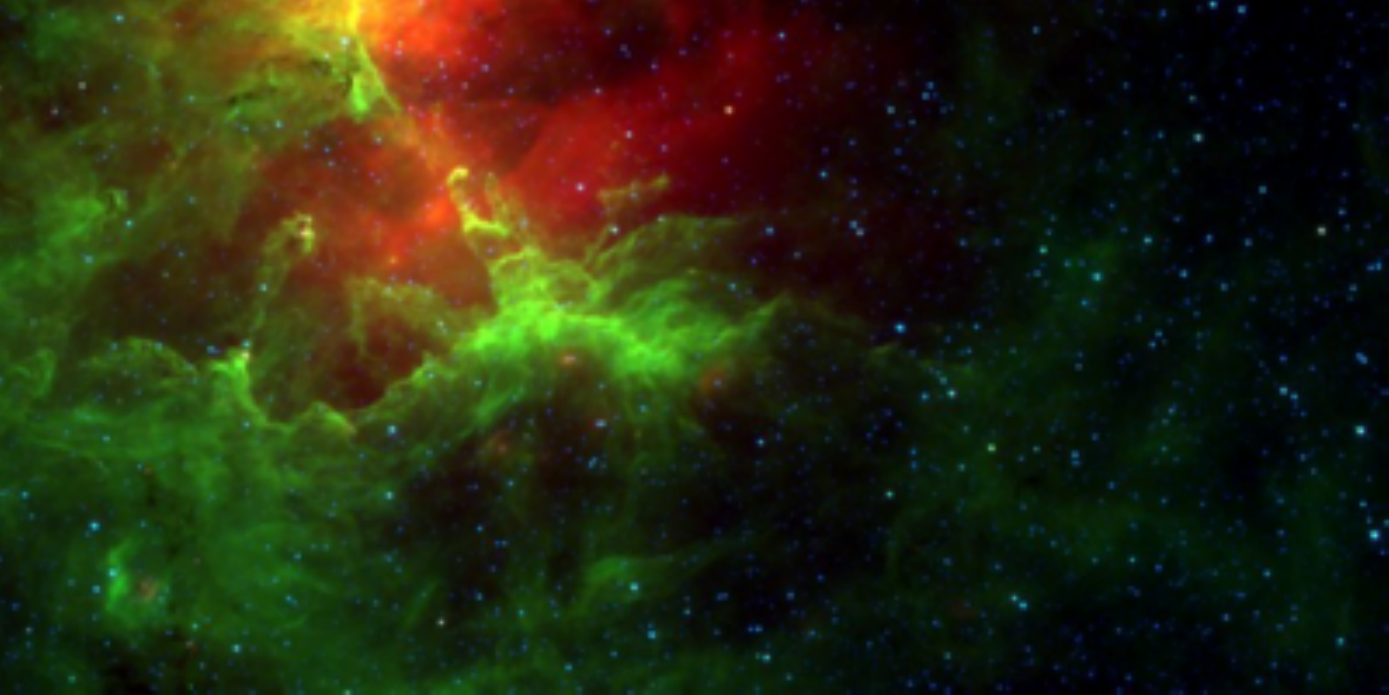}}
\subfigure[$l$=7.0$^{\circ}$, $b$=-0.28$^{\circ}$, $zoom$=3, the Trifid Nebula]{
\includegraphics[width=0.45\textwidth]{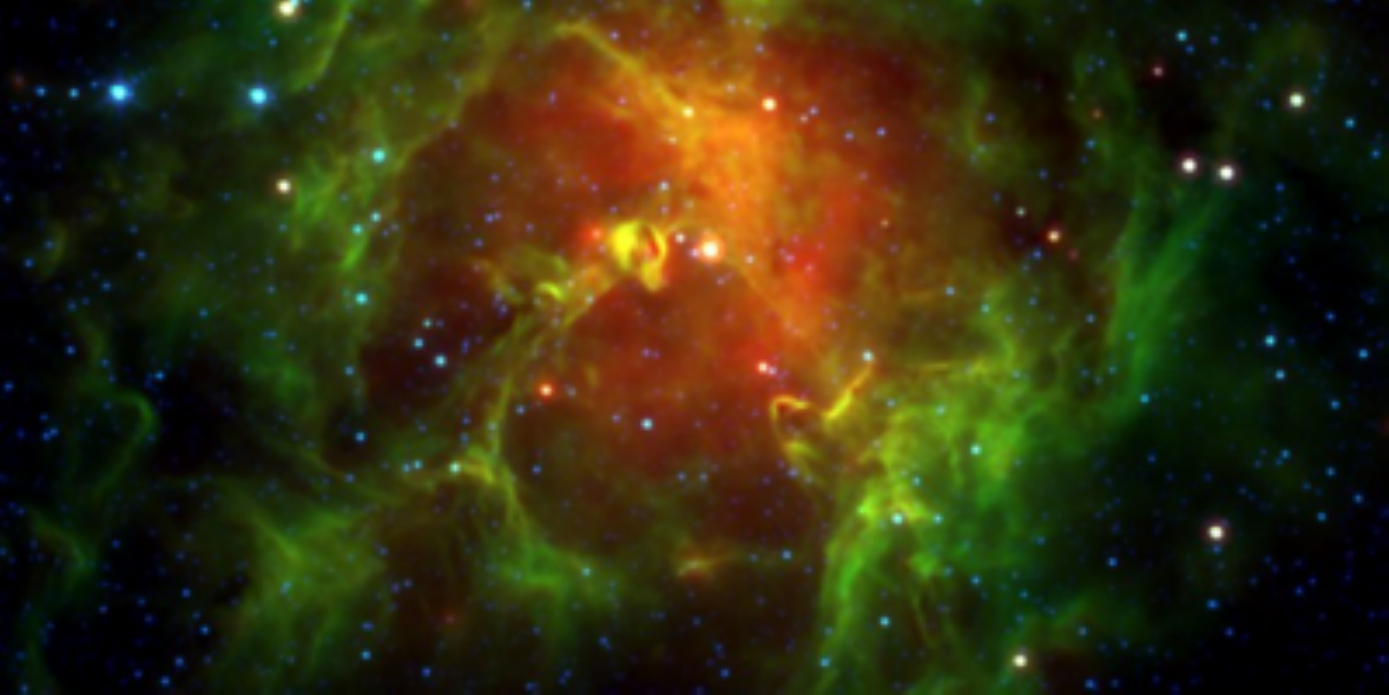}}
\subfigure[$l$=317.2$^{\circ}$, $b$=0.13$^{\circ}$, $zoom$=1]{
\includegraphics[width=0.45\textwidth]{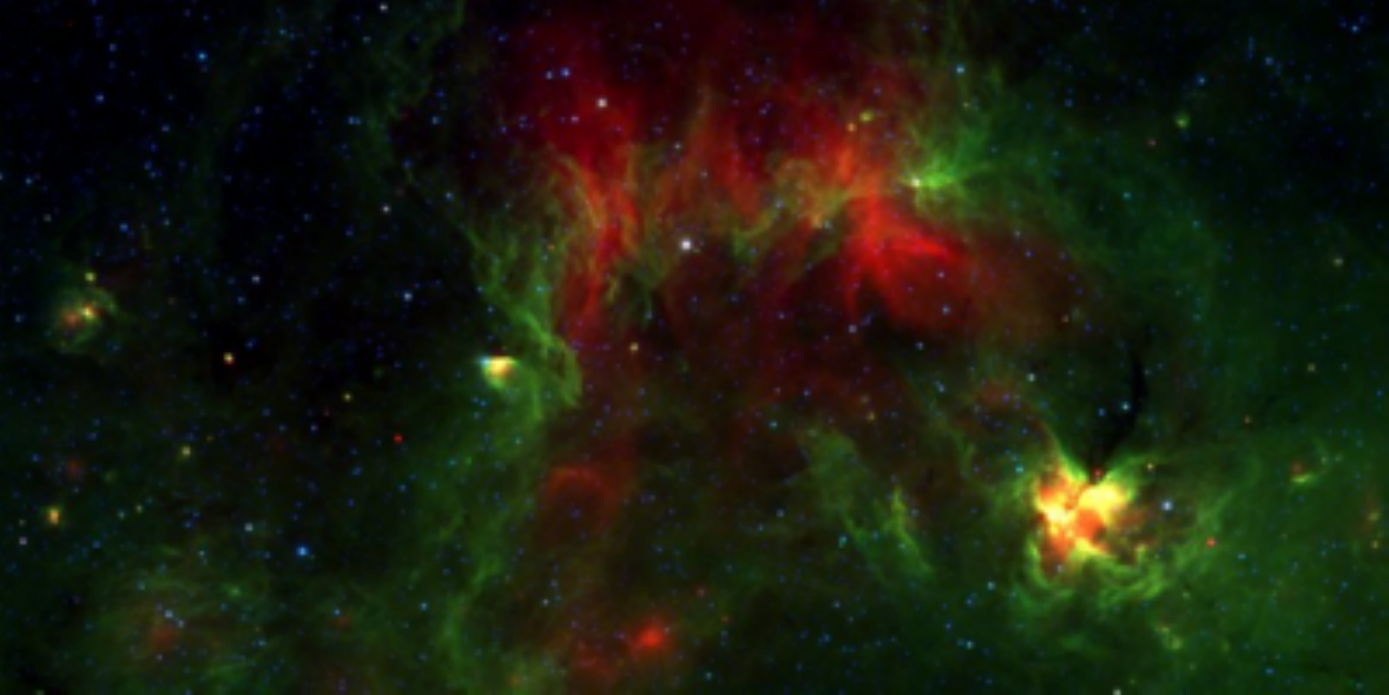}}
\subfigure[$l$=60.1$^{\circ}$, $b$=-0.28$^{\circ}$, $zoom$=3]{
\includegraphics[width=0.45\textwidth]{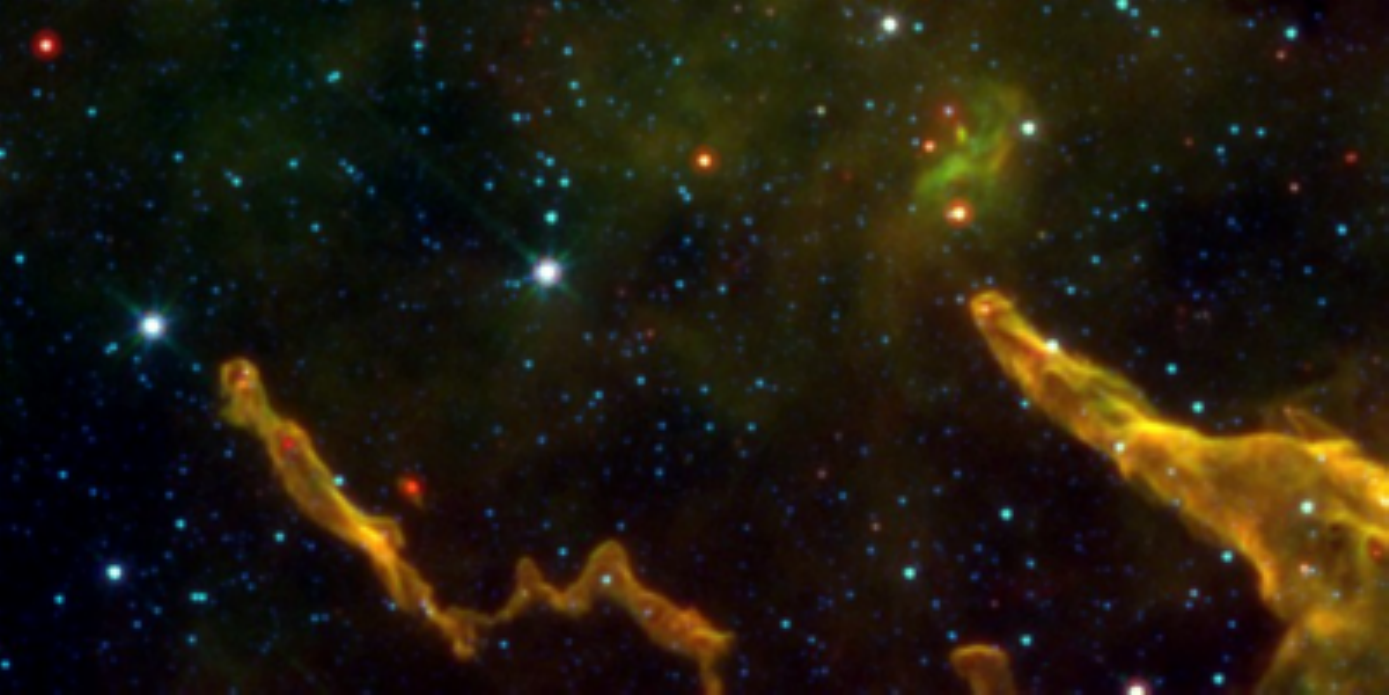}}
\subfigure[$l$=59.8$^{\circ}$, $b$=0.03$^{\circ}$, $zoom$=3]{
\includegraphics[width=0.45\textwidth]{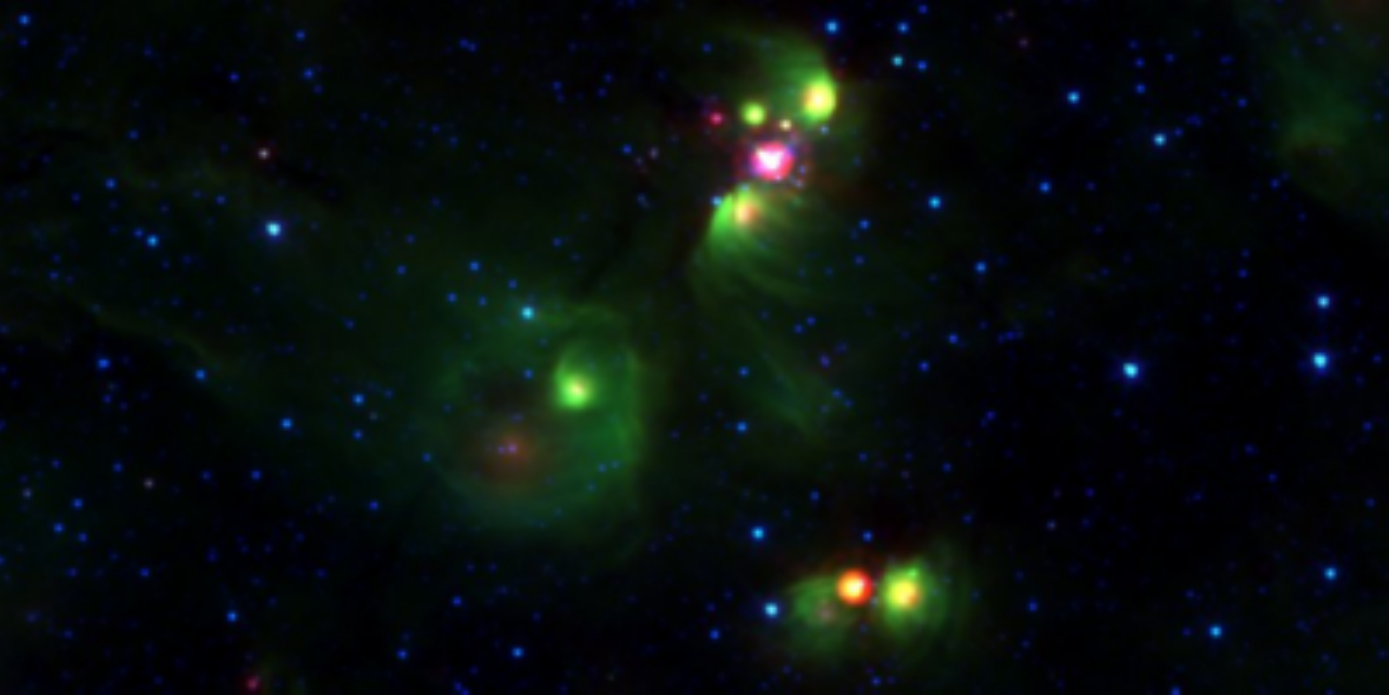}}
\subfigure[$l$=48.7$^{\circ}$, $b$=-0.43$^{\circ}$, $zoom$=3]{
\includegraphics[width=0.45\textwidth]{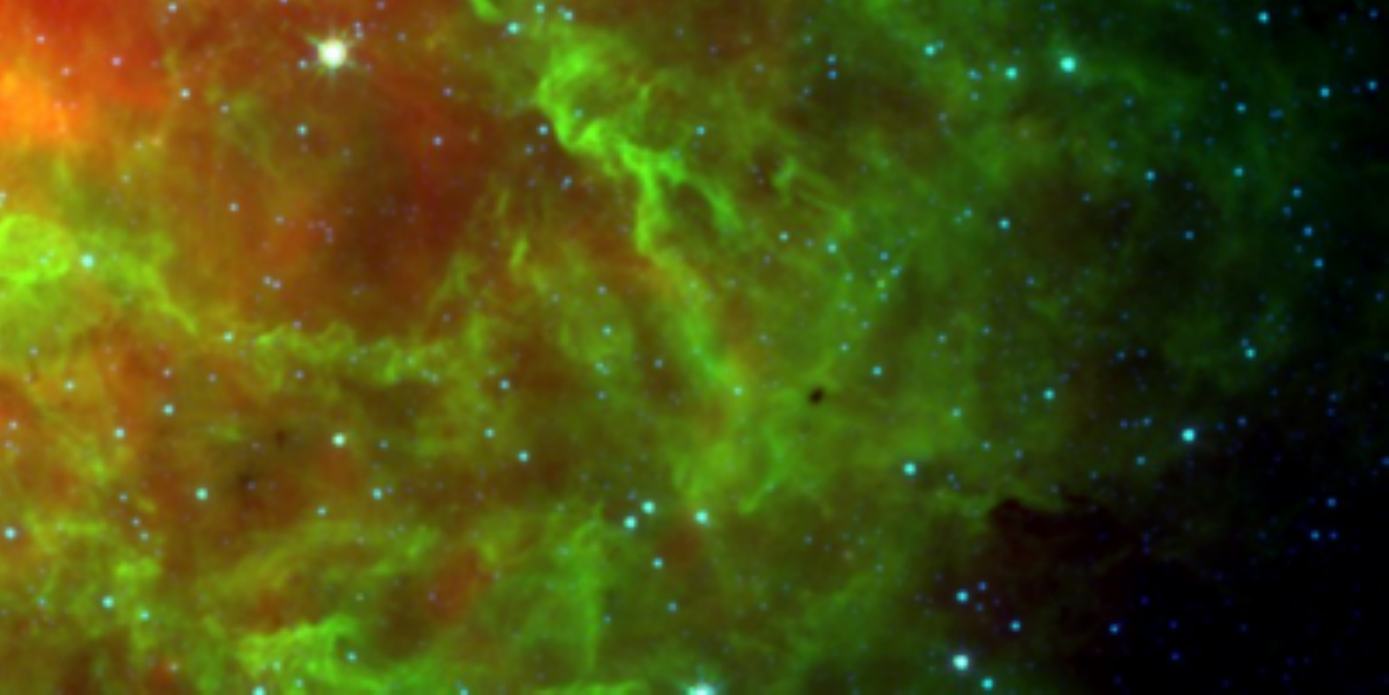}}
\subfigure[$l$=359.58$^{\circ}$, $b$=-0.06$^{\circ}$, $zoom$=2, towards Galactic centre]{
\includegraphics[width=0.45\textwidth]{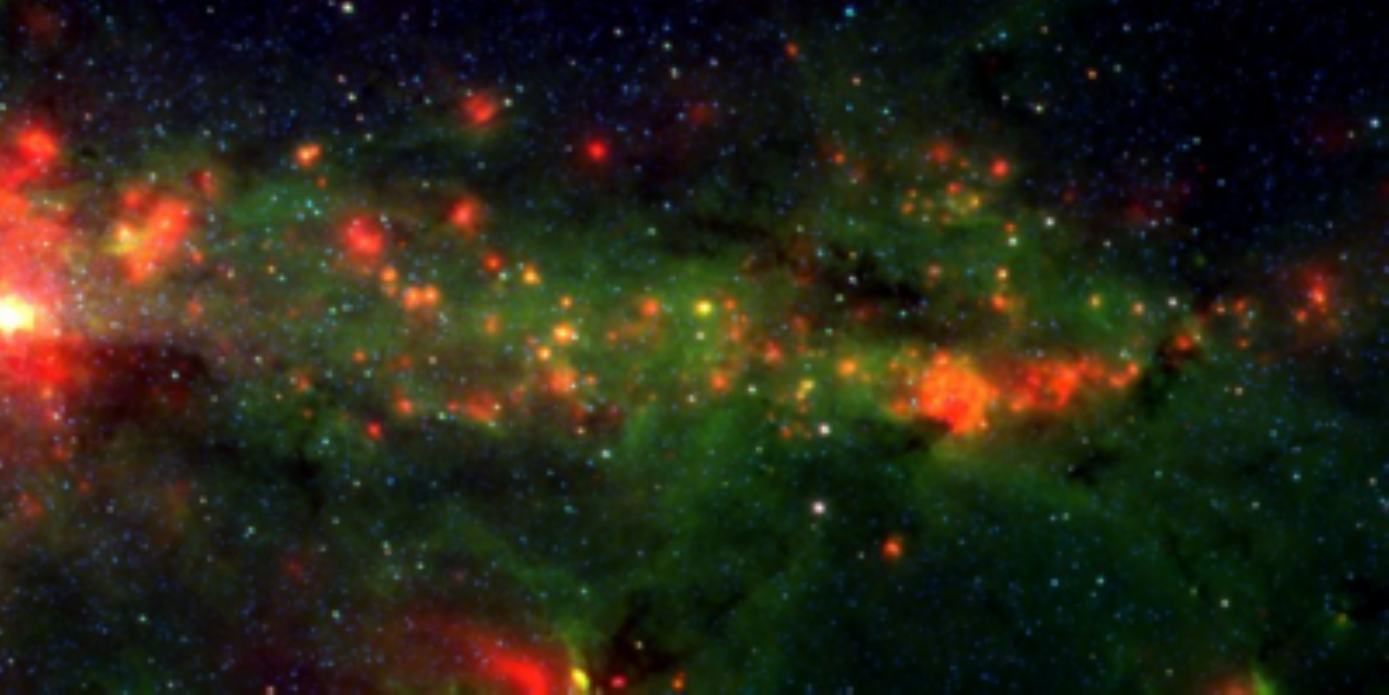}}
\subfigure[$l$=339.8$^{\circ}$, $b$=-0.13$^{\circ}$, $zoom$=1, contains Westerlund 1]{
\includegraphics[width=0.45\textwidth]{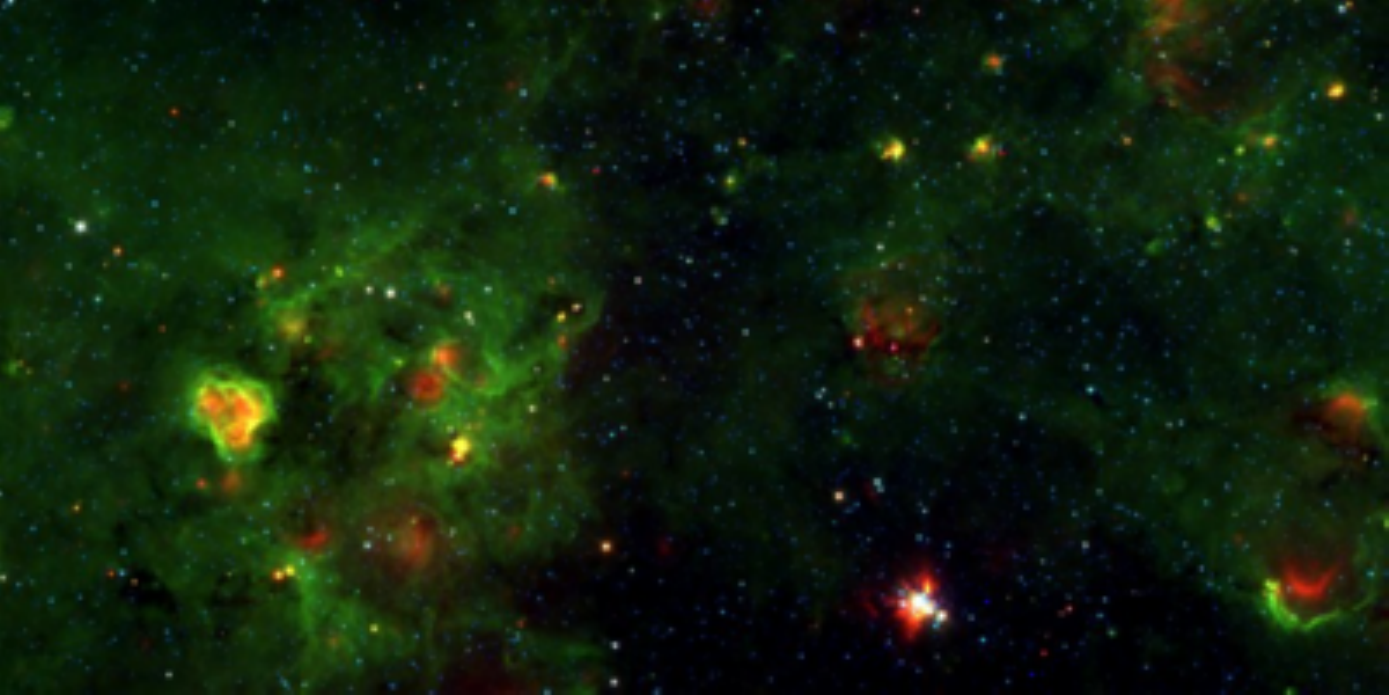}}
\subfigure[$l$=51.8$^{\circ}$, $b$=0.63$^{\circ}$, $zoom$=1]{
\includegraphics[width=0.45\textwidth]{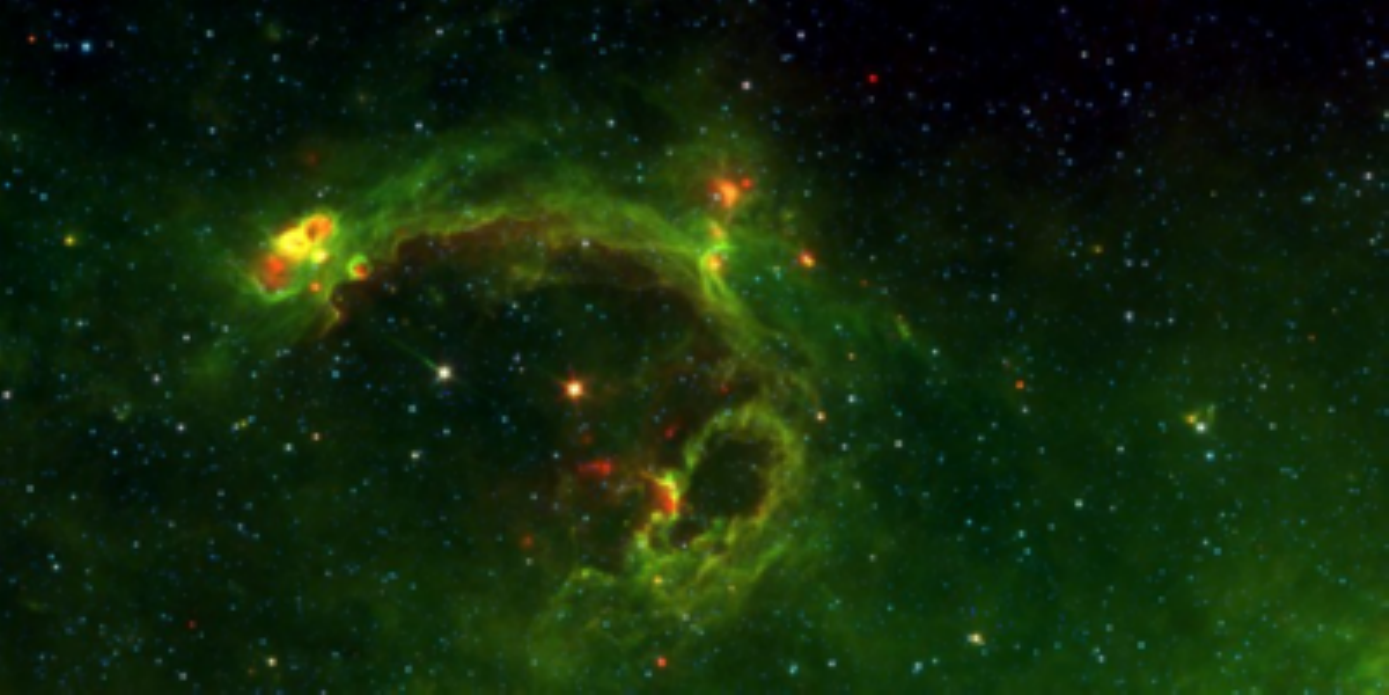}}
\caption{Ten Milky Way Project images most-favourited by volunteers, in no particular order. Coordinates are image centres, image sizes are indicated by the zoom level ($zoom$). Zoom levels 1, 2 and 3 refer to images of $1.5\times1^{\circ}$, $0.75\times0.375^{\circ}$ and $0.3\times0.15^{\circ}$ respectively. Colour figure available online.}
\label{favourite-images}
\end{center}
\end{figure*}

Users are able to mark images as `favourites' as they classify on the MWP. These images are often particularly beautiful, interesting or unusual. Ten of the most-favourited images are shown in Figure~\ref{favourite-images}. The list includes many notable and well-known objects that have been `rediscovered' by the MWP users, such as the Eagle Nebula ($b$) the Trifid Nebula ($c$), and the Galactic centre ($h$). Users not only select beautiful images but also those that contain interesting objects. Image $i$ contains the massive star cluster Westerlund 1 \citep{Westerlund+61}, which is much-discussed on Milky Way Talk (see Section \ref{mwp-desc}).

\section{Catalogue construction}

\subsection{MWP user statistics and scoring}
\label{scoring-section}

By October 2011, over 35,000 people had logged in to milkywayproject.org. 45\% of those users classified at least one of the 12,263 MWP images; 25\% classified 5 or more images and 5.7\% classified 50 or more. Figure~\ref{user-classifications} shows the distribution of users with the number of classifications they perform over the lifetime of their involvement with the project.

The top-five nations visiting the Milky Way Project have been the United States (42\%), United Kingdom (20\%),  Canada (4.3\%), Poland (3.8\%) and Germany (3.8\%). The remaining 26.1\% of project visitors have come from 173 other countries. The MWP site is predominately accessed by English-speaking countries with a high level of internet connectivity, in addition it has been translated into Polish and hence that country also provides a large proportion of visitors.

In order to assist in the data-reduction process, users are given scores according to how experienced they are at drawing bubbles. We treat the first 10 bubbles a user draws as practice drawings and these are not included in the final reduction. Users begin with a score of 0 and are given scores according to the number of precision bubbles they have drawn (see Table~\ref{user-scores}). Precision bubbles are those drawn using the full toolset, meaning they have to have adjusted the ellipticity, the thickness and the rotation. This is done to ensure that users' scores reflect their ability to draw bubbles well. While only precision bubbles are used to score volunteers, all bubbles drawn as included in the data reduction. The scores are used as weights when averaging the bubble drawings to produce the catalogue.

\begin{table}
\caption{Table showing a user's score given the number of precision bubbles they have drawn. Precision bubbles are those that require the use of multiple modifications to the default bubble parameters (see text for details). As such the number of precision bubbles drawn is a proxy for the care users are willing to take thus their experience with the tool.}
\begin{center}
\begin{tabular}{cc}
\hline
Precision Bubbles & Score\\
\hline
1 & 1 \\
5 & 2 \\
20 & 4  \\
50 & 6 \\
100 & 8 \\
500 & 10 \\
\hline
\end{tabular}
\end{center}
\label{user-scores}
\end{table}

\subsection{Combining the user-drawn bubbles}
\label{combine-section}

\begin{figure}
\includegraphics[width=0.48\textwidth]{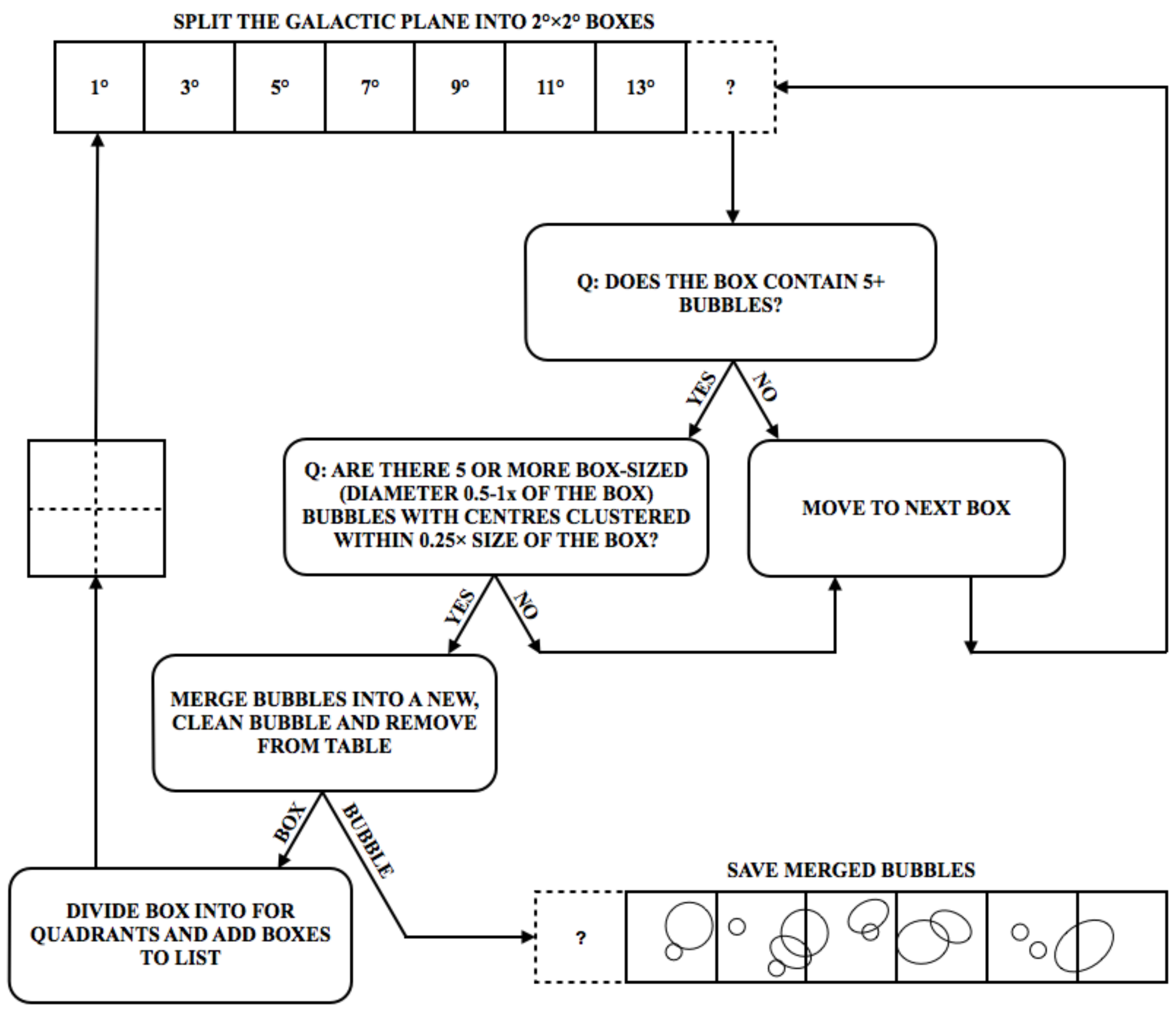}
\caption{Flow chart describing the way that similar bubbles are found and combined.}
\label{bubble-flow-chart}
\end{figure}

Combining all bubble drawings, at all zoom levels, the MWP has created a database of 520,120 user-drawn bubbles as of Oct 31st 2011. To identify bubbles with at least 5 user classifications for inclusion in our final catalogue, the data reduction process (shown as a flow chart in Figure~\ref{bubble-flow-chart}) begins by splitting the dataset into $2^{\circ}{\times}2^{\circ}$ boxes and treating each box in turn.

If a box contains 5 or more bubbles with a maximum outer ellipse that is between a half- and a whole-box, then a simple clustering algorithm picks out groups of these bubbles with dispersions in their positions of less than a quarter of the box size (i.e. less than the radius of the smallest bubble under consideration). If a cluster contains at least 5 bubble drawings, it is saved for additional processing and inclusion in the catalogue, and the bubble drawings are removed from the working list. Bubble drawings that are not clustered enough, or numerous enough, remain on the working list for potential inclusion in a later iteration. The box is then split into four and the process repeats until no more boxes containing 5 or more bubbles are found, or until the box size falls below the smallest bubbles drawn on the MWP -- the ellipse-drawing tool has a lower size limit of a diameter of 20 pixels ($0.45'$ at the highest image zoom level).

The same process is also run on an offset grid where the initial boxes are displaced by 1~degree in both galactic latitude and longitude. This catches bubbles that may fall on box boundaries. The two resultant lists of bubble-groups are combined later on by clustering bubbles that fall within 0.5 radii separation from each other and which both have radii within 50\% of each other.

\begin{figure}
\begin{center}
\subfigure[Image: GLIMPSE only, as in CP06]{
\includegraphics[width=0.45\textwidth]{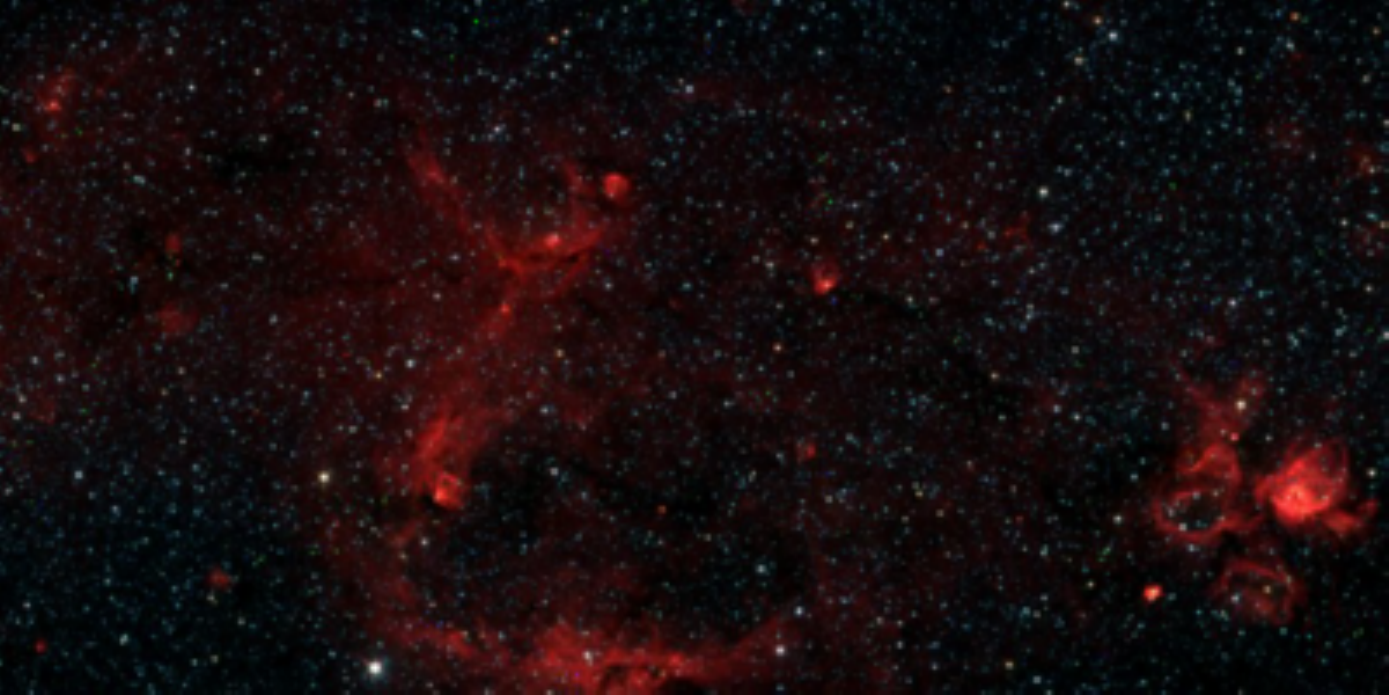}}
\subfigure[Image: GLIMPSE+MIPSGAL, from the MWP]{
\includegraphics[width=0.45\textwidth]{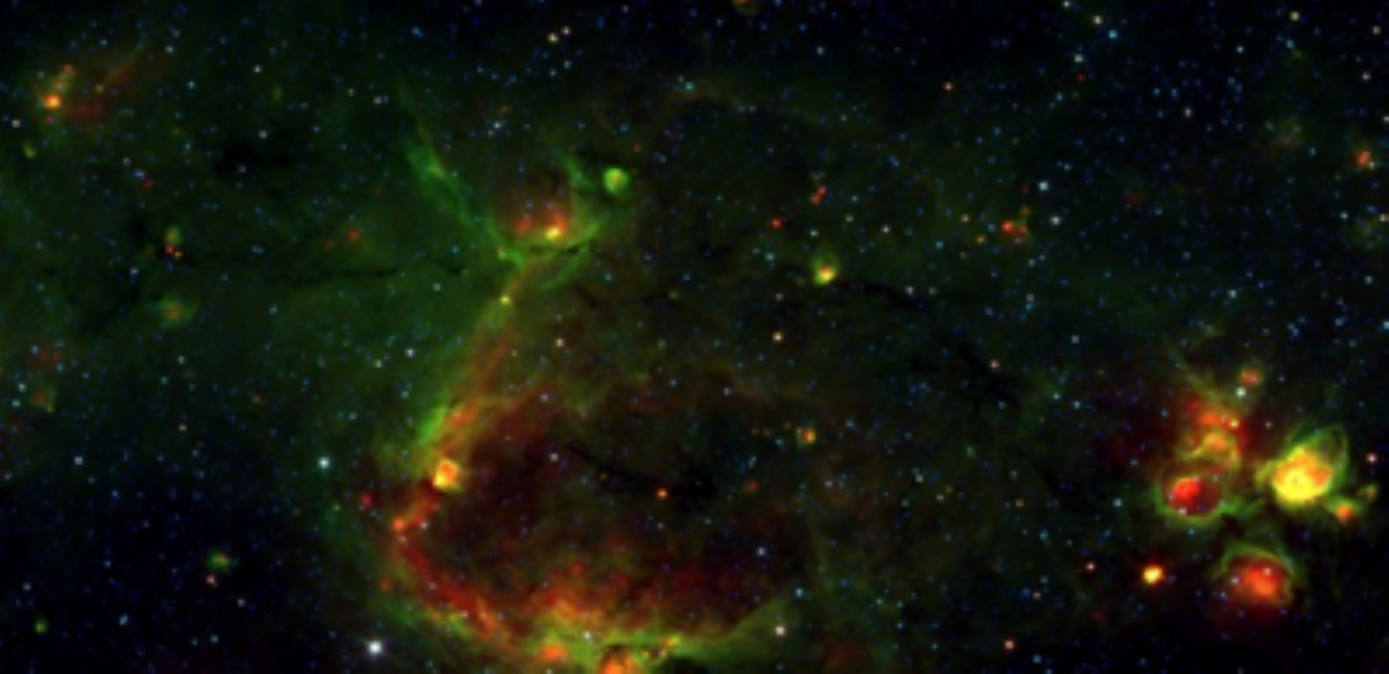}}
\subfigure[Raw User Drawings (`Heat Map')]{
\includegraphics[width=0.45\textwidth]{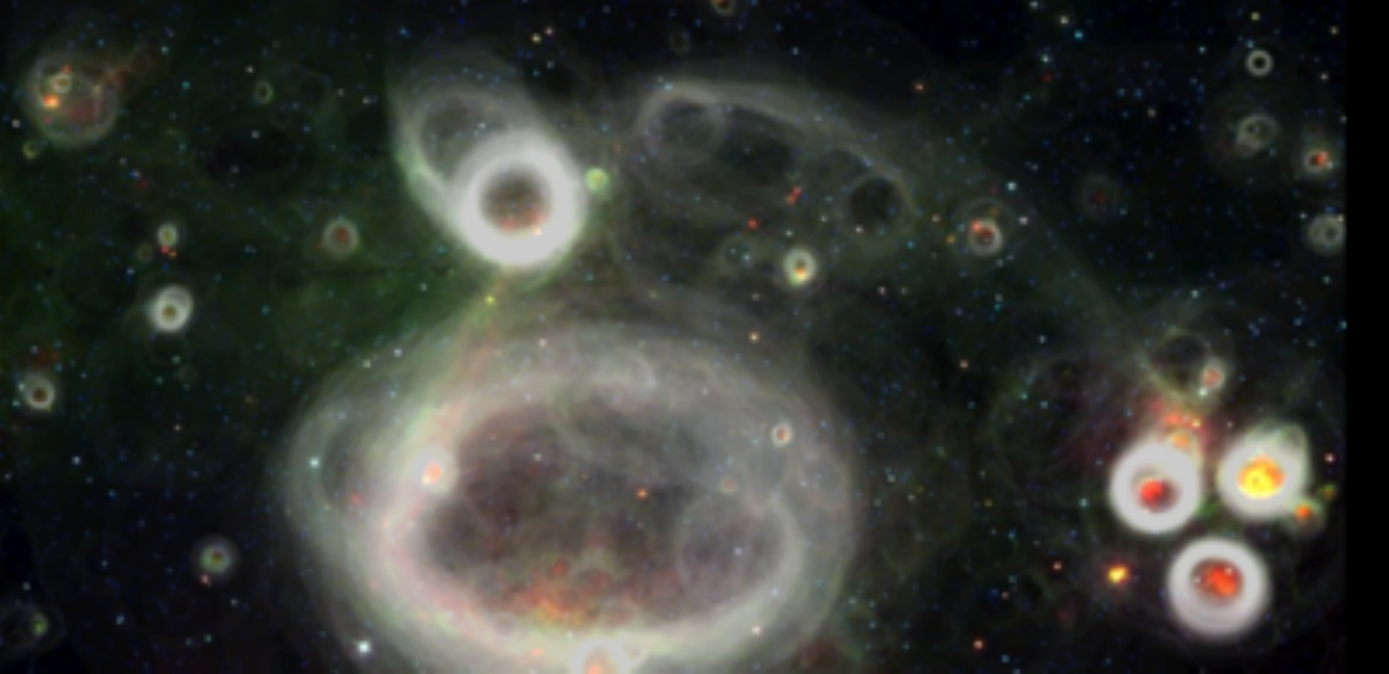}}
\subfigure[Reduced, `cleaned' bubbles.]{
\includegraphics[width=0.45\textwidth]{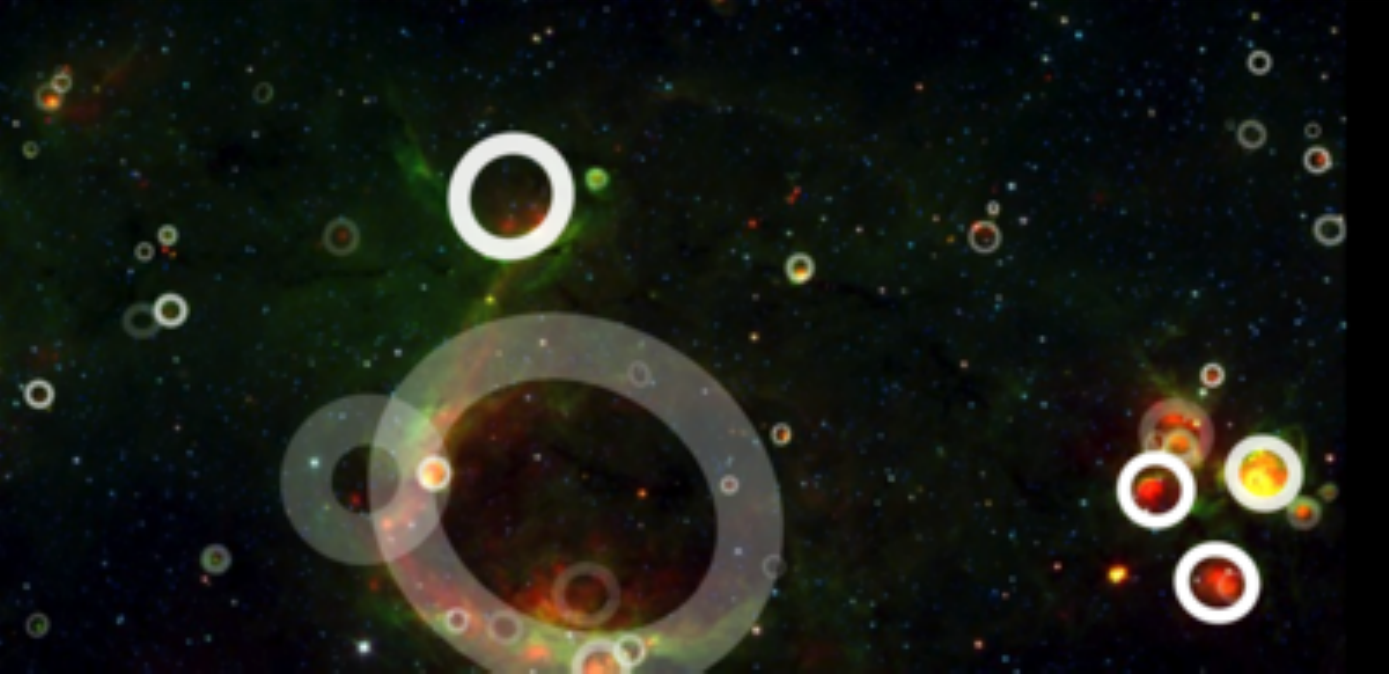}}
\caption{Example of raw user drawings and reduced, cleaned result using a sample MWP image. A GLIMPSE-only colour sample is included to illustrate the differences in the appearance of images inspected by CP06 and the MWP users. Image shown is centred at $l$=18.8$^{\circ}$, $b$=-0.125$^{\circ}$, with size 1.5$^{\circ} \times 0.75^{\circ}$. In image (c) all user-drawn bubbles are placed, from all zoom levels, with an opacity of 2.5\%. In image (d) reduced bubbles are placed with an opacity $2 \times$ their hit rate, such that bubbles with hit rates $\geq 50\%$ are drawn as a solid white bubbles. Colour figure available online.}
\label{reduction-example}
\end{center}
\end{figure}

Each resulting cluster of bubbles marked $\ge$ 5 times is combined into a single `clean' bubble using a weighted mean, where the weighting is provided by the score of the user that drew each bubble (see Section~\ref{scoring-section}). The bubble's mean size, position, angle (in degrees from North, in galactic coordinates) and thickness are all determined in this way (see Figure~\ref{reduction-example}). The cleaned bubble catalogue is given in Table~\ref{bubble-sample}.

The bubble's `hit rate' is the ratio of the number of qualifying bubbles drawn to the number of times the bubble was seen by users on the MWP website. A clean bubble produced from a cluster of 5 user drawings placed onto an asset that was seen 50 times would have a hit rate of 0.1. In cases where a bubble could be marked in more than one asset, for example across different zoom levels, the total view counts are summed, such that a cluster of 5 bubbles drawn onto two assets of 50 views each would have a hit rate of 0.05. The hit rate gives a measure of consensus among users that a bubble is present in the data.

The bubble's dispersion is also calculated as the spread in coordinates of the individual classifications ($\sqrt{\sigma_{b}^2 + \sigma_{l}^2}$, where $\sigma$ is the variance in the coordinate value).

Only bubbles that were seen 50 times or more, and which have hit rates of 0.1 or more, are included in the final catalogue. This ensures that each final, cleaned bubble is a combination of at least 5 individual users' drawings and was drawn by volunteers at least 10\% of the time when displayed on the website.

\subsection{Selection effects}

It is not known how many bubbles exist in the Galaxy, hence it is impossible to quantify the completeness of the MWP catalogue. There will be bubbles that are either not visible in the data used on the MWP, or that are not seen as bubbles. Distant bubbles may be obscured by foreground extinction. Faint bubbles may be masked by bright Galactic background emission or confused with brighter nebular structures. Fragmented or highly distorted bubbles present at high inclination angles may not appear as bubbles to the observer.

The MWP's `citizen science' approach creates its own biases, whilst overcoming others experienced in similar studies. By comparison with CP06, this study has many thousands of times more eyes scanning each section of the sky, and each section is broken down to an optimal colour stretch, thus improving the chances of seeing bubble-like structures. The majority of the MWP volunteers have no professional bias or expectation as to what constitutes a `good' bubble. MWP volunteers may experience measurement fatigue when classifying assets with many bubbles. They may also suffer bright neighbour bias, and fail to draw quite obvious bubbles that are adjacent to very prominent or beautiful examples.

\subsection{Small bubbles and other objects}
\label{other-objects}

In addition to marking elliptical bubbles on images, users are also encouraged to mark the locations of other interesting objects. Users can mark areas using a simple rectangle and are asked to label them as either a {\it small bubble, green knot, dark nebula, star cluster, galaxy, fuzzy red object} or {\it other}.

The ellipse-drawing tool of the MWP has a lower size limit of a diameter of 20 pixels ($0.45'$ at the highest image zoom level). The {\it Small Bubble} category allows users to mark bubbles which are too small to draw in detail but which can still be clearly made out. These small bubbles are reduced in a similar fashion to the more complex ellipses. To produce the catalogue of small bubbles listed in Table~\ref{small-bubble-sample} we use only the small bubbles drawn by users at the highest zoom level (this is the vast majority of those drawn). Small bubbles marked at lower zoom levels are equivalent to larger bubbles at the higher zoom levels. By rounding their locations to the nearest 20 pixels, the drawings are clustered. Our catalogue of 1,362 small bubbles is given in Table~\ref{small-bubble-sample}. Each of these small bubbles was drawn by at least five users and was drawn by at least 10\% of the volunteers who saw it -- as with the main, large-bubble catalogue.

Catalogues of {\it green knots, dark nebulae, star clusters, galaxies, fuzzy red objects} and objects in the {\it other} category are currently being prepared for later publication.

\section{Results}

\begin{table*}
\begin{center}
\caption{Table of bubble parameters (shown here are the 50 bubbles from the catalogue with the highest hit rates). The complete large bubble catalogue contains 3,744 visually identified bubbles. Where cross-correlation is possible, identifiers from CP06 and CWP07 are given. Hierarchy flags denote (1) bubbles identified as having smaller bubbles on their rim, and (2) bubbles located within a larger bubble. The complete bubble catalogue can be found online at http://data.milkywayproject.org.}
\begin{sideways}
\begin{tabular}{|ccccccccccc}
\hline
MWP & Churchwell & $l$ & $b$ & Radius & Thickness & Ecc. & Angle & Hit & Dispersion & Hierarchy \\
ID & ID & (deg) & (deg) & (arcmin) & (arcmin) & & ($^{\circ}$) & Rate & (arcmin) & Flag \\
\hline
MWP1G018261$-$02967 & N22 & 018.261 & $-$0.297 & 2.36 & 1.88 & 0.23 & 18 & 0.67 & 0.007 & 1\\
MWP1G312675+00465 & S128 & 312.675 & +0.046 & 1.71 & 1.27 & 0.41 & 4 & 0.64 & 0.004 & \\
MWP1G310982+04071 & S137 & 310.982 & +0.407 & 3.29 & 3.17 & 0.29 & 43 & 0.62 & 0.011 & 1,2\\
MWP1G307391$-$07207 &  & 307.391 & $-$0.721 & 1.90 & 1.29 & 0.29 & 69 & 0.61 & 0.007 & \\
MWP1G343918$-$06487 & S15 & 343.918 & $-$0.649 & 2.24 & 2.37 & 0.26 & 15 & 0.60 & 0.007 & \\
MWP1G316519$-$05993 & S114 & 316.519 & $-$0.599 & 1.53 & 1.67 & 0.29 & 36 & 0.60 & 0.006 & \\
MWP1G049699$-$01620 & N102,N103 & 049.699 & $-$0.162 & 2.23 & 1.73 & 0.10 & 26 & 0.60 & 0.007 & \\
MWP1G338712+06434 &  & 338.712 & +0.643 & 2.13 & 1.93 & 0.10 & 50 & 0.59 & 0.006 & 2\\
MWP1G354981$-$05283 & CS39 & 354.981 & $-$0.528 & 1.33 & 1.44 & 0.31 & 22 & 0.59 & 0.006 & 2\\
MWP1G320159+07984 & S96 & 320.159 & +0.798 & 1.49 & 1.51 & 0.12 & 23 & 0.58 & 0.009 & \\
MWP1G063163+04407 & N133 & 063.163 & +0.441 & 2.71 & 2.88 & 0.32 & 22 & 0.58 & 0.011 & \\
MWP1G323536$-$03512 & S86 & 323.536 & $-$0.351 & 1.29 & 1.57 & 0.28 & 15 & 0.57 & 0.005 & 2\\
MWP1G337683$-$03402 & S37 & 337.683 & $-$0.340 & 1.79 & 1.65 & 0.33 & 43 & 0.57 & 0.009 & 1\\
MWP1G354588+00038 & CS47 & 354.588 & +0.004 & 1.39 & 1.17 & 0.07 & 5 & 0.56 & 0.005 & \\
MWP1G045387$-$07155 & N95 & 045.387 & $-$0.715 & 1.95 & 1.25 & 0.11 & 27 & 0.56 & 0.009 & \\
MWP1G317463$-$03466 & S106 & 317.463 & $-$0.347 & 1.90 & 1.56 & 0.09 & 10 & 0.56 & 0.004 & \\
MWP1G028006+03153 &  & 028.006 & +0.315 & 2.07 & 1.66 & 0.10 & 19 & 0.55 & 0.004 & \\
MWP1G040421$-$00493 & N77 & 040.421 & $-$0.049 & 1.42 & 1.18 & 0.07 & 13 & 0.54 & 0.007 & \\
MWP1G339130$-$03706 &  & 339.130 & $-$0.371 & 3.54 & 2.31 & 0.09 & 23 & 0.54 & 0.017 & 1\\
MWP1G327988$-$00931 & S71 & 327.988 & $-$0.093 & 1.80 & 1.29 & 0.54 & 16 & 0.54 & 0.009 & 2\\
MWP1G340601+03413 &  & 340.601 & +0.341 & 0.52 & 0.44 & 0.37 & 33 & 0.54 & 0.002 & 2\\
MWP1G013733$-$00160 & N12 & 013.733 & $-$0.016 & 4.75 & 2.79 & 0.23 & 19 & 0.54 & 0.013 & \\
MWP1G354612+04841 & CS46 & 354.612 & +0.484 & 1.45 & 1.38 & 0.11 & 23 & 0.54 & 0.005 & \\
MWP1G044332$-$08343 & N92 & 044.332 & $-$0.834 & 2.07 & 1.65 & 0.55 & 77 & 0.54 & 0.009 & 1\\
MWP1G064010+07285 &  & 064.010 & +0.729 & 1.02 & 0.94 & 0.24 & 10 & 0.54 & 0.003 & \\
MWP1G057544$-$02820 & N123 & 057.544 & $-$0.282 & 1.44 & 1.21 & 0.24 & 37 & 0.54 & 0.009 & \\
MWP1G340299$-$01980 & S26 & 340.299 & $-$0.198 & 1.64 & 1.59 & 0.31 & 38 & 0.54 & 0.019 & 1\\
MWP1G345484+04011 & S11 & 345.484 & +0.401 & 2.41 & 2.05 & 0.24 & 18 & 0.54 & 0.011 & 1\\
MWP1G043775+00606 & N90 & 043.775 & +0.061 & 1.77 & 1.37 & 0.26 & 9 & 0.54 & 0.005 & \\
MWP1G017921$-$06843 & N20 & 017.921 & $-$0.684 & 1.40 & 1.04 & 0.18 & 23 & 0.54 & 0.007 & \\
MWP1G018193$-$04002 & N21 & 018.193 & $-$0.400 & 2.56 & 2.00 & 0.41 & 19 & 0.54 & 0.008 & 1\\
MWP1G311483+03982 & S133,S134,S135,S132 & 311.483 & +0.398 & 2.17 & 1.52 & 0.48 & 19 & 0.53 & 0.008 & 1\\
MWP1G054106$-$00635 & N116,N117 & 054.106 & $-$0.064 & 2.53 & 2.48 & 0.28 & 33 & 0.53 & 0.012 & 1\\
MWP1G044777$-$05495 & N93 & 044.777 & $-$0.549 & 0.81 & 0.79 & 0.33 & 15 & 0.53 & 0.003 & 2\\
MWP1G312975$-$04354 & S123 & 312.975 & $-$0.435 & 2.45 & 1.83 & 0.20 & 24 & 0.53 & 0.009 & \\
MWP1G354683+04723 & CS44 & 354.683 & +0.472 & 1.59 & 1.56 & 0.13 & 20 & 0.53 & 0.007 & \\
MWP1G350585$-$00744 & CS101 & 350.585 & $-$0.074 & 2.33 & 1.26 & 0.32 & 22 & 0.53 & 0.007 & \\
MWP1G354182$-$00528 & CS51 & 354.182 & $-$0.053 & 2.73 & 1.80 & 0.41 & 11 & 0.52 & 0.009 & \\
MWP1G311893+00861 &  & 311.893 & +0.086 & 2.42 & 1.85 & 0.18 & 22 & 0.52 & 0.008 & 1\\
MWP1G038907$-$04386 & N74 & 038.907 & $-$0.439 & 1.59 & 1.54 & 0.45 & 49 & 0.52 & 0.003 & \\
MWP1G052605+03708 &  & 052.605 & +0.371 & 0.42 & 0.71 & 0.26 & 24 & 0.52 & 0.001 & 2\\
MWP1G359737$-$04097 & CS2 & 359.737 & $-$0.410 & 1.98 & 1.28 & 0.26 & 14 & 0.52 & 0.008 & 1\\
MWP1G299701+00151 & S174 & 299.701 & +0.015 & 3.16 & 1.75 & 0.21 & 20 & 0.52 & 0.012 & \\
MWP1G054480+09282 &  & 054.480 & +0.928 & 2.60 & 3.96 & 0.22 & 19 & 0.52 & 0.010 & \\
MWP1G023089+05617 & N29,N30 & 023.089 & +0.562 & 2.33 & 1.89 & 0.23 & 30 & 0.52 & 0.030 & \\
\hline
\end{tabular}
\end{sideways}
\end{center}
\label{bubble-sample}
\end{table*}

\begin{table*}
\caption{Table of MWP small bubbles. Shown here are the 50 bubbles from the small bubble catalogue with the highest hit rates. The complete small bubble catalogue contains 1,362 visually identified bubbles. Where cross-correlation is possible, identifiers from CP06 and CWP07 are given. Hierarchy flags denote (1) bubbles identified as having smaller bubbles on their rim, and (2) bubbles located on the rim of a larger bubble. The complete bubble catalogue can be found online at http://data.milkywayproject.org.}
\begin{center}
\begin{tabular}{|lccccccccccc}
\hline
MWP & Churchwell & $l$ & $b$ & Mean Radius & Hit & Hierarchy \\
ID & ID & (deg) & (deg) & (arcmin) & Rate & Flag \\
\hline
MWP1G331470$-$01400S &  & 331.47 & $-$0.14 & 0.37 & 0.54 & \\
MWP1G018180+01100S &  & 018.18 & +0.11 & 0.34 & 0.50 & \\
MWP1G017690$-$00900S &  & 017.69 & $-$0.09 & 0.42 & 0.49 & \\
MWP1G336010$-$03700S &  & 336.01 & $-$0.37 & 0.37 & 0.45 & \\
MWP1G012200$-$01600S &  & 012.20 & $-$0.16 & 0.38 & 0.42 & \\
MWP1G013780+04900S &  & 013.78 & +0.49 & 0.42 & 0.42 & 2\\
MWP1G024920+00800S &  & 024.92 & +0.08 & 0.47 & 0.42 & \\
MWP1G338870+00200S &  & 338.87 & +0.02 & 0.45 & 0.42 & \\
MWP1G011020$-$03700S &  & 011.02 & $-$0.37 & 0.41 & 0.41 & \\
MWP1G345510+01600S &  & 345.51 & +0.16 & 0.40 & 0.41 & 2\\
MWP1G039430$-$01900S &  & 039.43 & $-$0.19 & 0.41 & 0.40 & \\
MWP1G339770+00100S &  & 339.77 & +0.01 & 0.43 & 0.40 & 2\\
MWP1G346040+00500S &  & 346.04 & +0.05 & 0.34 & 0.40 & \\
MWP1G331400$-$01800S &  & 331.40 & $-$0.18 & 0.44 & 0.39 & 2\\
MWP1G004060$-$00100S &  & 004.06 & $-$0.01 & 0.38 & 0.38 & \\
MWP1G008910+01700S &  & 008.91 & +0.17 & 0.38 & 0.38 & 2\\
MWP1G012810$-$03100S &  & 012.81 & $-$0.31 & 0.44 & 0.38 & 2\\
MWP1G052980$-$06200S &  & 052.98 & $-$0.62 & 0.43 & 0.38 & 2\\
MWP1G339780+00200S &  & 339.78 & +0.02 & 0.34 & 0.38 & 2\\
MWP1G343290+01600S &  & 343.29 & +0.16 & 0.37 & 0.38 & 2\\
MWP1G357970$-$01700S &  & 357.97 & $-$0.17 & 0.37 & 0.38 & 2\\
MWP1G010990$-$03700S &  & 010.99 & $-$0.37 & 0.45 & 0.37 & \\
MWP1G018850$-$04800S & N24 & 018.85 & $-$0.48 & 0.41 & 0.37 & 2\\
MWP1G049570$-$02700S &  & 049.57 & $-$0.27 & 0.44 & 0.37 & 2\\
MWP1G301760+02600S &  & 301.76 & +0.26 & 0.41 & 0.37 & \\
MWP1G331140+01100S &  & 331.14 & +0.11 & 0.34 & 0.37 & \\
MWP1G334350$-$06600S &  & 334.35 & $-$0.66 & 0.34 & 0.37 & \\
MWP1G352760$-$03500S &  & 352.76 & $-$0.35 & 0.39 & 0.37 & \\
MWP1G307500$-$08200S &  & 307.50 & $-$0.82 & 0.49 & 0.36 & 2\\
MWP1G332670$-$03300S &  & 332.67 & $-$0.33 & 0.38 & 0.36 & \\
MWP1G339240+00100S &  & 339.24 & +0.01 & 0.47 & 0.36 & 2\\
MWP1G005100+00000S &  & 005.10 & +0.00 & 0.49 & 0.35 & \\
MWP1G014210$-$01100S &  & 014.21 & $-$0.11 & 0.41 & 0.35 & \\
MWP1G032400$-$03300S &  & 032.40 & $-$0.33 & 0.40 & 0.35 & \\
MWP1G035720$-$09300S &  & 035.72 & $-$0.93 & 0.42 & 0.35 & \\
MWP1G335940$-$02900S &  & 335.94 & $-$0.29 & 0.46 & 0.35 & 2\\
MWP1G339700+03000S &  & 339.70 & +0.30 & 0.37 & 0.35 & \\
MWP1G002170+00100S &  & 002.17 & +0.01 & 0.39 & 0.34 & \\
MWP1G005630$-$02900S &  & 005.63 & $-$0.29 & 0.42 & 0.34 & \\
MWP1G009970$-$02100S &  & 009.97 & $-$0.21 & 0.41 & 0.34 & \\
MWP1G018170+03400S &  & 018.17 & +0.34 & 0.46 & 0.34 & 2\\
MWP1G025820$-$01900S &  & 025.82 & $-$0.19 & 0.36 & 0.34 & \\
MWP1G027610+00300S &  & 027.61 & +0.03 & 0.45 & 0.34 & 2\\
MWP1G046140$-$01400S &  & 046.14 & $-$0.14 & 0.39 & 0.34 & \\
MWP1G297720$-$07000S &  & 297.72 & $-$0.70 & 0.46 & 0.34 & \\
MWP1G311630$-$02600S &  & 311.63 & $-$0.26 & 0.34 & 0.34 & \\
MWP1G335470+02000S &  & 335.47 & +0.20 & 0.38 & 0.34 & \\
MWP1G342900$-$00900S &  & 342.90 & $-$0.09 & 0.50 & 0.34 & \\
MWP1G345480$-$02200S &  & 345.48 & $-$0.22 & 0.36 & 0.34 & \\
MWP1G358890+00800S &  & 358.89 & +0.08 & 0.34 & 0.34 & \\
\hline
\end{tabular}
\end{center}
\label{small-bubble-sample}
\end{table*}

\subsection{Catalogue description}
\label{results1}

\begin{figure*}
\includegraphics[width=0.98\textwidth]{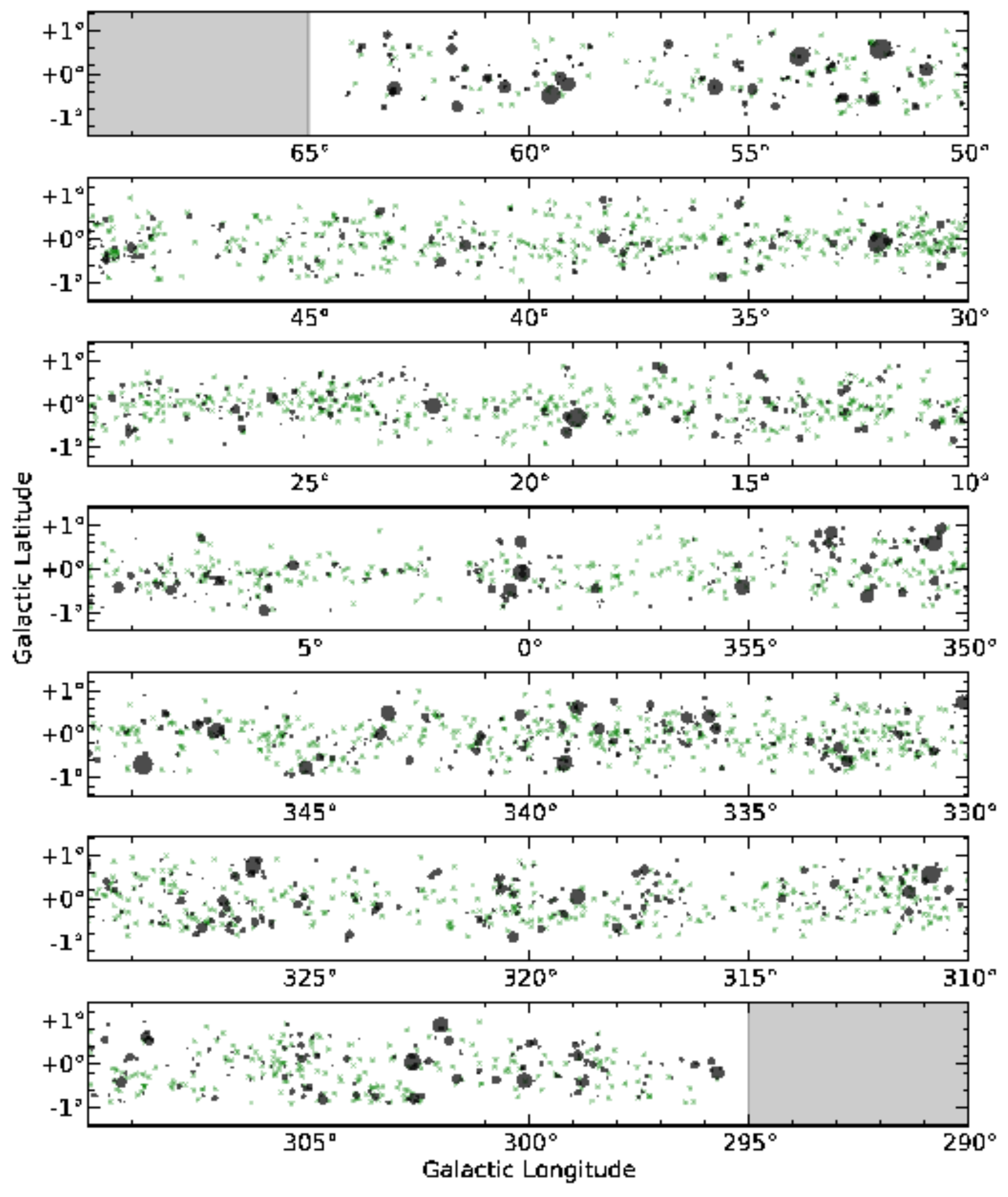}
\caption{All \bubblecount\ MWP bubbles plotted in Galactic coordinnates. The MWP large bubble catalogue is marked with outer radii as grey ellipses, small bubbles shown as green crosses.}
\label{bubble-map}
\end{figure*}

The final, reduced catalogue contains \bubblecount\ visually identified bubbles. These are split into a catalogue of 3,744 large bubbles drawn by users as ellipses, and a catalogue of 1,362 small bubbles drawn by users at the highest zoom level images in the MWP. These bubbles are plotted in Galactic coordinates in Figure~\ref{bubble-map}.

Each bubble in both lists has been drawn by at least five different individuals, and the listed parameters have been obtained from a weighted average based on each user's score (see Section~\ref{scoring-section}). The complete catalogue can be accessed at http://data.milkywayproject.org. Table~\ref{bubble-sample} gives the large bubbles, in order of hit rate. The large bubble catalogue includes values for the position in galactic longitude and latitude; mean geometric radius and mean thickness (as defined by CP06), position angle (given in degree from North), eccentricity (as defined by CP06); the hit rate (described in Section~\ref{combine-section}); and a hierarchy flag indicating whether bubbles are (1) identified as having further, smaller bubbles within their boundary, or (2) located on the rim of a larger bubble.

\subsection{Cross-Matching with Existing Catalogues}
\label{crossmatch}

For each bubble and small bubble produced by the MWP, cross-matching was performed with GLIMPSE bubbles (i.e. the CP06 and CWP07 catalogues) and the \citet{Paladini+03} and \citet{Anderson+11} catalogues of \hiirs. Sources are marked as coincident when the central coordinate of the catalogue object lies within the radius of the MWP bubble. 12\% of MWP bubbles matched GLIMPSE bubbles, and 86\% of GLIMPSE bubbles were re-discovered by MWP. Similarly 10\% and 7\% of MWP bubbles are coincident with \citet{Paladini+03} and \citet{Anderson+11} \hiirs , respectively. The MWP finds 86\% of the available Paladini sources and 96\% of the Anderson sources. The presence of 24\um\ emission coincident with 20~cm emission was a selection criterion for \citet{Anderson+11} and so there may be some overlap in selection methods with the MWP.

\citet{MIPSGAL-rings} catalogued 416 disk and ringlike structures seen in the MIPSGAL 24 \um\ images and suggested that the majority of these objects were produced by evolved stars. The \citet{MIPSGAL-rings} catalogue is dominated by small sources with radii ${<}20''$, and so these objects should be unlikely to overlap with the main MWP bubbles catalogue. In fact we rediscover only 9\%\ of the sources in this catalogue and they constitute less than 1\% of the combined MWP bubble catalogues. Of the 1,093 small MWP bubbles with a mean width of less than 1~arcmin, only five correspond to objects in the \citet{MIPSGAL-rings} catalogue (for reference, the IDs of these bubbles are: MWP1G031730+07000S, MWP1G314360+04900S, MWP1G319220+01600S, MWP1G334110+03800S, MWP1G358770+01100S).

The above crossover fractions (summarised in Table~\ref{cross-matching}) show that the MWP has excellent overlap with existing bubble catalogues and is also more complete, in terms of locating \hiirs, than the two Churchwell studies. The lack of agreement between the MWP and \citet{MIPSGAL-rings} is evidence that the small bubble catalogue is not contaminated by small, 24~\um\ disk- and ring-like structures.

\begin{table}
\caption{Crossover between the MWP and relevant catalogues of bubbles (CP06; CWP07), \hiirs\ \citep{Paladini+03, Anderson+11}, and MIPSGAL ring-like structures \citep{MIPSGAL-rings}.}
\begin{center}
\begin{tabular}{lcc}
\hline
Catalogue & Fraction MWP & Fraction\\
 & Bubbles  Matched & Rediscovered \\
\hline
CP06 and CWP07 & 0.12 & 0.85 \\
\citet{Paladini+03} & 0.10 & 0.86 \\
\citet{Anderson+11} & 0.07 & 0.96 \\
\citet{MIPSGAL-rings} & 0.01 & 0.09 \\
\hline
\end{tabular}
\end{center}
\label{cross-matching}
\end{table}

\subsection{Errors}

\begin{figure*}
\begin{center}
\subfigure{
\includegraphics[width=0.33\textwidth]{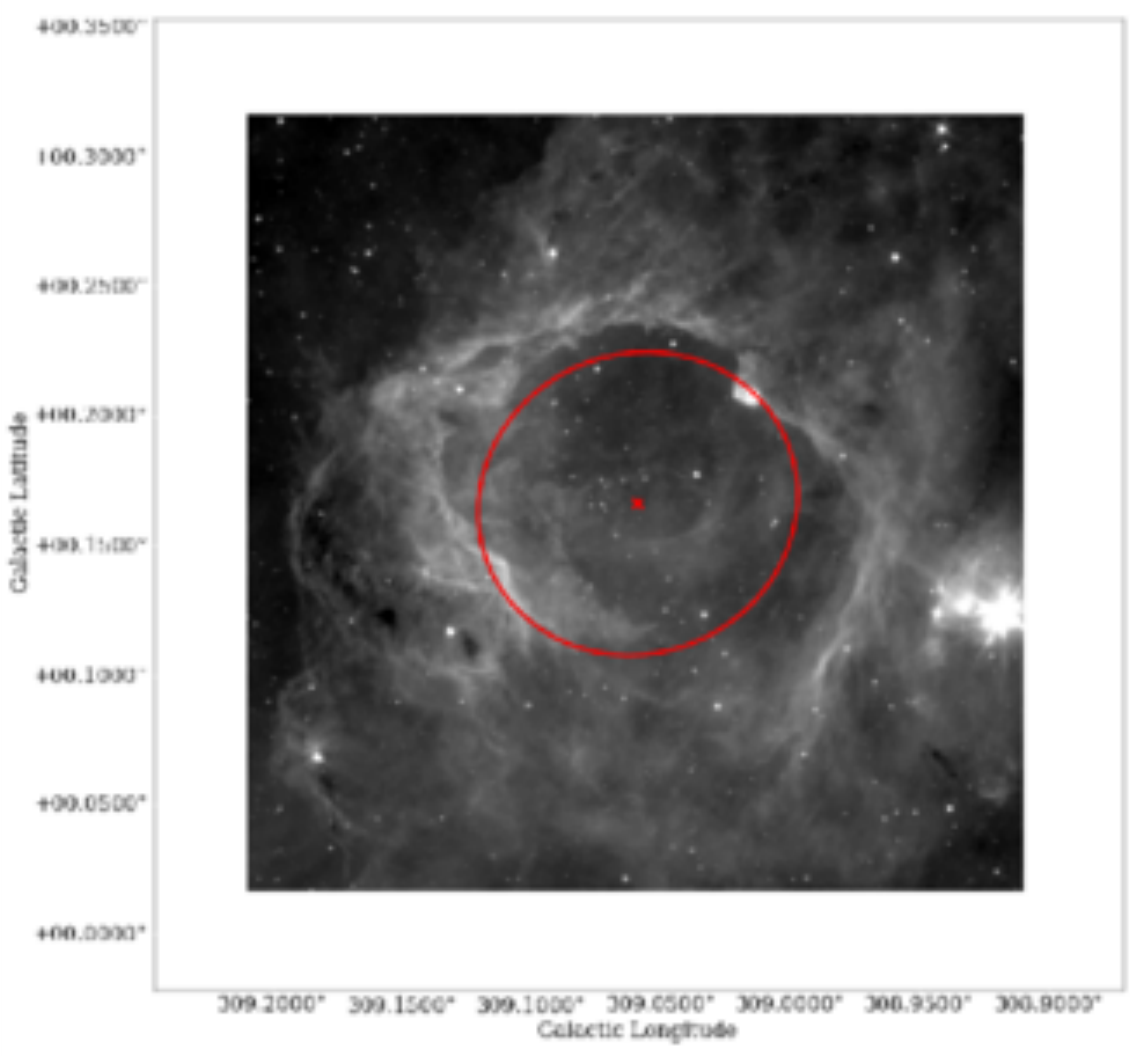}}
\subfigure{
\includegraphics[width=0.33\textwidth]{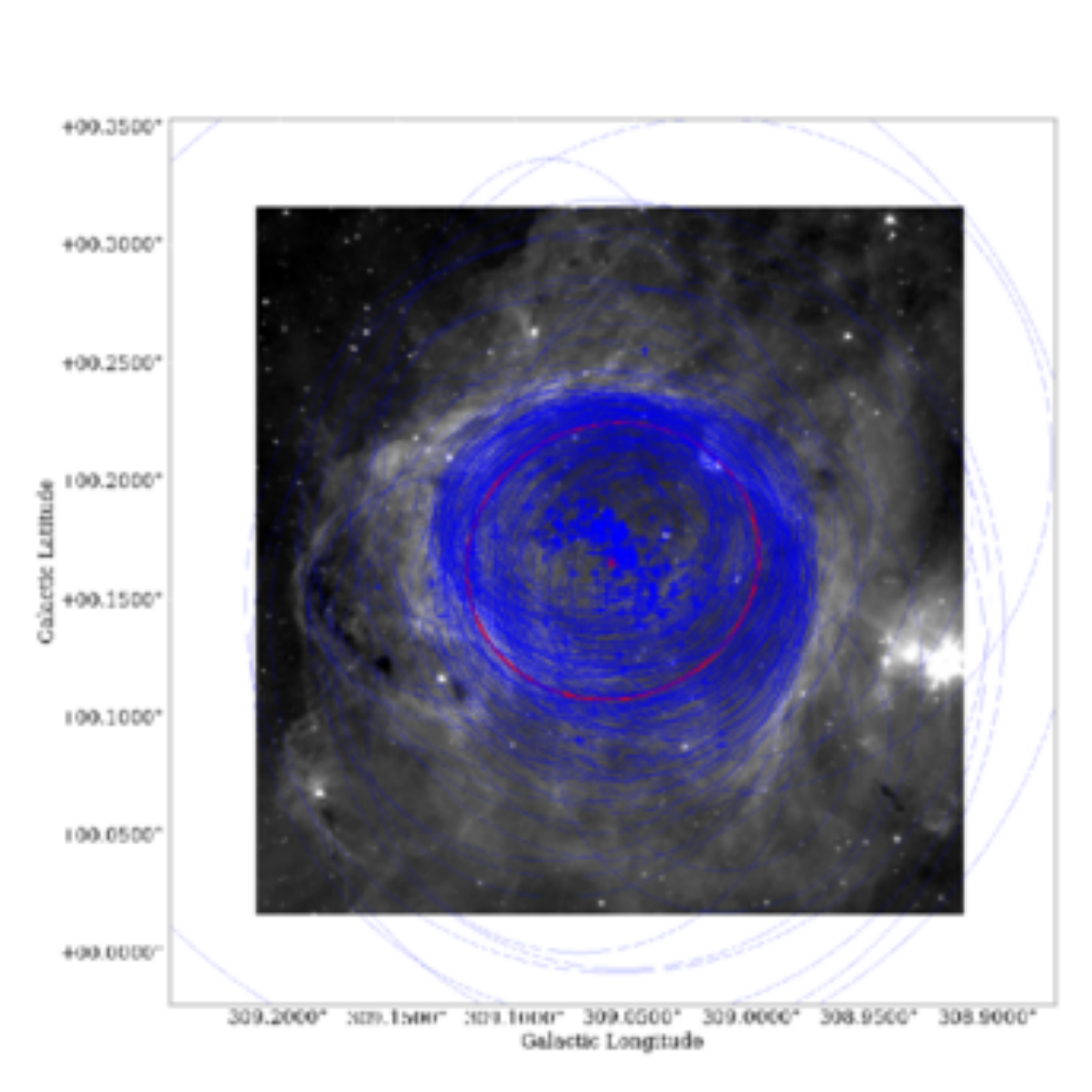}}
\subfigure{
\includegraphics[width=0.75\textwidth]{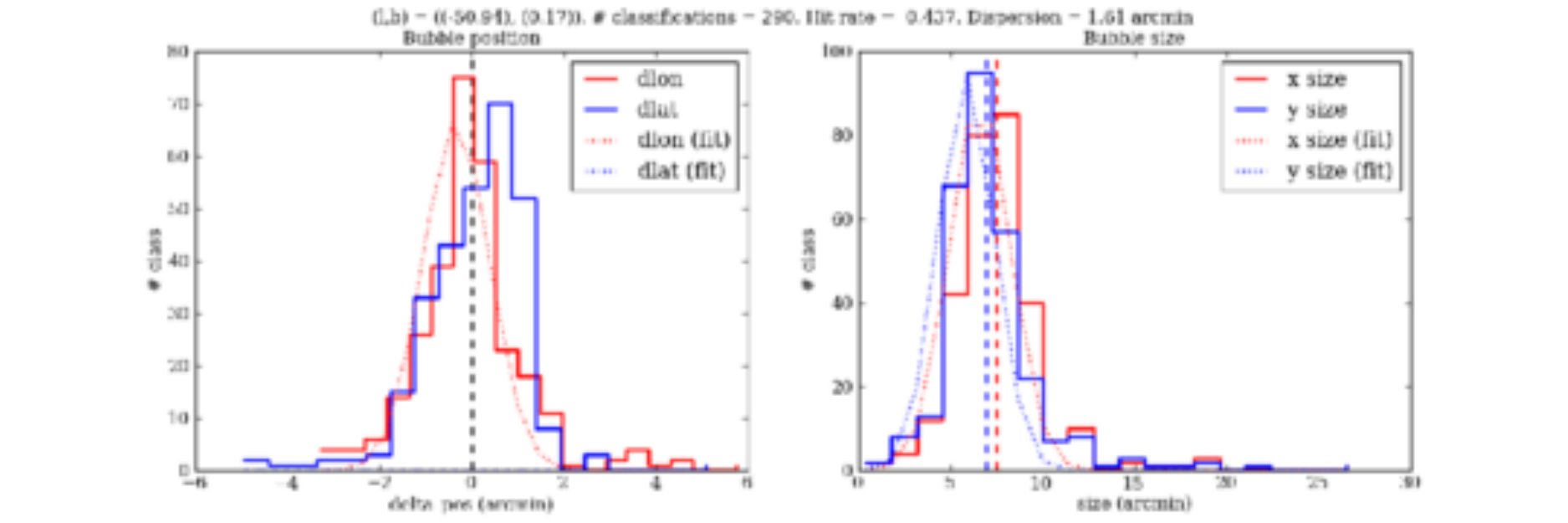}}
\caption{Error measurements for MWP bubble MWP1G309059+01661. This bubble has a hit rate of 0.437, and a dispersion of 1.61'. Top figures show reduced and raw bubble drawings. Bottom figures show dispersions in measurements of position and size. Colour figure available online.}
\label{MWP1G309059+01661}
\end{center}
\end{figure*}

\begin{figure*}
\begin{center}
\subfigure{
\includegraphics[width=0.33\textwidth]{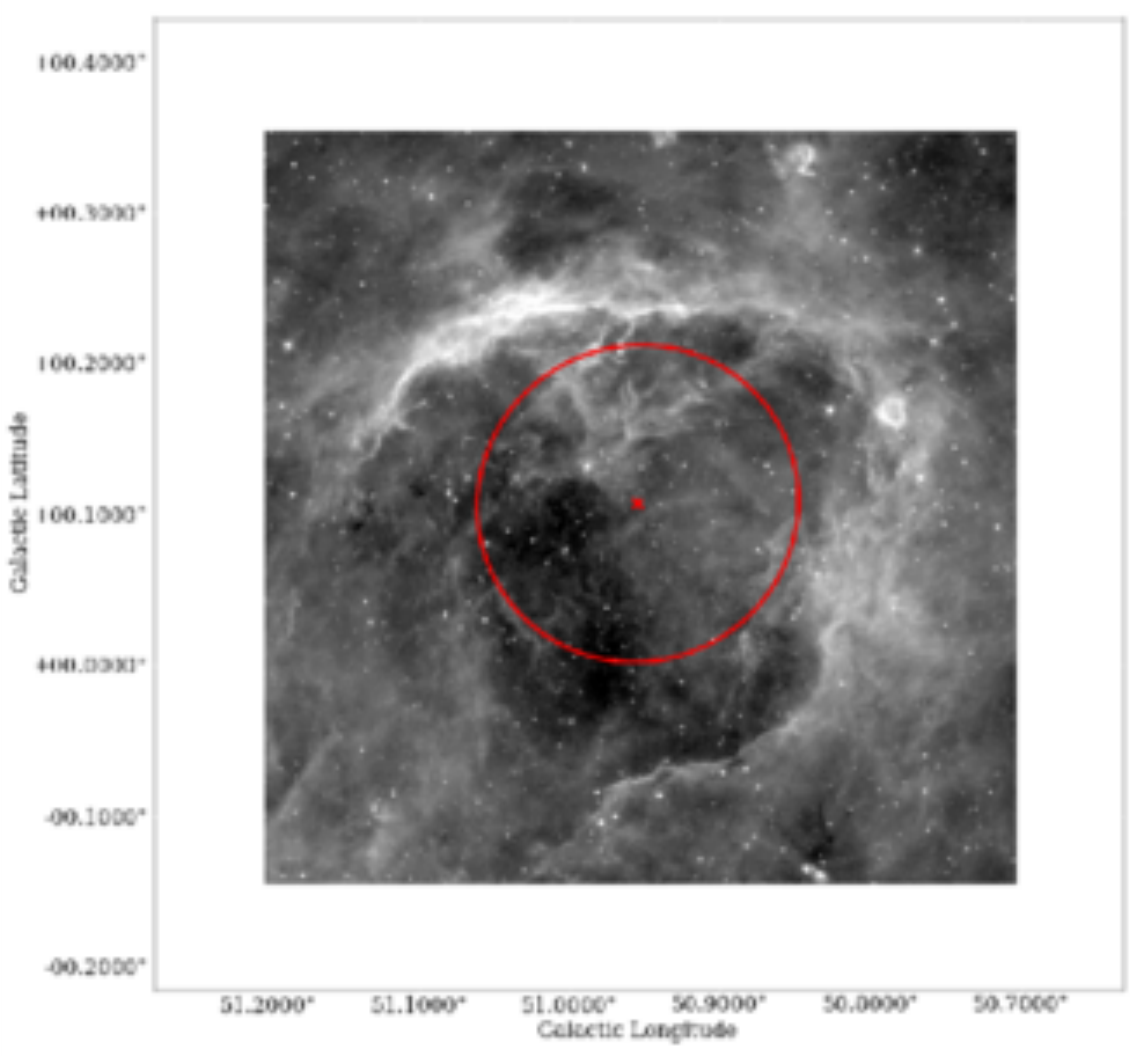}}
\subfigure{
\includegraphics[width=0.33\textwidth]{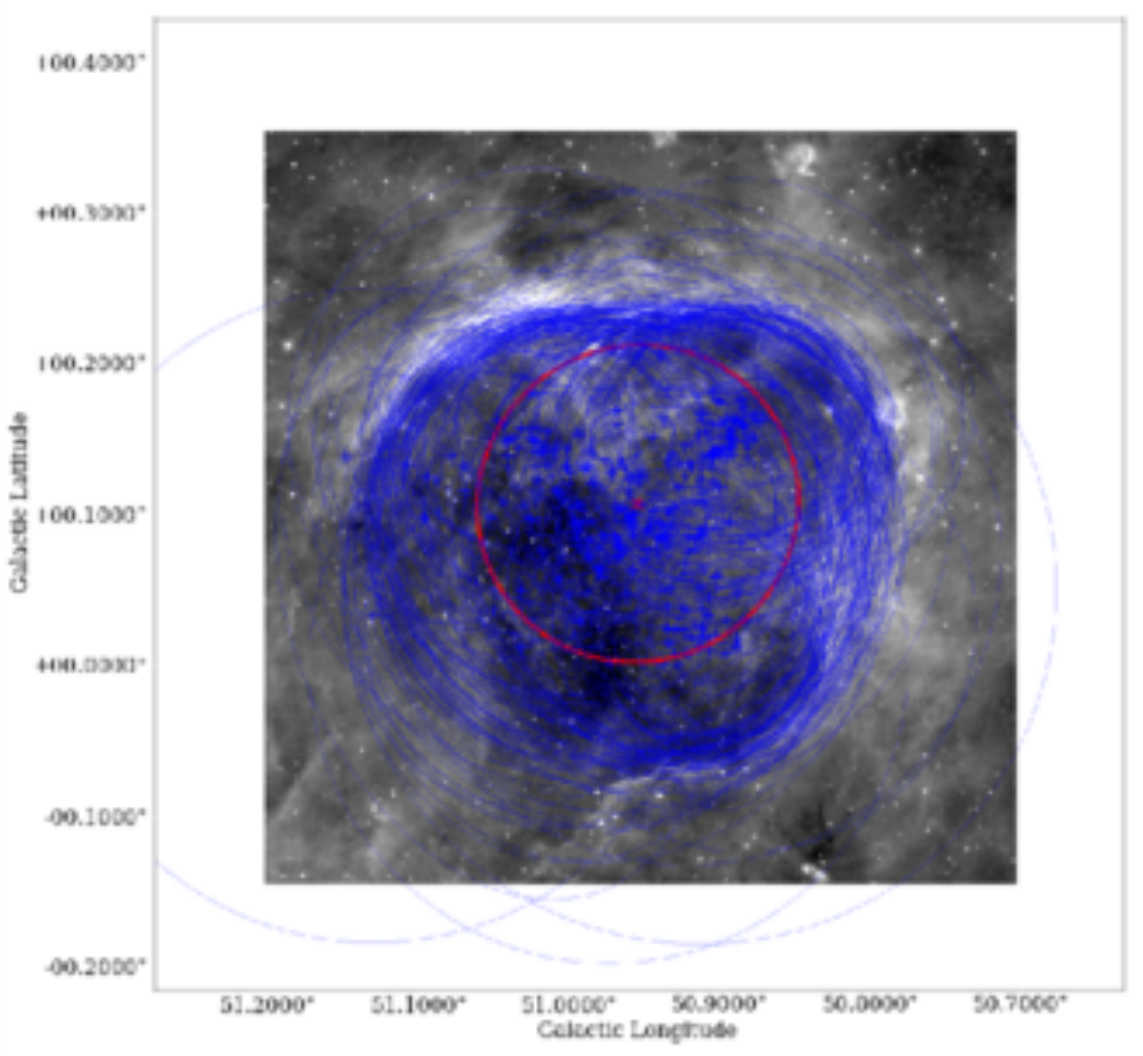}}
\subfigure{
\includegraphics[width=0.75\textwidth]{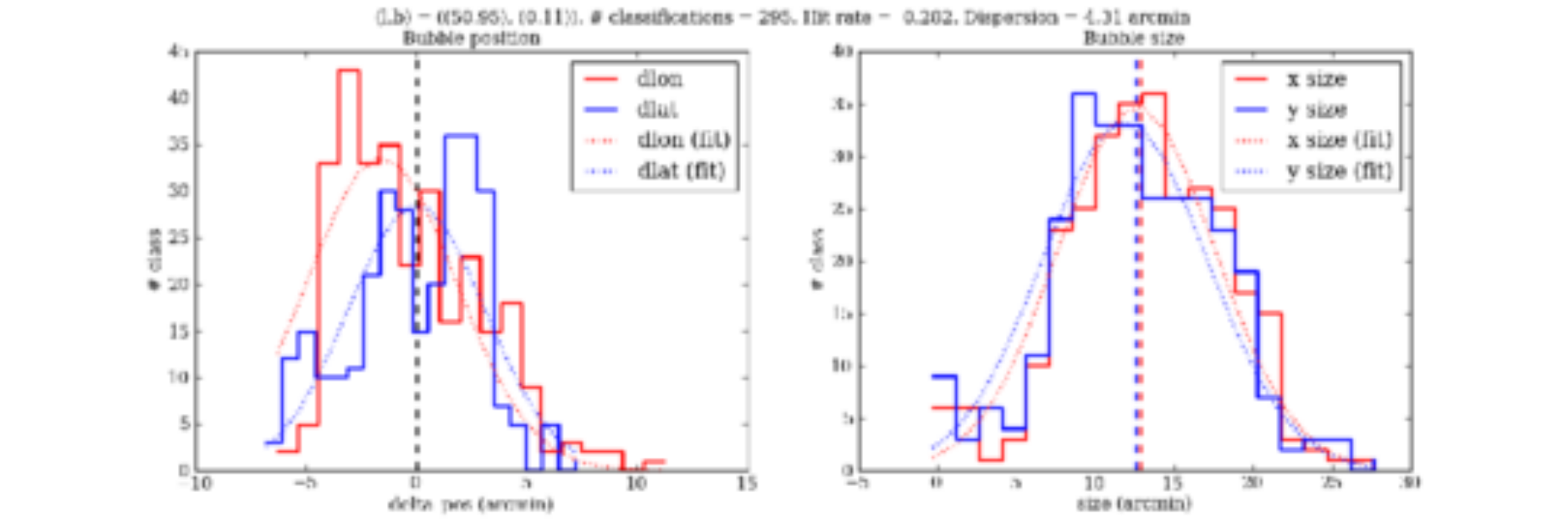}}
\caption{Errors for bubble MWP1G050955+01074. This bubble has a hit rate of 0.282, and a dispersion of 4.31'. See Figure~\ref{MWP1G309059+01661} for more information. Colour figure available online.}
\label{MWP1G050955+01074}
\end{center}
\end{figure*}

\begin{figure*}
\begin{center}
\subfigure{
\includegraphics[width=0.33\textwidth]{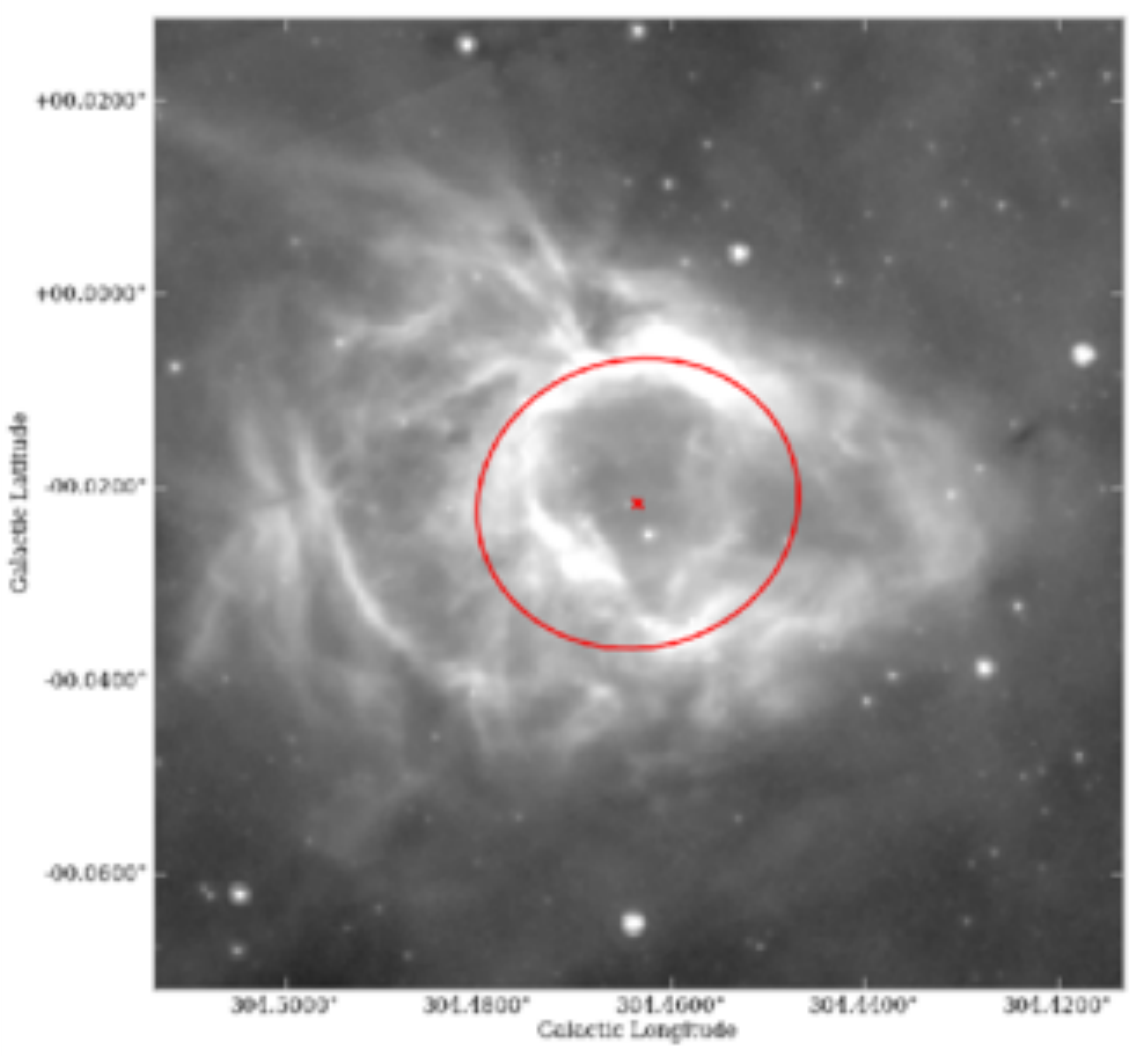}}
\subfigure{
\includegraphics[width=0.33\textwidth]{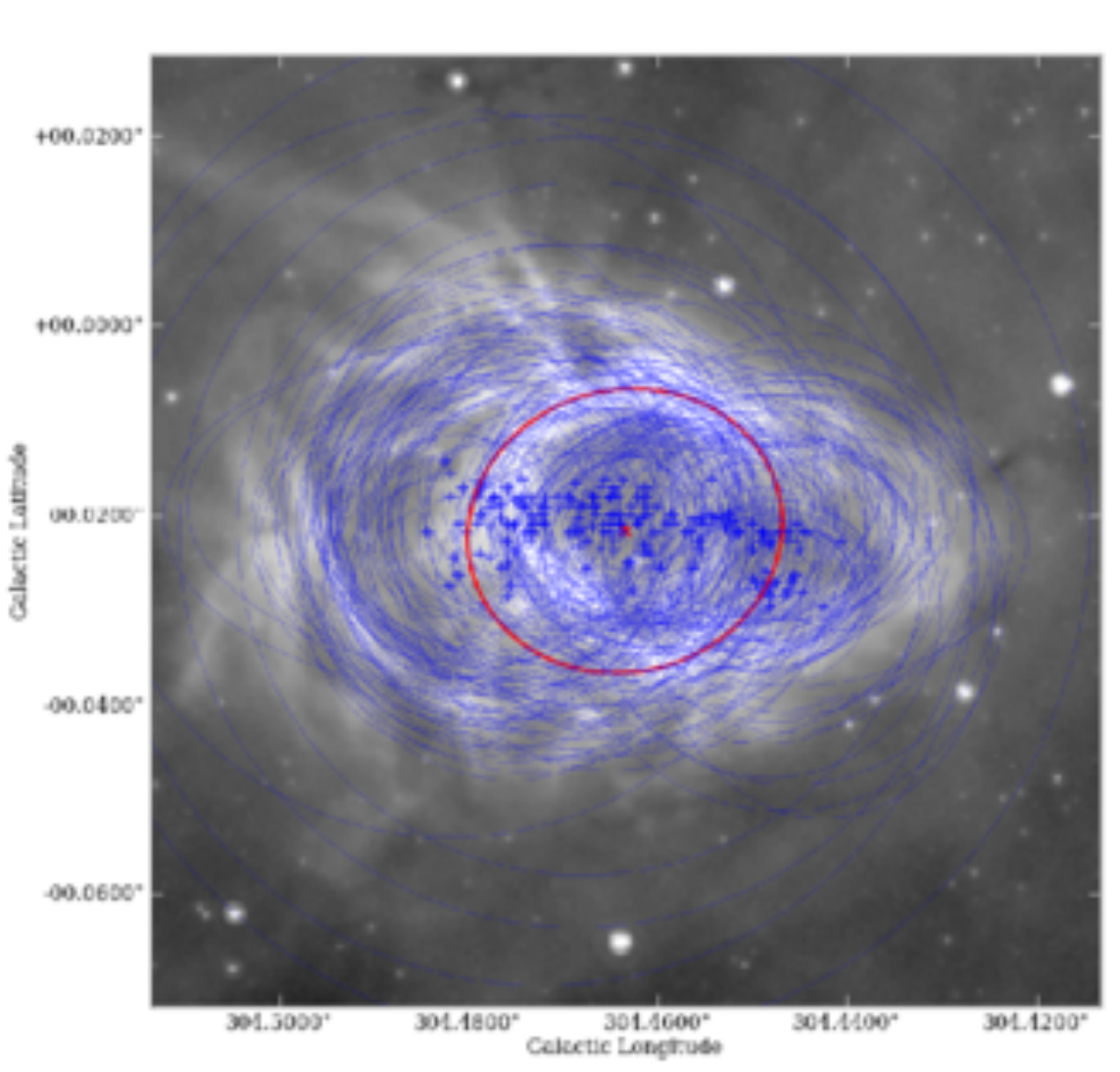}}
\subfigure{
\includegraphics[width=0.75\textwidth]{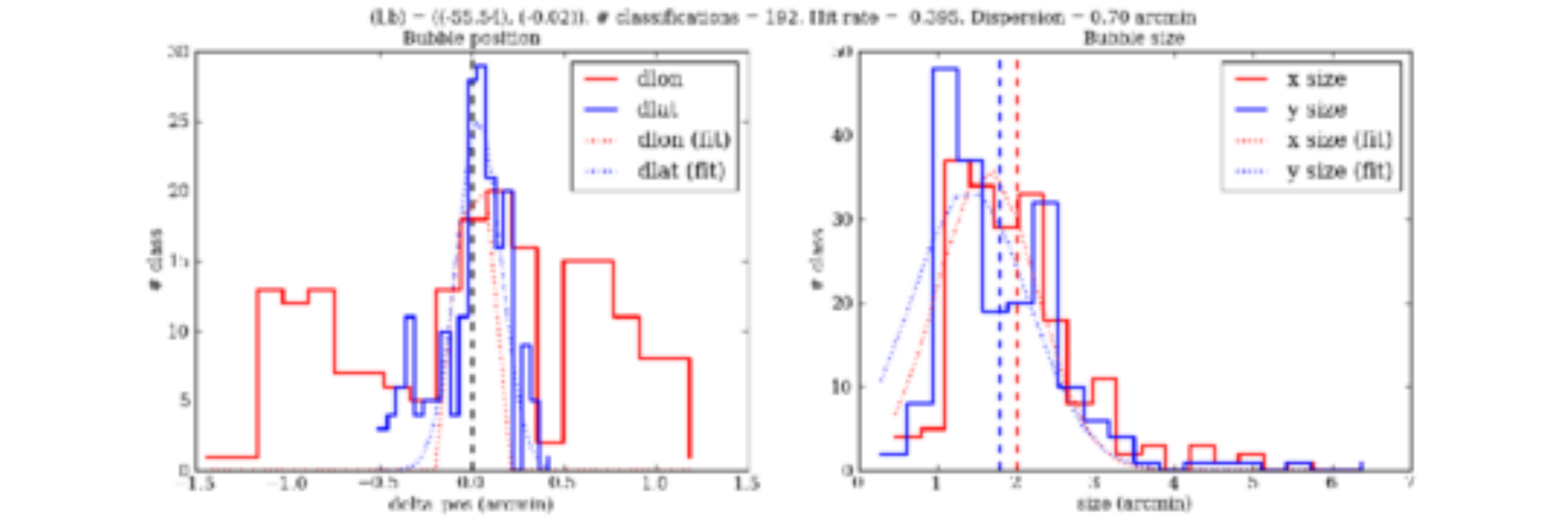}}
\caption{Errors for bubble MWP1G304463$-$00217. This bubble has a hit rate of 0.395, and a dispersion of 0.70'. See Figure~\ref{MWP1G309059+01661} for more information. Colour figure available online}
\label{MWP1G304463-00217}
\end{center}
\end{figure*}

\begin{figure*}
\begin{center}
\subfigure{
\includegraphics[width=0.33\textwidth]{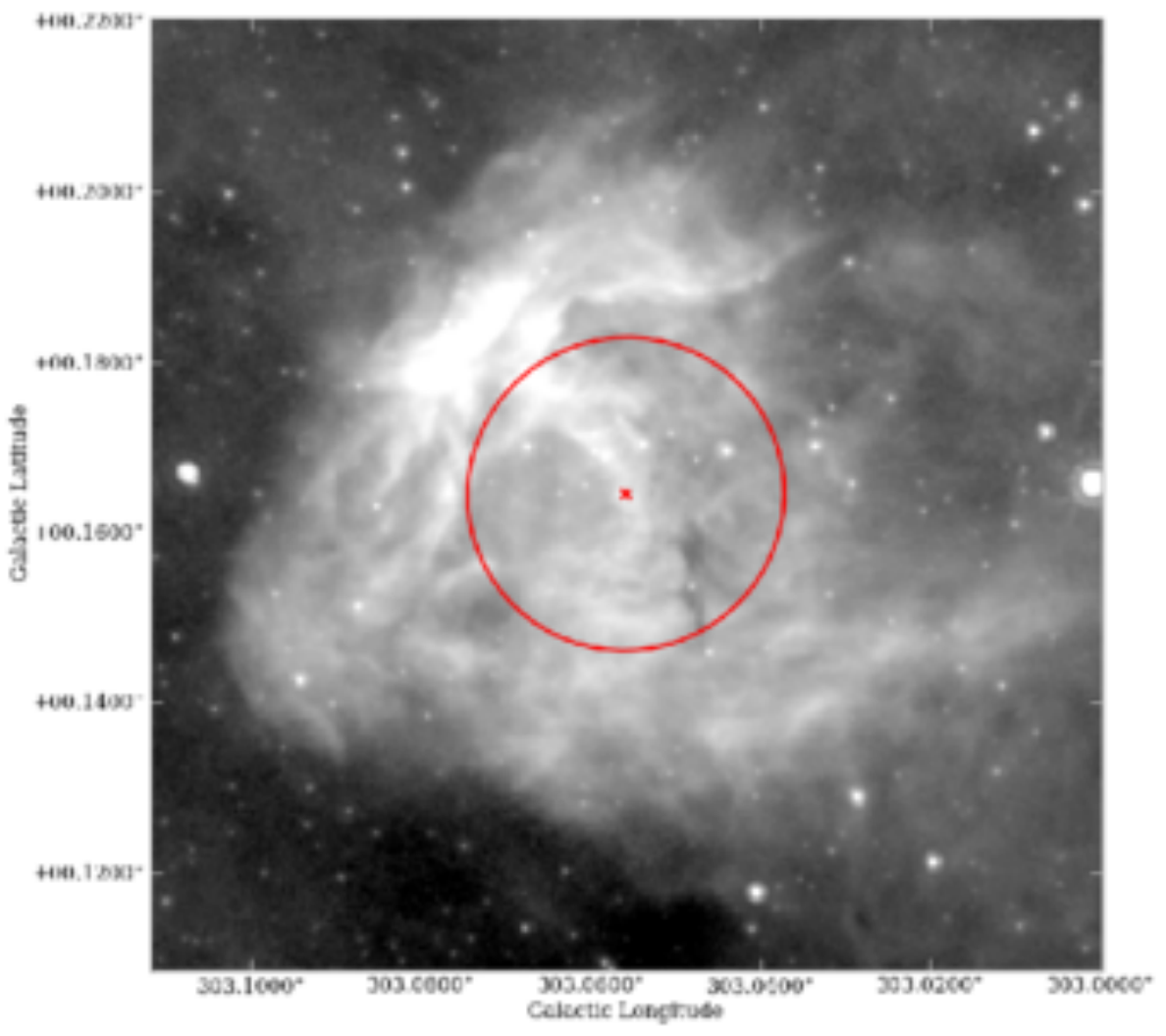}}
\subfigure{
\includegraphics[width=0.33\textwidth]{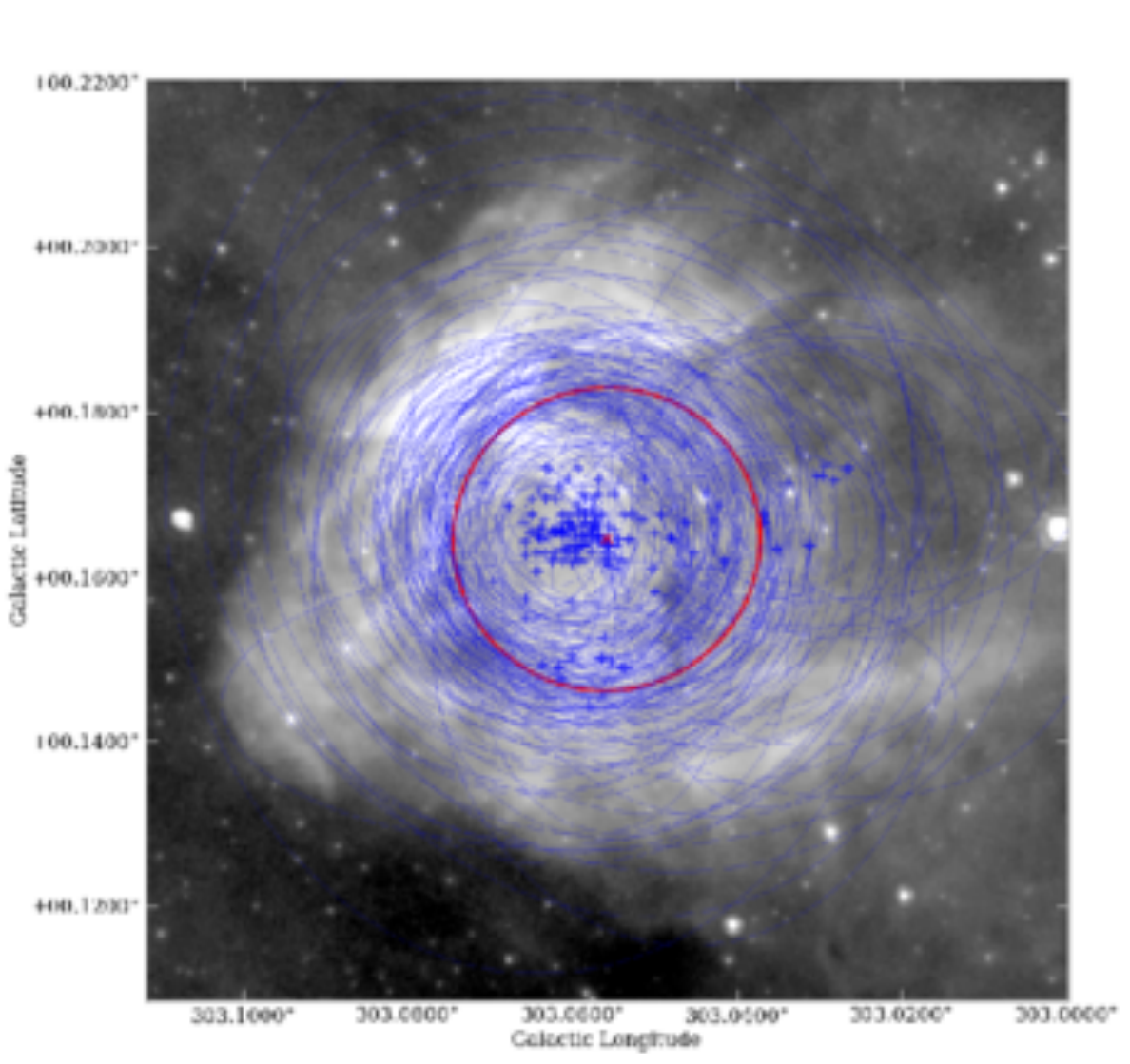}}
\subfigure{
\includegraphics[width=0.75\textwidth]{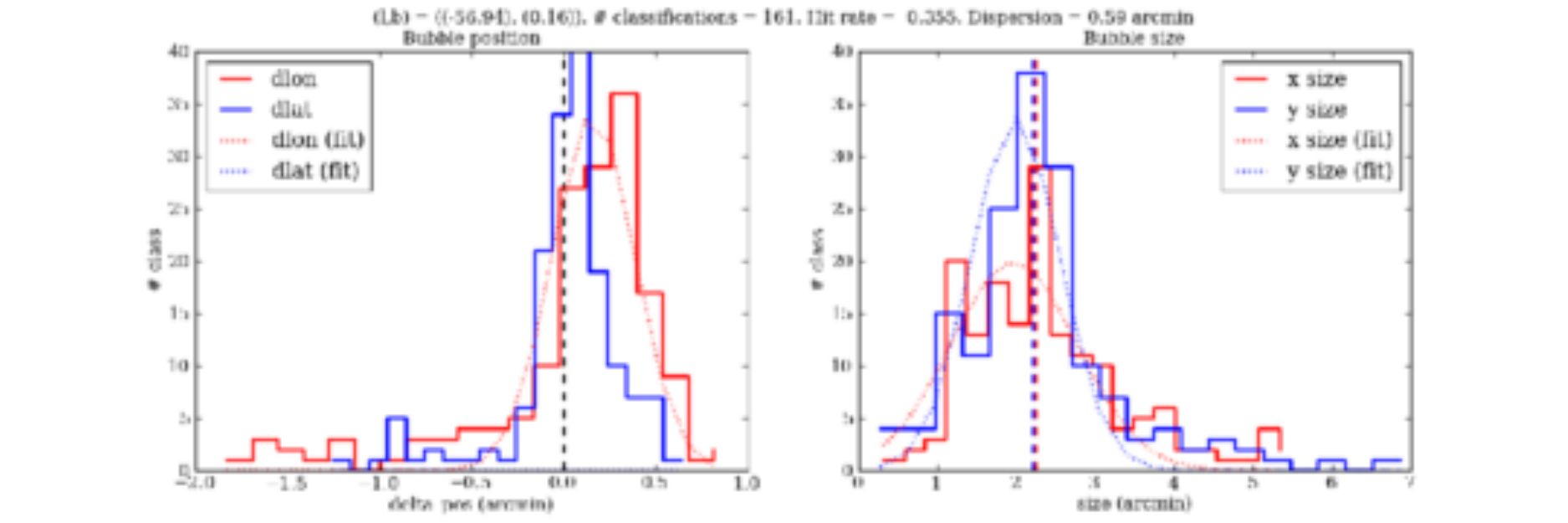}}
\caption{Errors for bubble MWP1G303056+01645. This bubble has a hit rate of 0.355, and a dispersion of 0.59'. See Figure~\ref{MWP1G309059+01661} for more information. Colour figure available online.}
\label{MWP1G303056+01645}
\end{center}
\end{figure*}

A number of quality control measures, such as the user weighting scheme, were adopted from the outset; these are described in Section~\ref{scoring-section}. To assess the performance of the processing procedure, we examine the four bubbles in the catalog with the highest classification scores,  i.e. those drawn by the largest number of users. These bubbles were identified by an average of 243.2 users. Three of these were previously identified as bubbles and HII regions by other authors (CP06; Misanovic et al (2002); Lockman et al (1996); Kuchar \& Clark (1997)). Bubble MWP1G303056+01645 (Figure~\ref{MWP1G303056+01645}) has no associated CP06 bubble.

In bubbles MWP1G309059+01661 (Figure~\ref{MWP1G309059+01661}) and MWP1G303056+01645 (Figure~\ref{MWP1G303056+01645}) the computed weighted averages for the bubble position and size parameters, indicated with the dashed lines, accurately follow the distribution of individual classifications, and the distribution is well approximated by a gaussian distribution. In these cases, each user classification can be reasonably considered to be an independent measurement of the same quantity. In other cases, such as with MWP1G304463$-$00217 (Figure~\ref{MWP1G304463-00217}), there are skews or multiple peaks in the distribution of classifications, indicating a relative lack of consensus among the users over the location, size, shape or multiplicity of the bubble. Indeed, multiple peaks in the distribution may suggest the presence of more than one bubble at a given location.

These variations can be attributed to the definition of our clustering algorithm. Using a tighter clustering threshold will do better at separating out closely spaced bubbles, but will also artificially fragment single bubbles into multiples where the bubble's positional coordinates have a high uncertainty value. Our data reduction could be modified to track the level of dispersion in bubble-drawing clusters and dynamically split or reject clusters based on this value. We aim to address this issue in a subsequent data release.

To assess the reliability of the data presented in a quantitative way, two reliability metrics were defined:

\begin{enumerate}
\item{Hit rate: the fraction of all users that were presented with the region of sky, who drew a bubble at this location (described in Section~\ref{combine-section}). This is a measure of the reliability of the existence of a bubble; and }
\item{Dispersion: the spread in coordinates of the individual classifications ($\sqrt{\sigma_{b}^2 + \sigma_{l}^2}$, where $\sigma$ is the variance in the coordinate value). This describes the uncertainty on a bubble's location.}
\end{enumerate}

Note that these metrics do not offer any insight into the physical nature of the bubbles -- i.e. how likely a feature with the form of a bubble is to actually be an \hiir\ -- rather they reflect the level of consensus gathered from the users that a bubble appears at this location or with this form.

The dispersion measure shows the varying levels of uncertainty on the bubble positions gathered from the classifications. In addition, the clustering threshold chosen in our data processing algorithm may merge bubbles that closer inspection suggest to consist of several bubble components. Thus the dispersion metric offers one way of assessing how many bubbles may have been `lost' in data processing.

\begin{figure}
\includegraphics[width=0.48\textwidth]{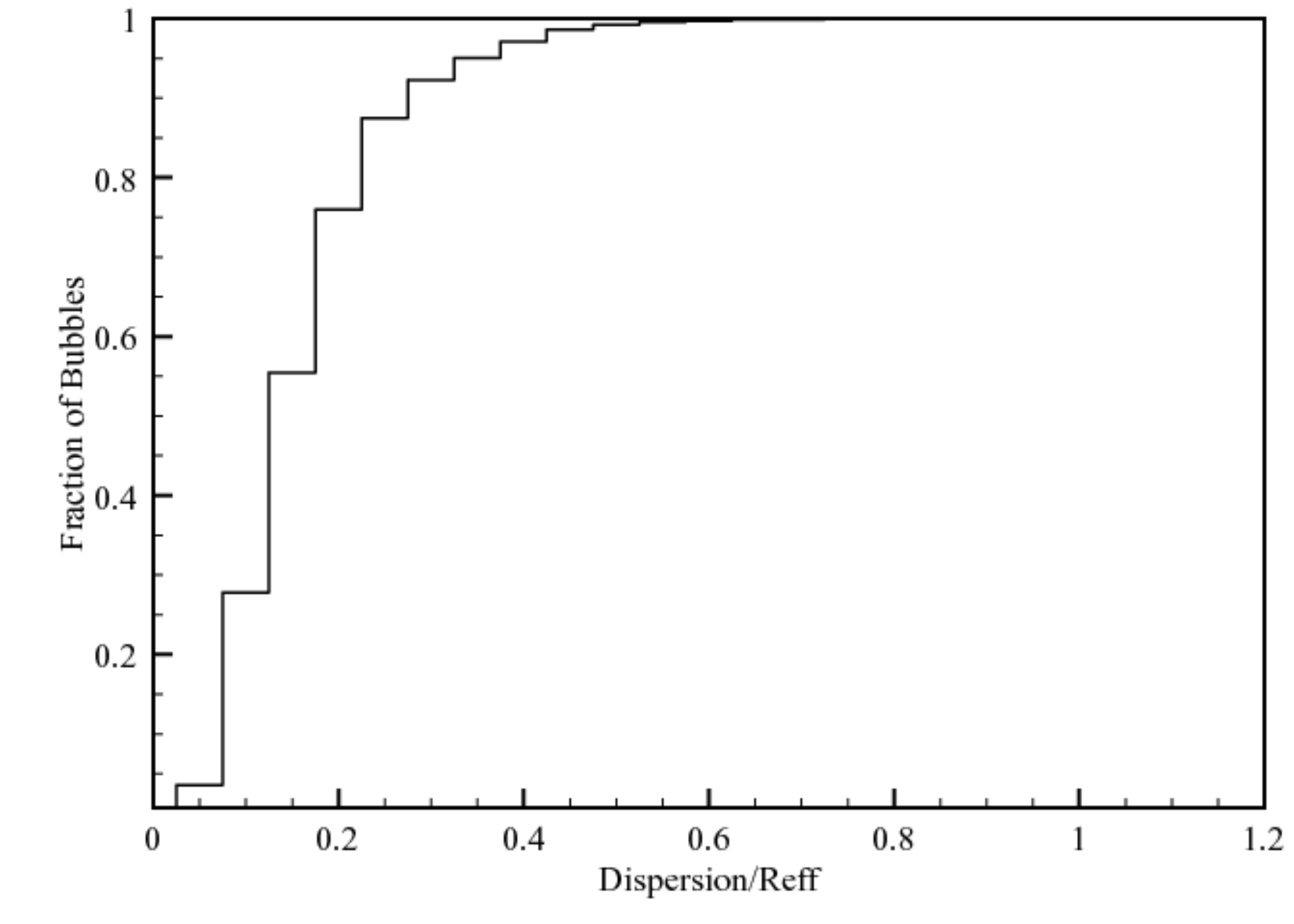}
\caption{Cumulative distribution function of all large MWP bubbles normalised to their effective radii.}
\label{dispersion}
\end{figure}

Figure~\ref{dispersion} shows the cumulative distribution function of the dispersion of all large bubbles, normalised to their effective radii. If we posit that a dispersion on a bubble's central position covering much of the bubble's effective radius is highly suggestive of multiplicity, we can estimate how many bubbles may be lost. While 89\% of large bubbles have a dispersion $< 0.5$ R$_{eff}$, 2.4\% (88 bubbles) show a dispersion $> 0.75$ R$_{eff}$. Thus assuming that each of these is highly likely to host at least one additional bubble around its rim, we can estimate that around 80-100 bubbles were artificially merged by the algorithm.

\subsection{Bubble catalogue properties}

\begin{figure*}
\includegraphics[width=0.99\textwidth]{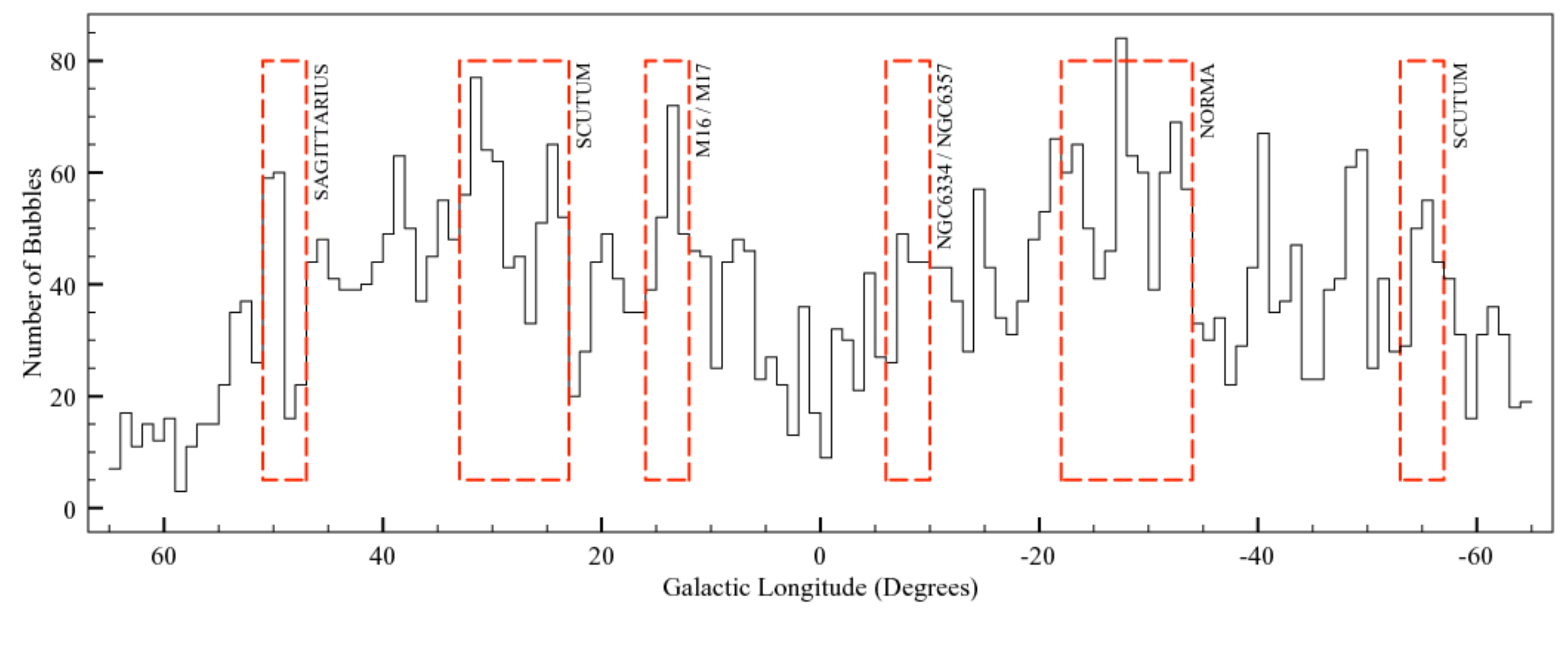}
\caption{Histogram showing distribution of combined MWP small- and large-bubble catalogues with Galactic longitude $l$. Notable line-of-site features (including Galactic spiral arms) marked as dashed boxes.}
\label{longitudearms}
\end{figure*}

The longitudinal distribution of the MWP bubbles is shown in Figure~\ref{longitudearms}. This figure shows a broad rise and fall either side of the Galactic centre, with the number of bubbles beginning low near the Galactic centre, rising, and then diminishing toward the edge of the survey at $l\pm65^{\circ}$. The notable lack of bubbles around $l=60^{\circ}$ is partly due to reaching the survey's edge at $l=65^{\circ}$ and also due to an absence of any filaments or bubbles around $l=58^{\circ}$. Figure~\ref{longitudearms} marks several notable Galactic line-of-sight features in red, adopted from \citet{ATLASGAL}. Many of the rises and falls in the number of bubbles across the range of Galactic longitude appear to derive from the large-scale structure of the Milky Way.

The drop in bubble count near the Galactic centre may be a physical effect but could also arise due to confusion from background emission toward the centre of the Milky Way. \citet{ATLASGAL} studied the distribution of submm clumps from ATLASGAL -- The APEX Telescope Large Area Survey of the GALaxy \citep{Schuller+09}. They find a large peak in the number of sources toward the Galactic centre and note that this is in contrast with current surveys of \hiirs\ \citep[e.g.][]{Anderson+11} and recent surveys of $H_{2}O$ and $CH_{3}OH$ masers \citep{Green+11,Walsh+11}.

The distribution of large bubbles with latitude is shown in Figure~\ref{latitude}. We divide the GLIMPSE/MIPSGAL survey area by longitude into northern ($l= 0^{\circ}$ to $65^{\circ}$) and southern ($l= 295^{\circ}$ to $360^{\circ}$) regions to facilitate comparison of possible morphological, positional, and size differences with CP06 and CWP07. The same profile is shown in CP06 Figure~5 and the plots share the same asymmetry toward lower latitudes.

Figures~\ref{angular-radii}, \ref{thickness} and \ref{eccentricity} show the distributions of the MWP bubble radii, thicknesses and eccentricities. Comparing these properties with those derived in CP06 reveals some differences. The MWP bubbles show a greater range in radius and thickness, though both plots show broadly similar features to the equivalents in CP06. Most strikingly, the MWP eccentricities peak at $\sim 0.35$, compared to a value of $\sim 0.65$ in CP06. This could be a result of averaging the parameters of multiple ellipses to create the MWP large bubbles, i.e. merging a large number of elliptic annuli tends toward circularity as the number and variety increases.

Figure~\ref{size-thickness} shows the distribution of the ratio of bubble thickness to outer diameter. Compared to CP06 Figure~10, the MWP catalogue displays a wider range of values, and in general the MWP bubbles are slightly thicker relative to their size (peak value at $\sim 0.4$) than those in CP06 (peak value at $\sim 0.25$). This may be partially due to the addition of 24~$\mu$m MIPSGAL data in the MWP, which often gives bubbles a `fuzzier' rim -- see Figure~\ref{reduction-example}a and b.

\subsection{Bubble Distances}

Cross-matching with \citet{Anderson+09} provides distances to 185 of the larger MWP bubbles (cross-matching as described in Section~\ref{crossmatch}). Figure~\ref{size-distance} shows these 185 bubbles plotted in terms of their diameters against (a) their distance from the Sun and (b) their distance from the Galactic centre. Figure~\ref{size-distance}a shows a slight tendency for more distant bubbles to be larger -- most likely a reflection of the selection effect whereby only very large distant bubbles are easily seen in the GLIMPSE/MIPSGAL images used in the MWP. Figure~\ref{size-distance}b shows little correlation other than reflecting the fact that fewer bubbles are seen at greater distances from the Galactic centre - this is to be expected given the longitudinal range of the MWP and the confusion effect where nearer bubbles and dust obscure more distant ones, looking toward the Galactic centre.

Figure~\ref{distance-thickness} shows that larger bubbles tend to have thinner shells relative to their diameters. This could be the effect of material cooling and condensing as the bubble expands. The apparent relationship warrants further investigation but we note caution that this relationship may be biased by the effect of a minimum size and thickness of a bubble in the MWP drawing tool.

\subsection{`Heat maps'}

In addition to the reduced bubble catalogue, a crowd-sourced `heat map' of bubble drawings has also been produced. This simple map reflects the full range of classifications placed onto the MWP images. All 520,120 bubbles drawn by all users are placed onto the sky with an opacity of 2.5\%\, meaning that 40 individuals need to have drawn over the same region for it to become fully opaque (white). Examples of these images are shown in Figures~\ref{reduction-example} and \ref{trigger-examples}. These MWP `heat maps' are available at http://data.milkywayproject.org as FITS and DS9 region files.

The MWP `heat maps' allow the bubble drawings to be explored without them needing to be reduced to elliptical annuli. Rather, the `heat maps' allow contours of overlapping classifications to be drawn over regions of the Galactic plane reflecting levels of agreement between independent classifiers. In most cases the structures outlined in these maps are photo-dissociation regions traced by 8~\um\ emission, but more fundamentally they are regions that multiple volunteers agree reflect the rims of bubbles.

\begin{figure}
\includegraphics[width=0.48\textwidth]{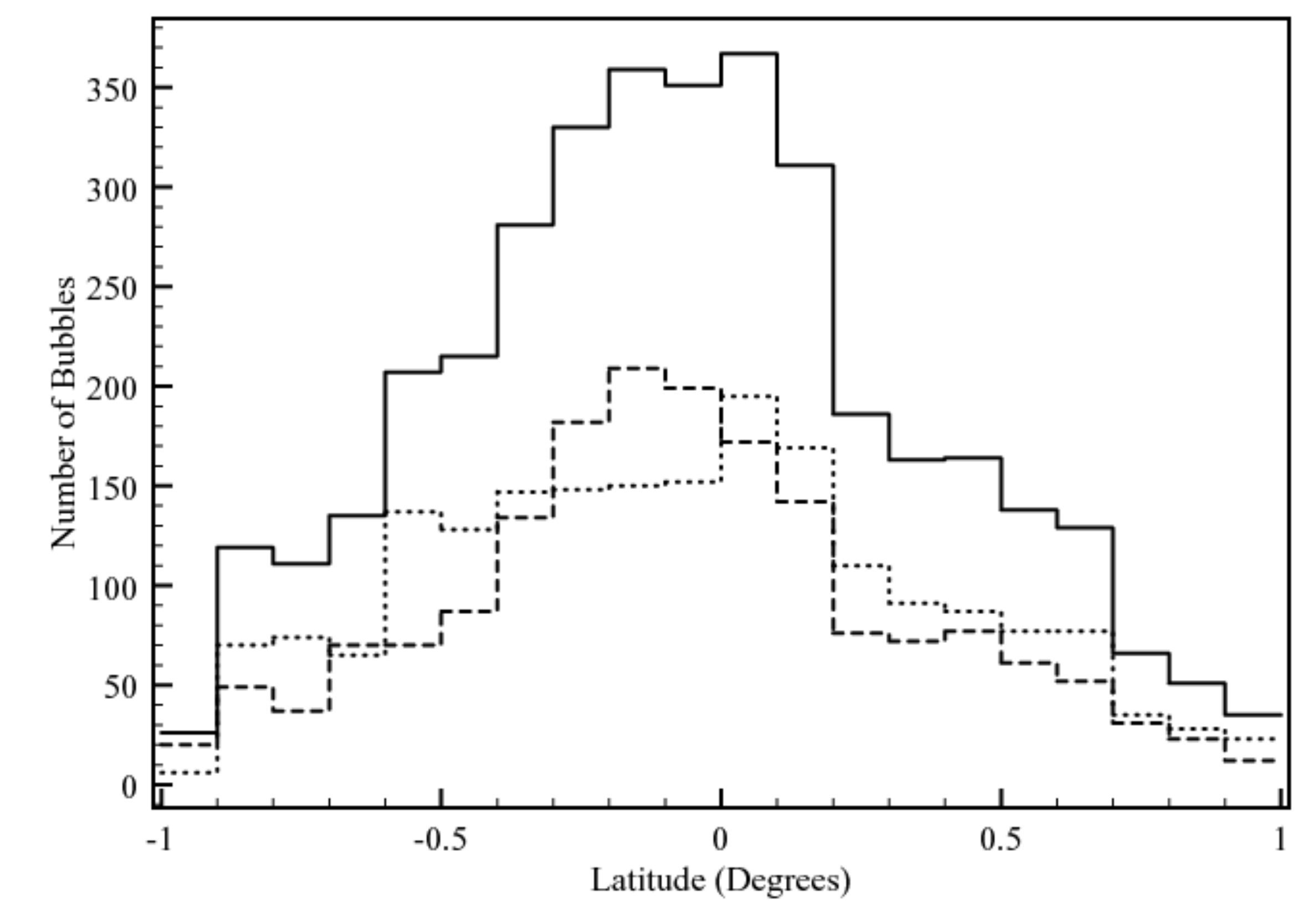}
\caption{Distribution of MWP large bubbles with $b$ (see figure 5 of CP06). In additional to the overall distribution (drawn as a solid line) the dashed line shows just southern bubbles ($l= 295^{\circ}$ to $360^{\circ}$) and the dotted line shown just northern bubbles ($l= 0^{\circ}$ to $65^{\circ}$).}
\label{latitude}
\end{figure}

\begin{figure}
\includegraphics[width=0.48\textwidth]{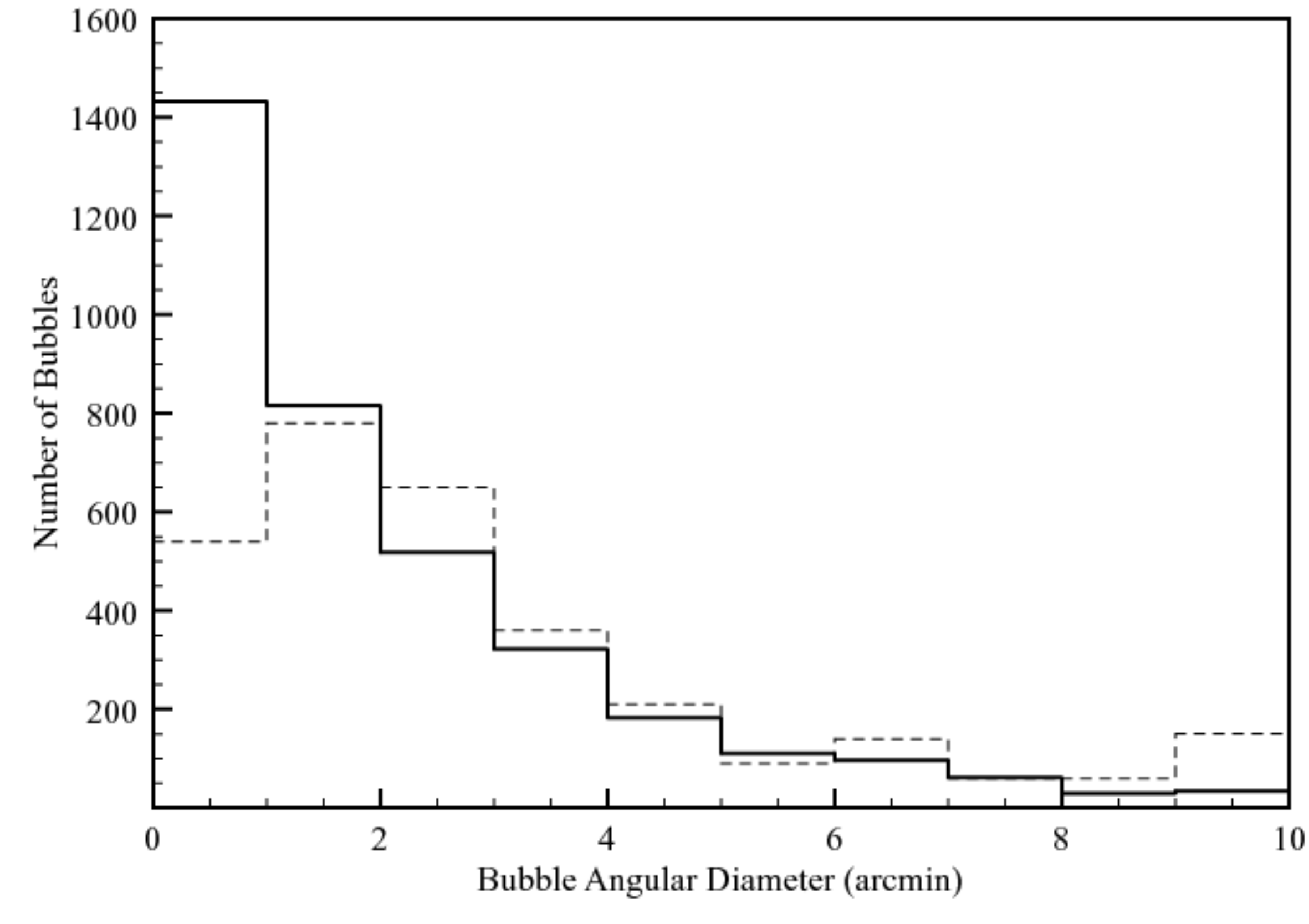}
\caption{Distribution of angular diameter of MWP large bubbles -- values from figure 6 of CP06, scaled to 10$\times$, shown as dashed line. The smallest radius users can draw is 0.45'. The largest image served to users is $1.5\times1^{\circ}$.}
\label{angular-radii}
\end{figure}

\begin{figure}
\includegraphics[width=0.48\textwidth]{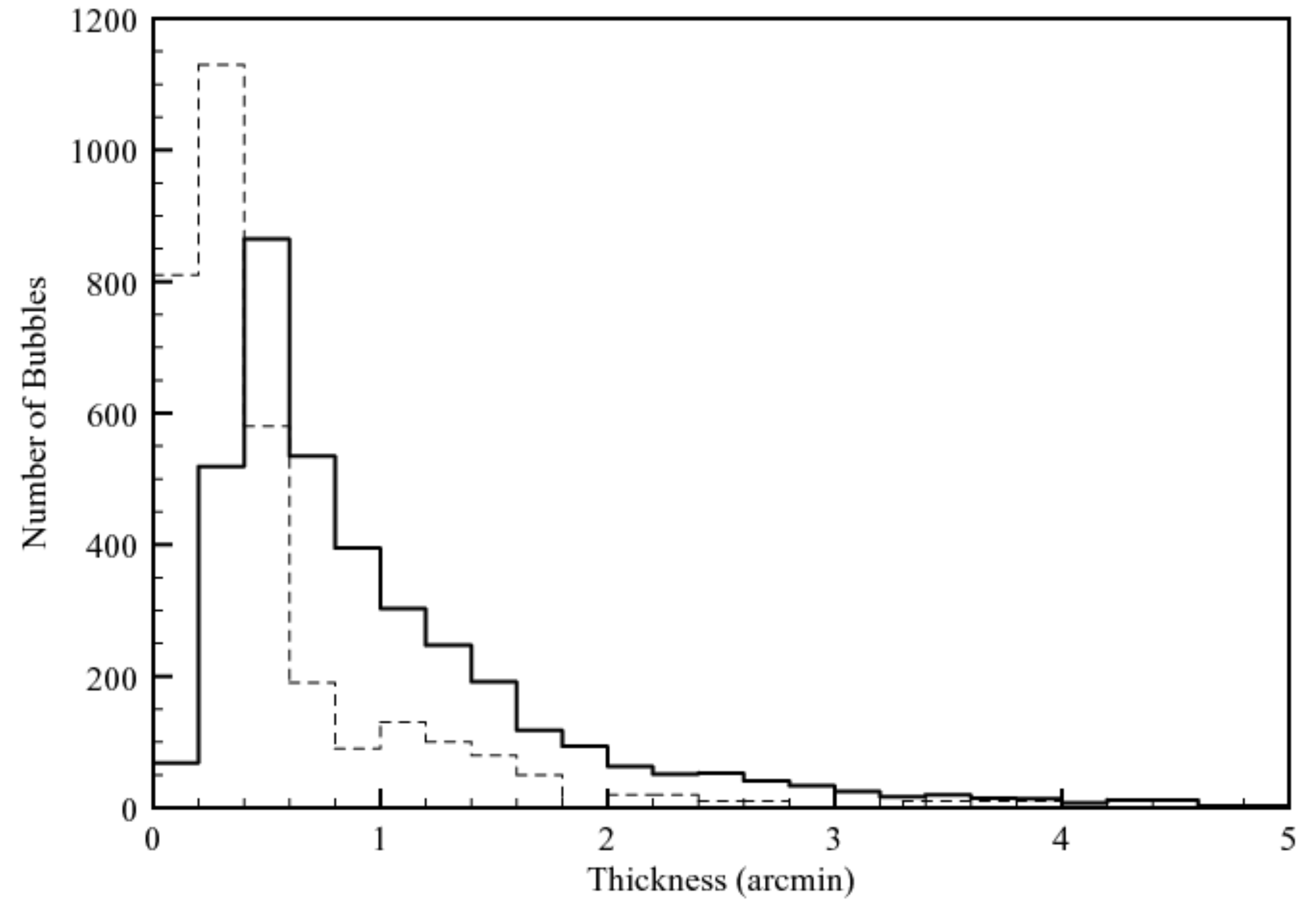}
\caption{Distribution of thicknesses of MWP large bubbles -- values from figure 9 of CP06, scaled to 10$\times$, shown as dashed line}
\label{thickness}
\end{figure}

\begin{figure}
\includegraphics[width=0.45\textwidth]{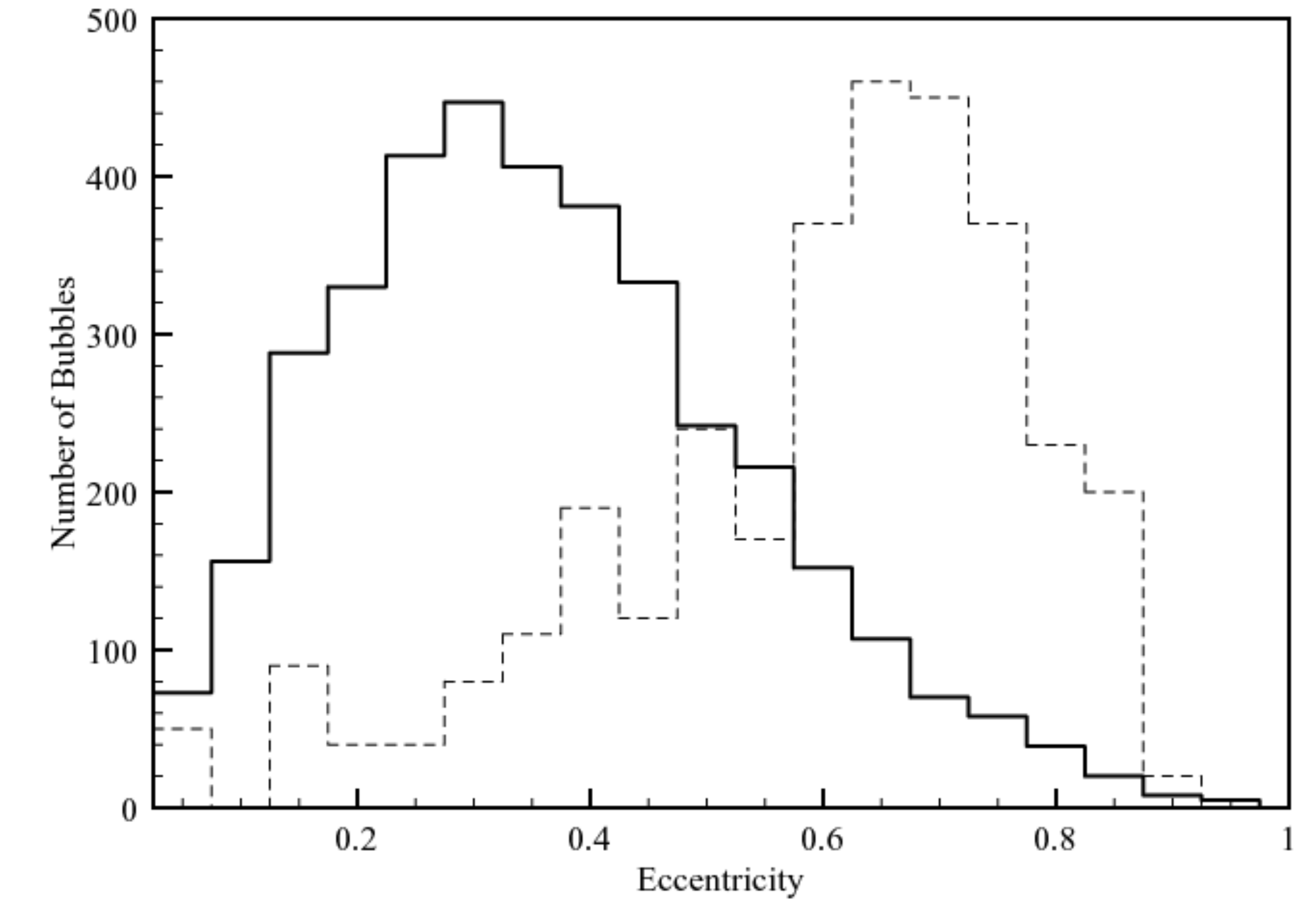}
\caption{Distribution of eccentricities of MWP large bubbles -- values from figure 13 of CP06, scaled to 10$\times$, shown as dashed line.}
\label{eccentricity}
\end{figure}

\begin{figure}
\includegraphics[width=0.48\textwidth]{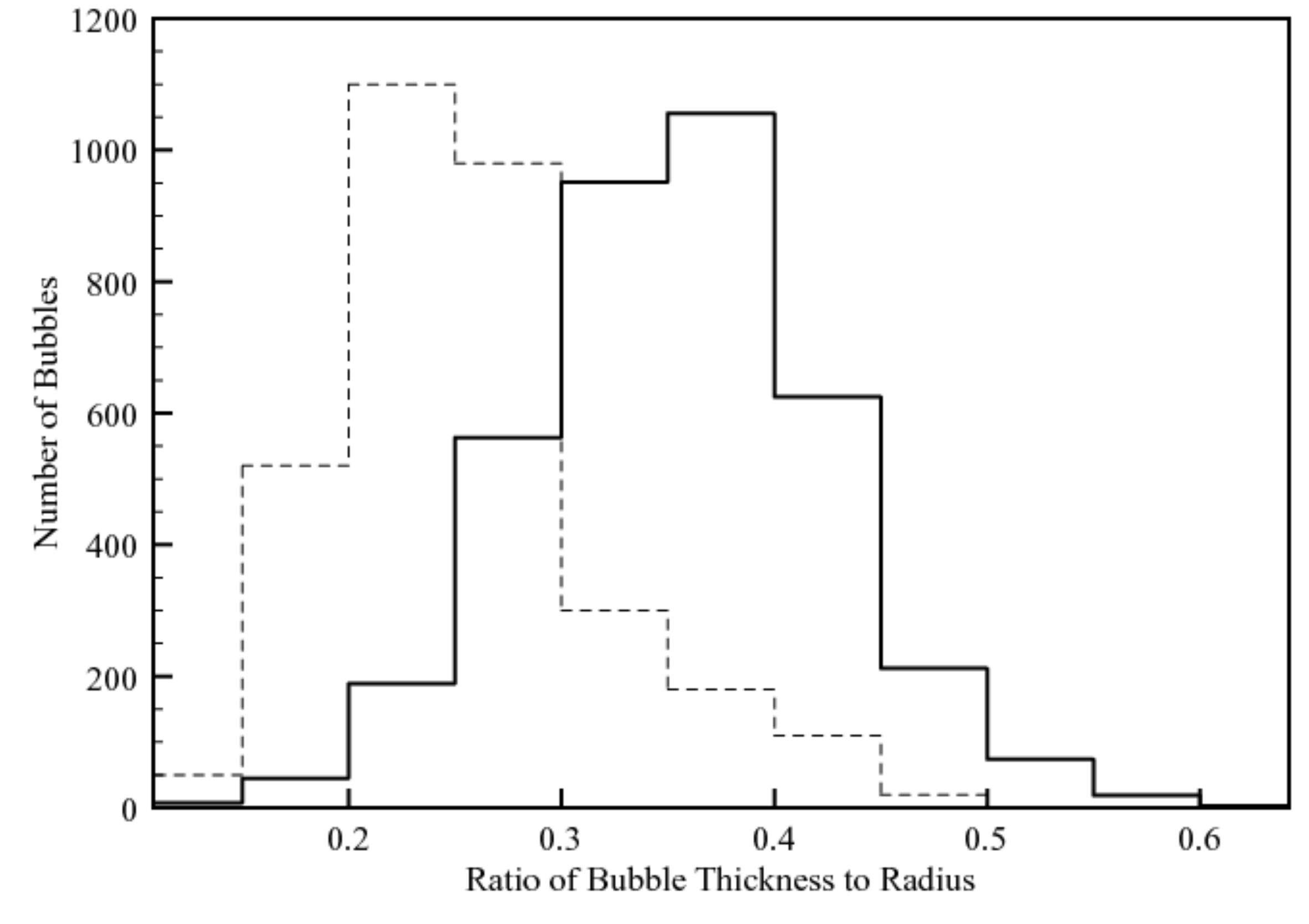}
\caption{Distribution of ratio of thickness to outer diameter  for MWP large bubbles -- values from figure 10 of CP06, scaled to 10$\times$, shown as dashed line.}
\label{size-thickness}
\end{figure}

\section{Discussion}

CP06 and CWP07 performed visual classification, using a handful of experts, of GLIMPSE data only. The MWP images include data from the MIPSGAL 24~\um\ survey, which can enhance the shape and definition of bubble-like structures (see Figure~\ref{reduction-example}). Most likely the biggest difference between the MWP and GLIMPSE studies is that tens of thousands of classifiers are involved in the MWP and most begin with no grounding in the astrophysical phenomena they are told to locate. This results in a set of unbiased classifications, sourced from multiple independent classifiers  who are able to explore and annotate the entire dataset without fatigue. 

\subsection{The nature of MWP bubbles}

Bubbles were identified in the MWP based on ringlike or arclike shapes observed by multiple independent classifiers in infrared images of the Galactic plane. Many astrophysical phenomena can give rise to such features. As CP06 noted, aside from young, massive stars several classes of evolved objects can produce bubbles in the interstellar medium (ISM), including supernova remnants (SNRs), planetary nebulae (PNe), and asymptotic giant branch (AGB) stars. SNRs and AGB stars don't excite PAH emission, thus these objects generally lack the 8~\um\ shells characteristic of the GLIMPSE bubbles. Regardless, 3 SNRs were included in the CP06 catalogue and a few bright, round 24~\um\ SNRs may be included in the MWP catalogue.

Other lines of evidence also suggest that the MWP bubbles, like the GLIMPSE bubbles, predominantly trace massive star formation. The distribution of bubbles with Galactic latitude reflects a low scale height (Figure~\ref{latitude}), similar to molecular clouds and the known Galactic OB population (see CP06). The CP06 bubble catalogue contained only 12\% of available \citet{Paladini+03} \hiirs. The MWP catalogue contains 86\% of the Paladini sources, indicating that the new catalogue is more complete.

\begin{figure*}
\subfigure[]{
\includegraphics[width=0.95\textwidth]{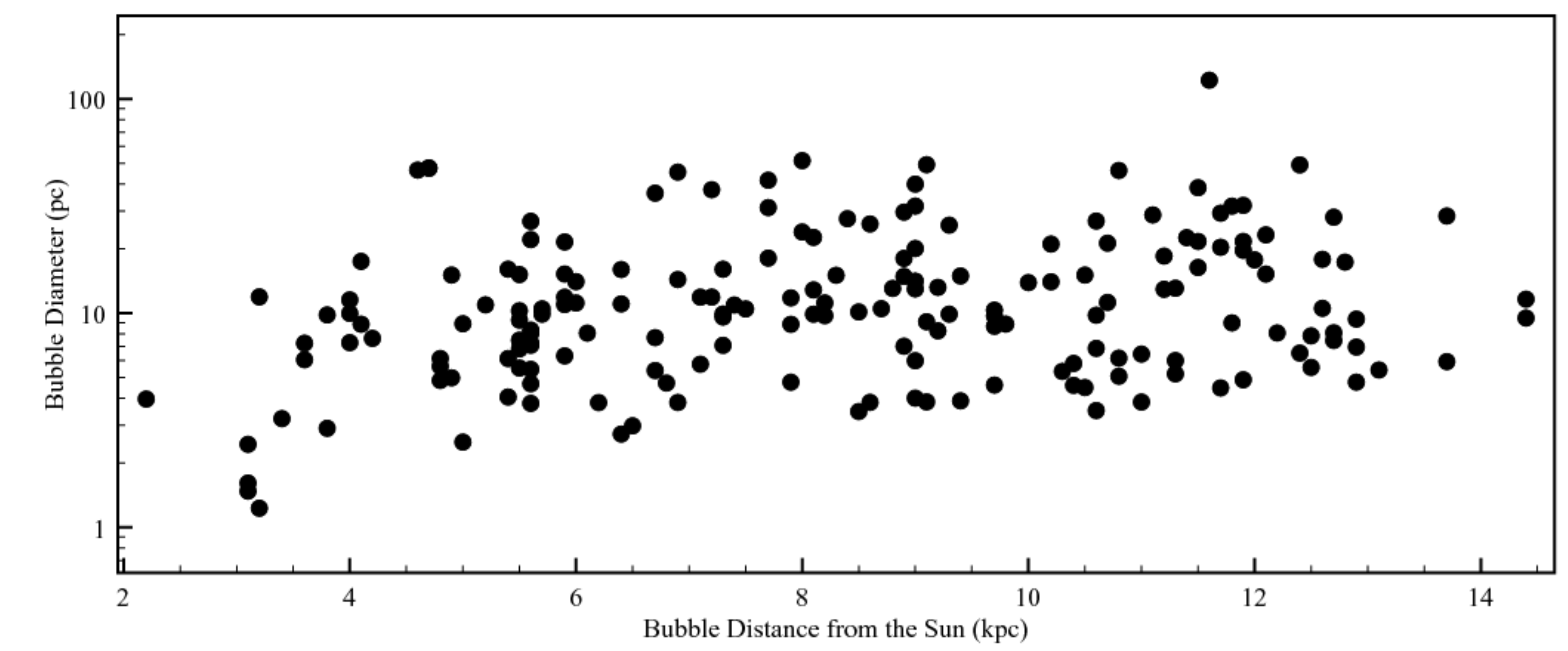}}
\subfigure[]{
\includegraphics[width=0.95\textwidth]{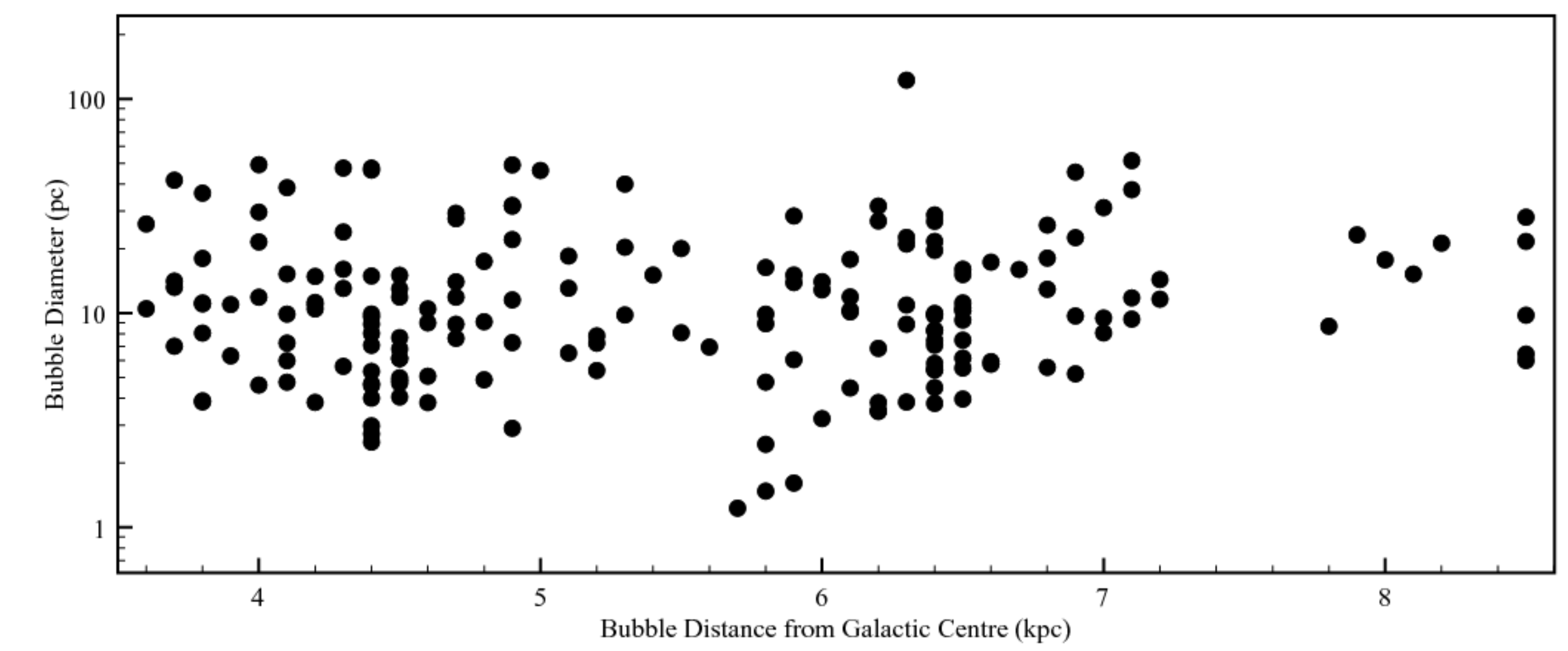}}
\caption{Bubble diameters against (a) distance from the Sun and (b) distance from the Galactic centre, for 185 MWP bubbles cross-matched with \citet{Anderson+09}.}
\label{size-distance}
\end{figure*}

A bubble is produced around a massive star when an \hiir, driven by thermal overpressure, stellar winds, radiation pressure, or a combination of these feedback mechanisms, expands into the surrounding cold ISM, sweeping up gas and dust into a dense shell surrounding a low-density, evacuated cavity \citep{Weaver+77,Garcia-Segura+Franco96,Arthur+11,Draine11}. The relative contributions of different feedback mechanisms likely depend on the properties of the driving star(s), with the most massive, early O-type stars combining powerful stellar winds with high UV luminosities producing `wind-blown bubbles,' while lower-mass, late-O and B dwarfs give rise to `classical' \hiirs\ powered by UV photons alone \citep[CWP07;][]{WP08}. 
\citet{1975ApJ...200L.107C} and \citet{Weaver+77} derived analytic solutions for the expansion of stellar wind-blown bubbles into a uniform low-density medium. More recent modelling efforts have included the effects of ionising radiation and the stellar winds (e.g. \citealt{2001PASP..113..677C,Draine11}).

The wind-blown bubbles around massive stars produced by these models display the following general structure: 

\begin{itemize}
\item An inner cavity cleared rapidly by a freely flowing hypersonic stellar wind;
\item A high-temperature region of shocked stellar wind material ($T > 10^6$ K);
\item A shell of shocked, photo-ionised gas ($T\sim 10^4$~K);
\item An shell of non-shocked, ionised gas ($T\sim 10^4$~K).
\item An outer shell of neutral material.
\end{itemize}

The bright PAH emission in the PDRs surrounding \hiirs\ produces the bright 8~\um\ bubble rims, while dust mixed with the ionised gas and heated by the hard radiation field produces 24~\um\ emission interior to the bubbles \citep{Povich+07,WP08,2010ApJ...713..592E,WHM+10}.  These considerations guided our choice of multiband colour combination for the MWP images, which visually bias identification toward \hiirs . This is consistent with the cross-matching results in Section~\ref{crossmatch}.

\citet{WP08} proposed that wind-blown bubbles with cleared, central cavities tend to produce a toroidal or arc-shaped morphology in the 24~\um\ emission, while this emission is more likely to be centrally peaked in classical \hiirs. \citet{Draine11} found that radiation pressure alone can produce cleared central cavities in \hiirs, but suggested cases where winds must also play a role. A wide range of 24~\um\ morphologies is apparent among the MWP bubbles.

\begin{figure}
\includegraphics[width=0.48\textwidth]{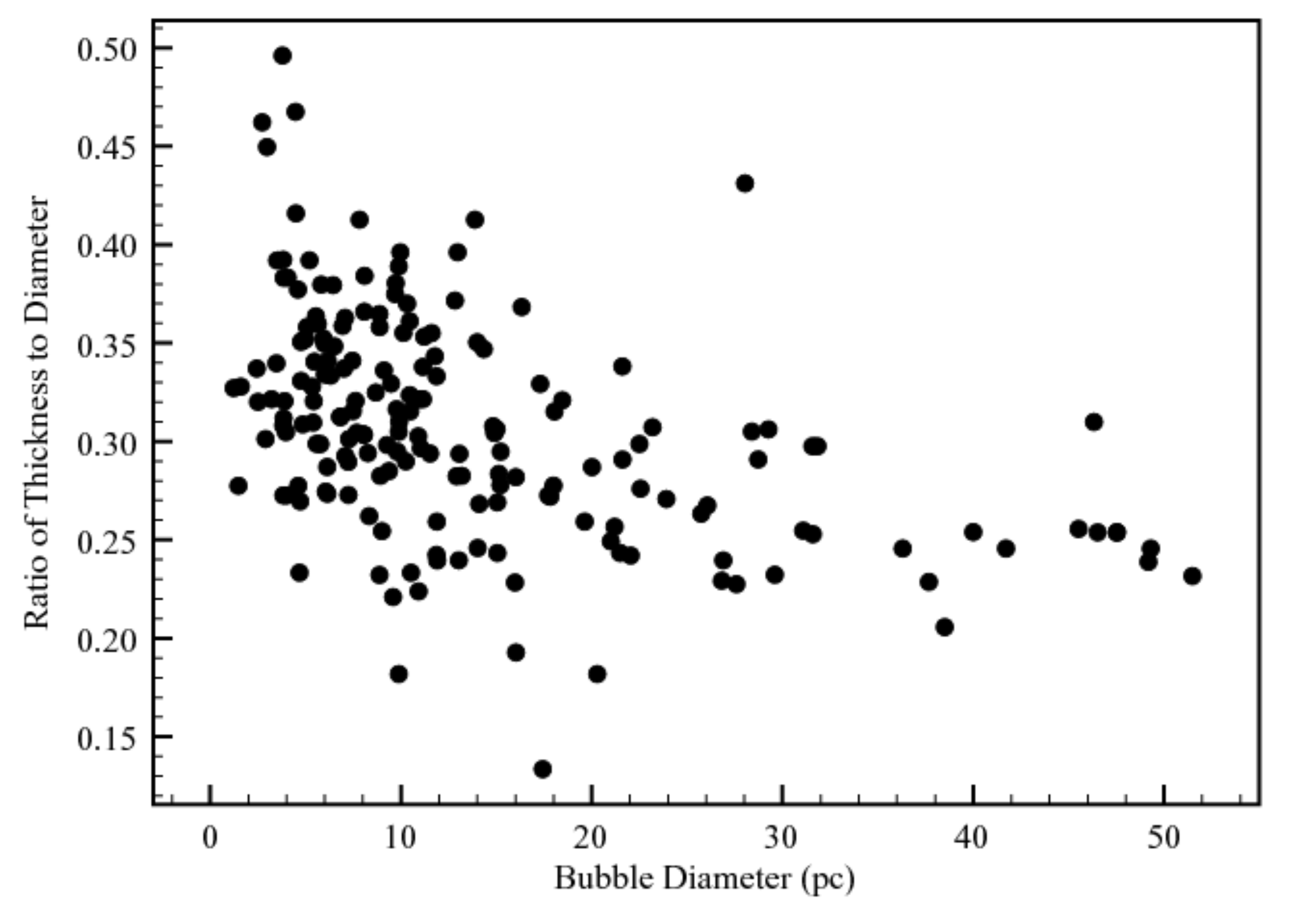}
\caption{Variation of the thickness-diameter ratio of large MWP bubbles (Table~\ref{bubble-sample}) with their physical size, for 185 MWP bubbles cross-matched with \citet{Anderson+09}.}
\label{distance-thickness}
\end{figure}

The models described above consider the effects of a single star on its surrounding uniform-density medium. Massive stars form preferentially in clusters within molecular clouds with highly non-uniform densities. \citet{2003ApJ...594..888F} performed two-dimensional radiation hydrodynamics simulations to study the combined influence of a massive star's stellar wind and its ionising radiation on the evolution of the circumstellar cloud material.  \citet{2006ApJ...647..397M} simulated the evolution of an \hiir\ surrounding a single O star, focusing on the interaction of the ionisation front with the turbulent cloud medium. More recently, \citet{Arthur+11} performed magnetohydrodynamic (MHD) simulations of \hiirs\ around OB stars expanding into a turbulent, magnetised medium. All of these simulations show strong inhomogeneities around the boundary of the expanding shell, reproducing the clumps, globules and filaments observed in \hiirs\ at optical or IR wavelengths and seen in the MWP bubbles.

\subsection{Multiple bubbles and potentially triggered star formation}

A major application of the CP06 and CPW07 bubble catalogs has been triggering studies \citep[e.g.][and references therein]{Deharveng+08, WP08, WHM+10, Zavagno+10}, as the Churchwell catalogs provided a large sample of (mostly previously unknown) regions of potential triggering by massive stars.  In any individual source, difficulties with establishing cause-and-effect are significant, and the Churchwell catalogs provided a basis for triggering investigations to move beyond case studies of individual objects to statistics. 

Bubbles serve as laboratories to test theories of sequential, massive star formation triggered by massive star winds and radiation pressure \citep{EL77, Whitworth+94}. CP06 and CWP07 noted that a (small) fraction of bubbles exhibited hierarchical structure, meaning that one or more small, `daughter' bubbles were found on the rims of, or projected inside, larger, `parent' bubbles. The prevalence of triggering is a key unresolved question in the study of massive star formation, with important implications for extragalactic studies as well as detailed star formation physics.  Quantifying triggering allows the formulation of galaxy-scale star formation rate prescriptions for use in simulations, particularly if the physical conditions in triggered regions affect the IMF, as suggested by \citet{Whitworth+94} and \citet{Dale09}.

\citet{Rahman+Murray10}, however, observed that Galactic giant \hiirs\ were arranged on the rims of very large (up to 100 pc diameter), mid-IR bubble structures associated with the most luminous sources of free-free emission in the Galaxy, suggesting large-scale triggering driven by very massive clusters and OB associations. 

However, whether these structural features are a direct result of the expanding shell's interaction with the molecular cloud, or whether the clearing of material simply reveals the underlying turbulent cloud structure, is not yet clear. \citet{Beaumont+Williams10} proposed that the 3-dimensional geometry of bubbles in the CP06 catalogue, when traced by molecular gas, resembles flattened rings rather than spherical shells, and they noted that this geometry could reduce the efficiency of triggering.

From results of smoothed particle hydrodynamic (SPH) simulations of the ionising radiation from massive stars or small protoclusters, \citet{2011MNRAS.414..321D} argue that ISM bubbles are features of the turbulent nature of the molecular clouds rather than shells created by feedback from massive stars and clusters. Their simulation involves multiple clusters within a very massive molecular cloud. In this regard their simulation is more analogous to a region such as the Carina Nebula than to the kind of bubbles predominantly seen in the MWP, where the ionisation is dominated by one or two stars.

Most studies to date have asked the question, `what percentage of bubbles show evidence for triggered star formation?' Understanding Galactic-scale star formation in the Milky Way, however, requires answering not this question but its corollary: `what fraction of all (massive) star formation is triggered?'. \citet{Thompson+11} have recently made the first effort to address this question, using the Red MSX Source (RMS) database of massive YSOs and the bubbles from CP06 and CWP07, but \emph{the major caveat of their analysis is that the Churchwell bubble catalogs are incomplete.}  By identifying an order of magnitude more bubbles, and providing a reliability indicator in the form of the hit rate, the MWP bubble catalog greatly ameliorates this problem.  

In addition, \citet{Oey+05} have suggested that the most convincing candidates for triggered star formation are regions where three-generation hierarchies can be established. The MWP has identified a much larger number of broken, old bubbles and small bubbles than the CP06 and CWP07 catalogs. As such, it is an excellent dataset for searching for multiple hierarchies. Indeed, 29\% of MWP bubbles have hierarchy flags of 1 or 2, indicating bubbles that are located on or within larger bubbles and bubbles that have smaller bubbles situated on or within them.

Additional analysis to assess triggering related to MWP bubbles (using multiwavelength datasets) is ongoing and will be the subject of a subsequent paper.

\begin{figure*}
\begin{center}
\subfigure[$l$=332.6$^{\circ}$, $b$=-0.68$^{\circ}$, $zoom$=2 (Heat Map)]{
\includegraphics[width=0.45\textwidth]{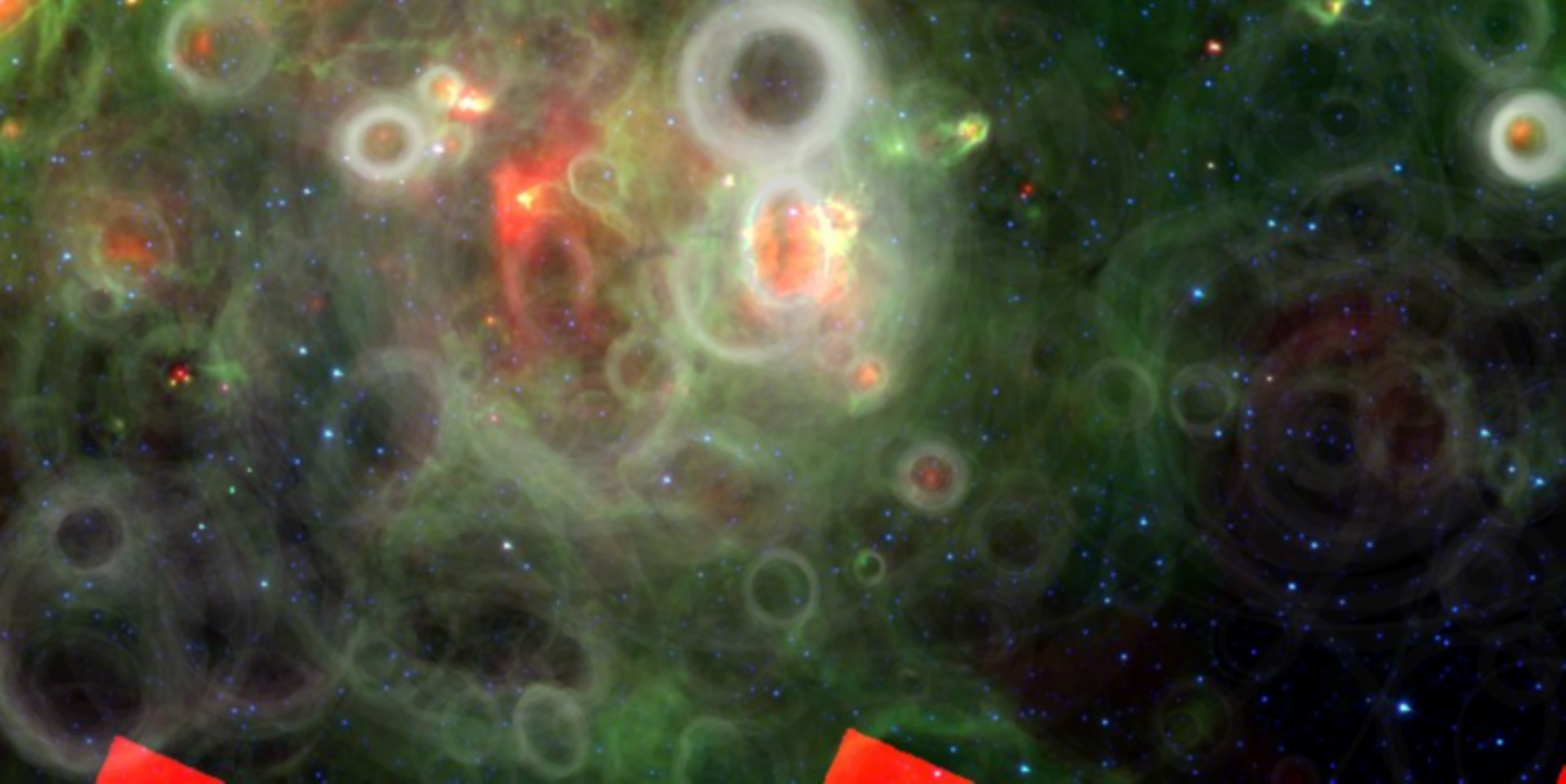}}
\subfigure[$l$=332.6$^{\circ}$, $b$=-0.68$^{\circ}$, $zoom$=2 (Catalogue)]{
\includegraphics[width=0.45\textwidth]{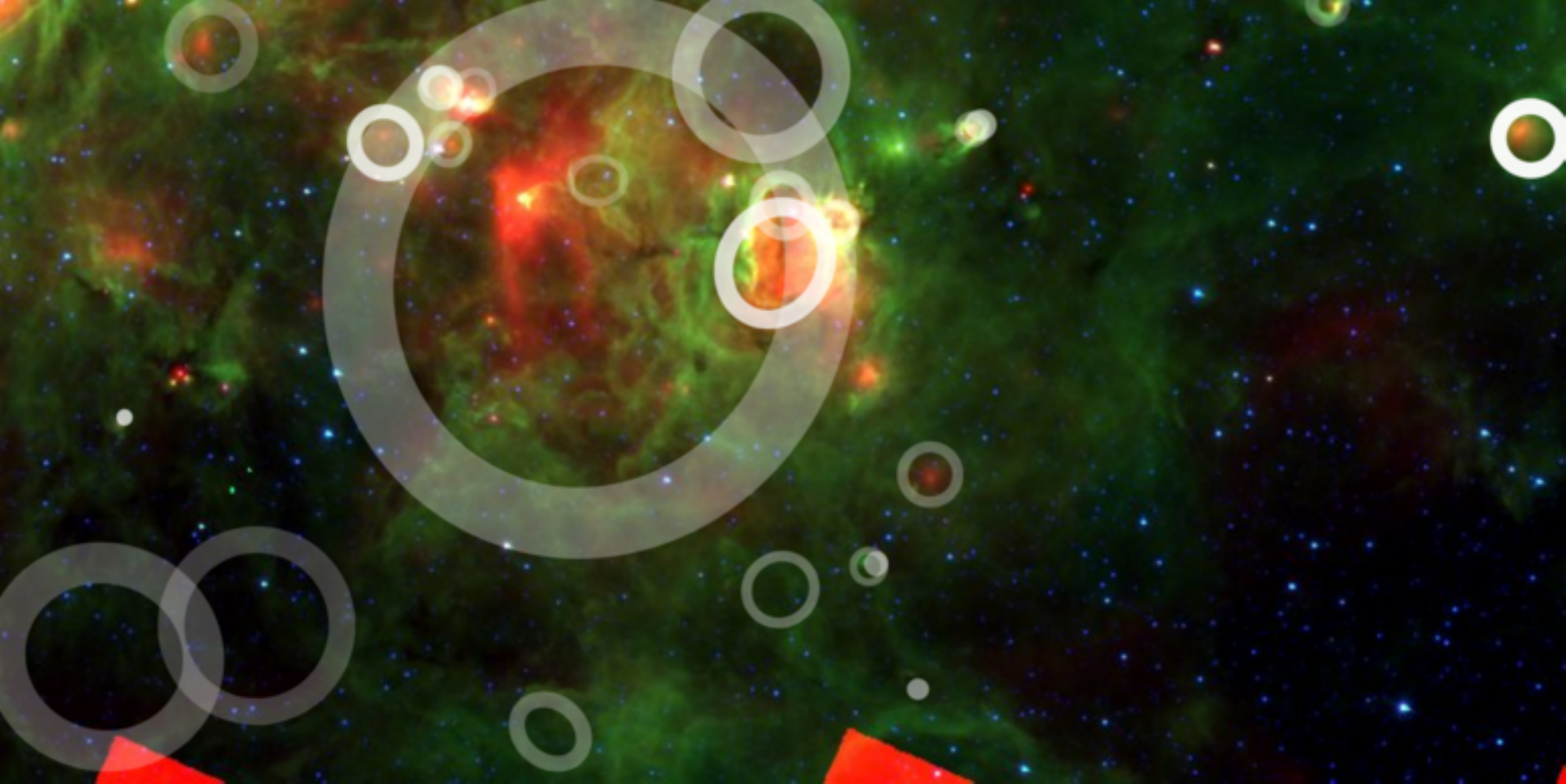}}
\subfigure[$l$=19.1$^{\circ}$, $b$=-0.44$^{\circ}$, $zoom$=2 (Heat Map)]{
\includegraphics[width=0.45\textwidth]{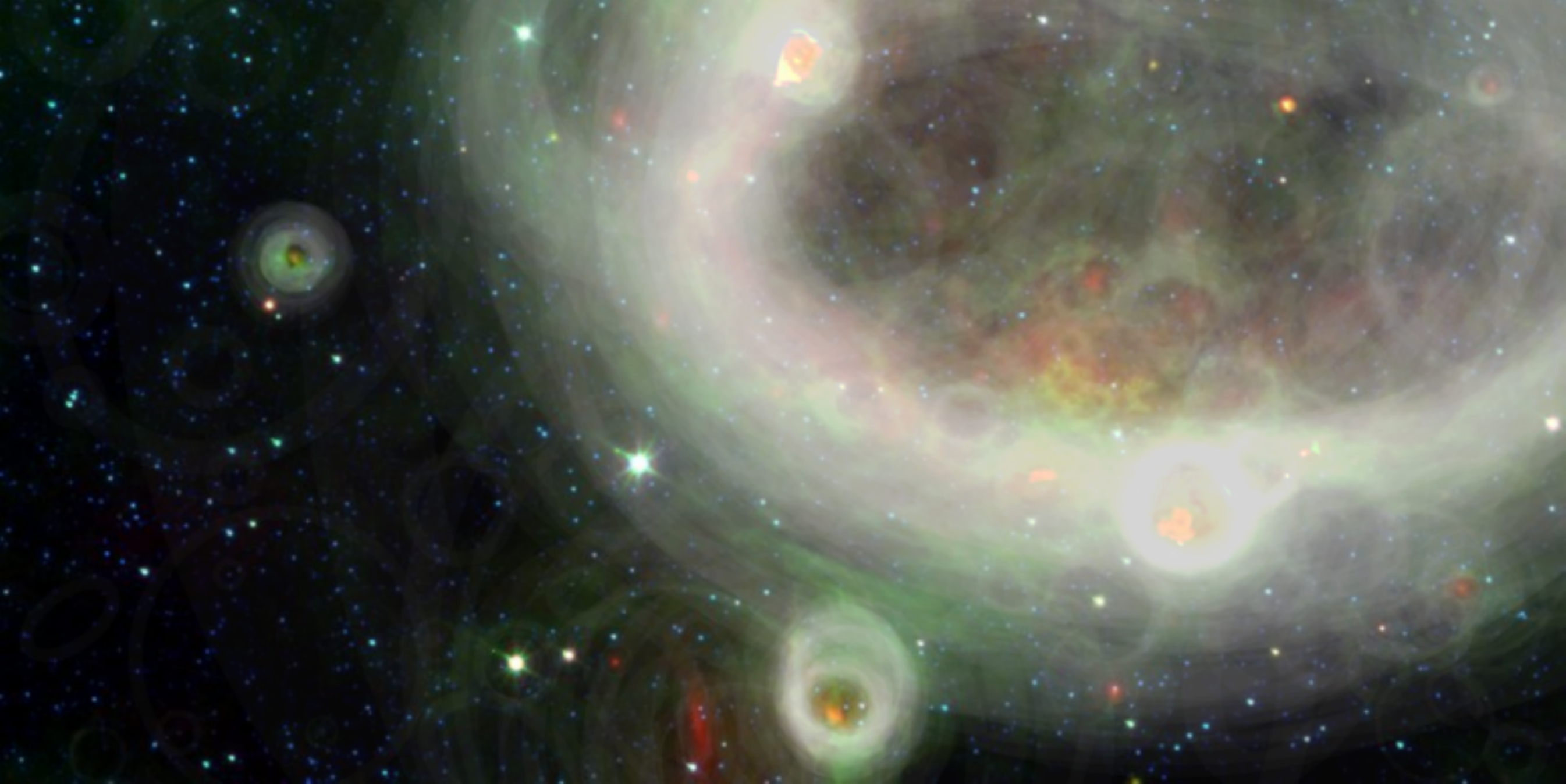}}
\subfigure[$l$=19.1$^{\circ}$, $b$=-0.44$^{\circ}$, $zoom$=2 (Catalogue)]{
\includegraphics[width=0.45\textwidth]{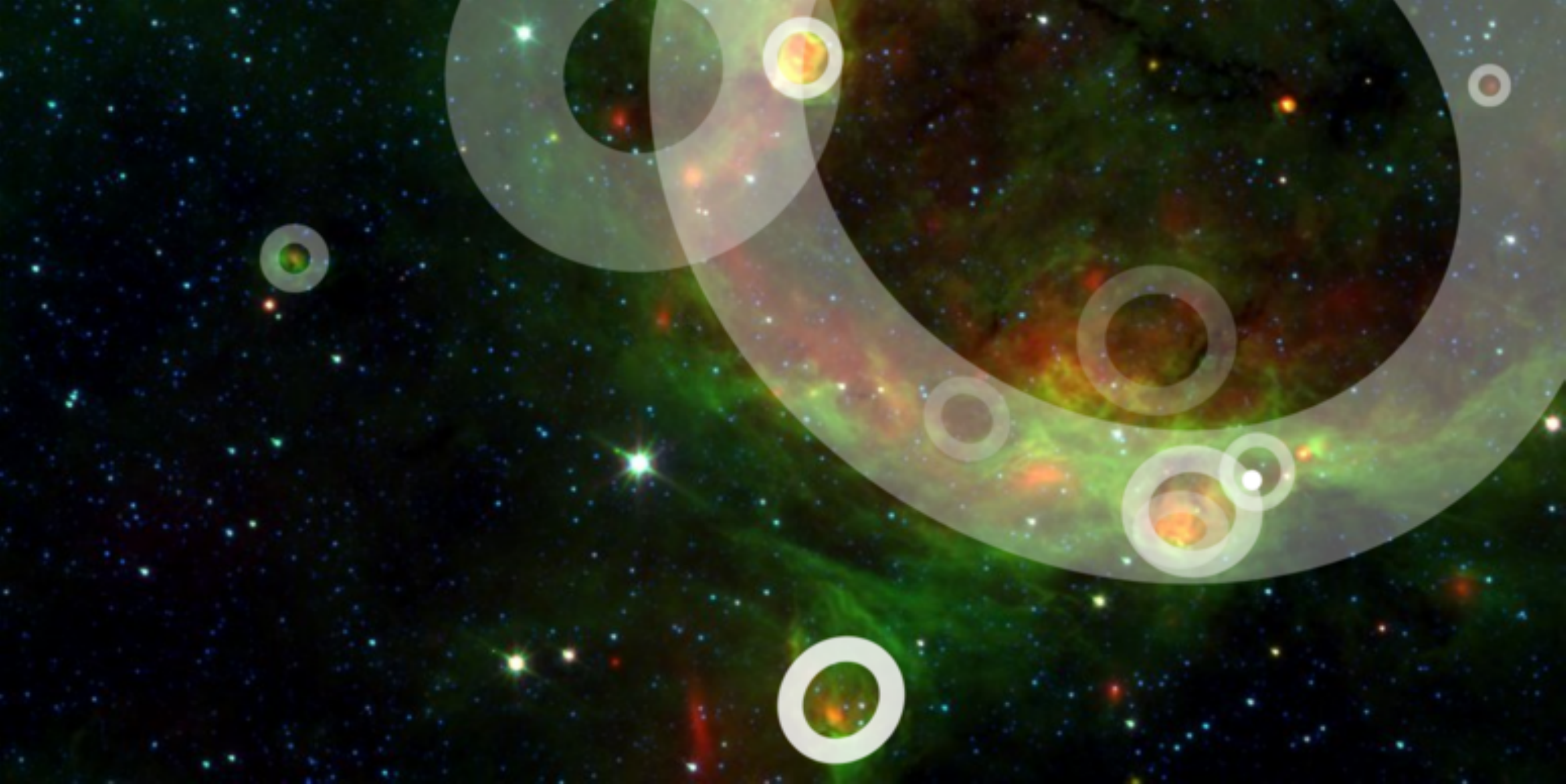}}
\subfigure[$l$=31.8$^{\circ}$, $b$=-0.06$^{\circ}$, $zoom$=2 (Heat Map)]{
\includegraphics[width=0.45\textwidth]{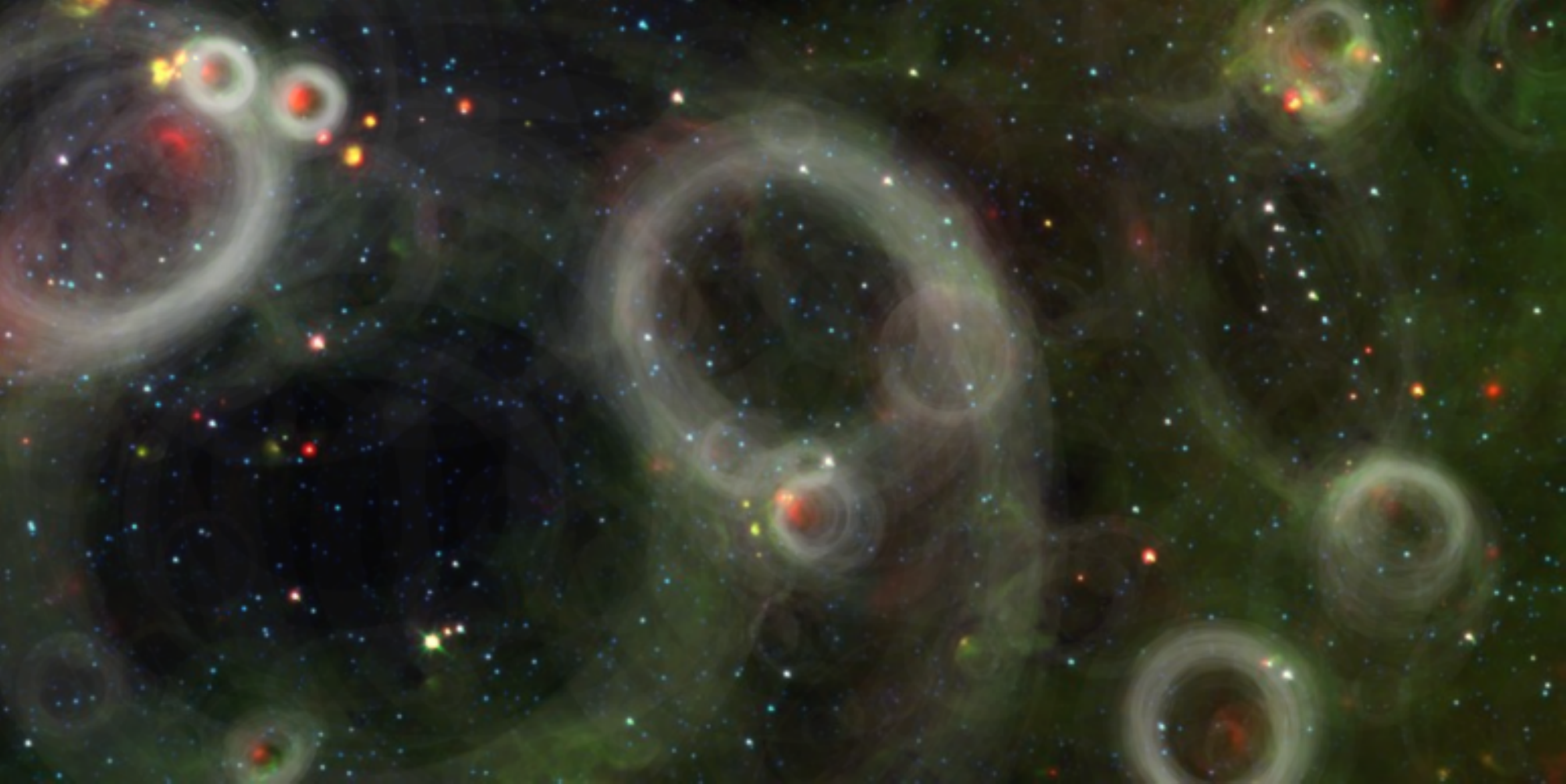}}
\subfigure[$l$=31.8$^{\circ}$, $b$=-0.06$^{\circ}$, $zoom$=2 (Catalogue)]{
\includegraphics[width=0.45\textwidth]{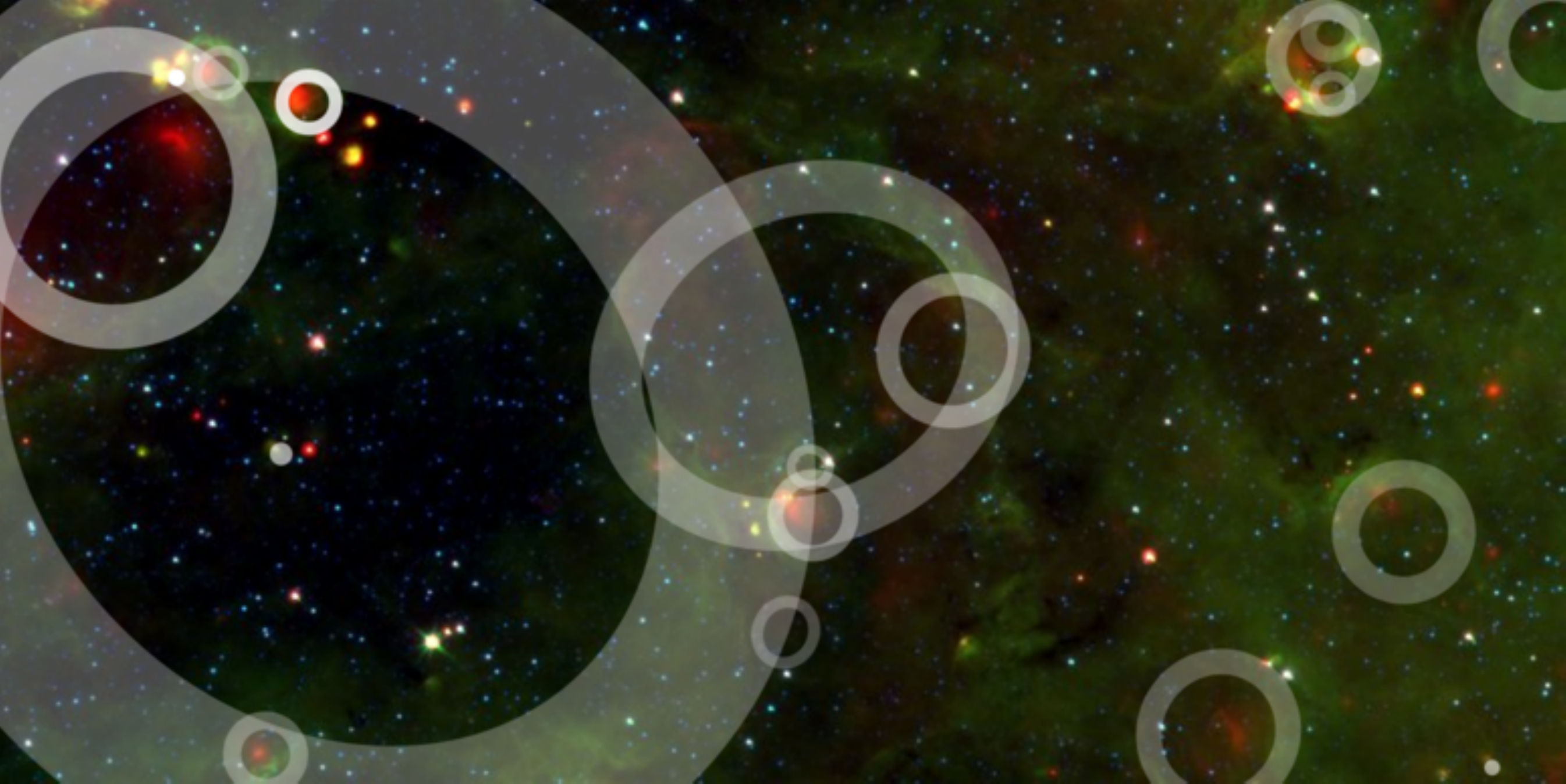}}
\caption{Examples of sites showing potential evidence of triggering. In each case the `heat map' and 	reduced data is shown overlaid on the MWP image. Coordinates are image centres, image sizes are indicated by the zoom level ($zoom$). Zoom levels 1, 2 and 3 refer to images of $1.5\times1^{\circ}$, $0.75\times0.375^{\circ}$ and $0.3\times0.15^{\circ}$ respectively. Colour figure available online.}
\label{trigger-examples}
\end{center}
\end{figure*}

\section{Summary and future work}

A new catalogue of \bubblecount\ infrared bubbles has been created through visual classification via the MWP website. Bubbles in the new catalogue have been independently measured by at least 5 individuals, producing averaged parameters for their position, radius, thickness, eccentricity and position angle. Citizen scientists have independently rediscovered the locations of 86\% of the \citet{CP06} and \citet{CWP07} bubble catalogues and 96\% of the \citet{Anderson+11} \hiir\  catalogue, whilst finding an order of magnitude more objects.

The MWP bubble catalog constitutes a resource that, in combination with other recent and ongoing Galactic Plane surveys of star-formation tracers (including the Methanol Multibeam Survey (MMB), the Bolocam Galactic Plane Survey (BGPS), ATLASGAL, and HiGAL), has the potential to provide sufficient statistics to address the question of how prevalent, and important, triggered star formation really is. In addition we hope that this new resource will complement these surveys as a tracer of massive star formation on Galactic scales.

Also outlined is the creation of a `heat map' of star-formation activity in the Galactic plane. This online resource provides a crowd-sourced map of bubbles and arcs in the Milky Way, and should enable better statistical analysis of nearby star-formation sites.

Additonal papers are currently being prepared to outline catalogues of `green knots', dark nebulae, star clusters, galaxies and `fuzzy red objects' that have also been created by the MWP's community of citizens scientists. Similarly, we anticipate a second, refined bubble catalogue incorporating not only better data-reduction techniques but also 100,000s more bubble drawings by volunteers.

\section*{Acknowledgments}

This publication has been made possible by the participation of more than 35,000 volunteers on the Milky Way Project. Their contributions are acknowledged individually at http://www.milkywayproject.org/authors. We would like to thank Bob Benjamin for his helpful suggestions.

The Milky Way Project, and R.J.S. were supported by The Leverhulme Trust. The 'Talk' discussion tool used in the MWP was developed at the Adler Planetarium with support from the National Science Foundation CDI grant : DRL-0941610. M.S.P. was supported by a National Science Foundation Astronomy \& Astrophysics Postdoctoral Fellowship under award AST-0901646.

C.J.C. is supported by an NSF Astronomy and Astrophysics Postdoctoral Fellowship under award AST-1003134. Support for the work of K.S. was provided by NASA through Einstein Postdoctoral Fellowship grant number PF9-00069 issued by the Chandra X-ray Observatory Center, which is operated by the Smithsonian Astrophysical Observatory for and on behalf of NASA under contract NAS8-03060.

This work is based on observations made with the Spitzer Space Telescope, which is operated by the Jet Propulsion Laboratory, California Institute of Technology under a contract with NASA.

\bibliographystyle{apj}
\bibliography{mwpref}

\clearpage

\appendix{}

\end{document}